\documentclass[letterpaper]{article} % DO NOT CHANGE THIS
\usepackage[]{aaai25}  % DO NOT CHANGE THIS
\usepackage{times}  % DO NOT CHANGE THIS
\usepackage{helvet}  % DO NOT CHANGE THIS
\usepackage{courier}  % DO NOT CHANGE THIS
\usepackage[hyphens]{url}  % DO NOT CHANGE THIS
\usepackage{graphicx} % DO NOT CHANGE THIS
\usepackage{natbib}  % DO NOT CHANGE THIS AND DO NOT ADD ANY OPTIONS TO IT
\usepackage{caption} % DO NOT CHANGE THIS AND DO NOT ADD ANY OPTIONS TO IT
\usepackage{algorithm}
\usepackage{algorithmic}
\usepackage{booktabs} 
\usepackage{multirow}
\usepackage{dcolumn}
\usepackage{multicol}   % multiple newspaper‑style columns
\usepackage{enumitem}   % list cosmetics
\usepackage{caption}    % captions without floats
\usepackage{longtable}  % for the alternative 3‑column table

\usepackage[font={small}]{subfig} % 
\usepackage{soul}

\usepackage{newfloat}
\usepackage{listings}
\DeclareCaptionStyle{ruled}{labelfont=normalfont,labelsep=colon,strut=off} % DO NOT CHANGE THIS
\floatstyle{ruled}
\newfloat{listing}{tb}{lst}{}
\floatname{listing}{Listing}

\usepackage{xcolor}

\title{Disagreement is Disappearing on U.S. Cable Debate Shows}

\begin{document}

\author{
S M Mehedi Zaman, 
Kiran Garimella
}

\affiliations{
Rutgers University, USA \\
sm.mehedi.zaman@rutgers.edu, kiran.garimella@rutgers.edu
}

%%
%% end of the preamble, start of the body of the document source.

%\setlength\titlebox{1.5in}

\maketitle

\begin{abstract}
Prime-time cable news programs are a highly influential part of the American media landscape, with top-rated opinion shows attracting millions of politically attentive viewers each night. In an era of intense political polarization, a critical question is whether these widely-watched “debate” shows foster genuine discussion or have devolved into partisan echo chambers that deepen societal divides. While these programs claim to air competing viewpoints, no large-scale evidence exists to quantify how often hosts and guests actually disagree. Measuring these exchanges is a significant challenge, as live broadcasts contain overlapping speakers, sarcasm, and billions of words of text.
To address this gap, we construct the first speaker-resolved map of agreement and disagreement across U.S. cable opinion programming. Our study assembles over 21,000 episodes from 24 flagship shows on Fox News, MSNBC, and CNN from 2010–2024, segmenting them into host-guest turns and labeling 2.13 million turn-pairs using a high-fidelity large-language-model classifier.
%Prime‑time “debate” shows claim to air competing viewpoints, yet no large‑scale evidence quantifies how often hosts and guests actually disagree. Measuring such exchanges is difficult: live broadcasts contain overlapping speakers, sarcasm, and billions of captioned words. 
%We construct the first speaker‑resolved map of agreement and disagreement across U.S. cable opinion programming. The study assembles 17,000 episodes spanning from 2010–2024 from 24 Fox News, MSNBC, and CNN shows, segments them into host–guest turns, and labels 2.13 million turn‑pairs using a large‑language‑model classifier.
We present three findings: (1) the proportion of disagreement/debate on prime time shows a consistent downward trend, dropping by roughly one‑third between 2017 and 2024; (2) on‑air challenge is partisan and asymmetric—conservatives seldom face push‑back on Fox, liberals seldom on MSNBC, with CNN declining toward the midpoint; (3) polarizing issues such as abortion, gun rights, and immigration attract the least disagreement. The work contributes a public corpus, an open‑source stance pipeline, and the first longitudinal evidence that televised “debate” is retreating from genuine discussion. By transforming into platforms for partisan affirmation, these shows erode the cross-cutting cleavages essential for a pluralistic society, thereby intensifying affective polarization.

\end{abstract}

\section{Introduction}

Televised debate shows---programmes built around a host who interviews or spars with one or more guests about the political story of the day—are advertised as democracy’s front‑row seats to an argument. These broadcasts occupy the most valuable real estate in cable schedules: the prime‑time block (8p.m.–11p.m. Eastern), when total viewership peaks and advertising rates spike. In 2024, Fox News, MSNBC, and CNN together delivered roughly forty hours of such programming each week, and seven of the ten highest‑rated cable slots were Fox opinion shows, each drawing well over two million nightly viewers~\cite{deadline2024cable}. Because the audience is both large and politically attentive, elites treat these discussions as signals---Presidents quote them, lawmakers schedule hearings around them, and social‑media influencers amplify their clips~\cite{gertz2020study}. If the “debate” has quietly become one‑sided affirmation, the entire information chain downstream is fed from an ideological echo.

The puzzle, then, is whether prime-time debate still fulfills its deliberative promise or whether it has degenerated into what researchers call an echo chamber: an environment where political communication overwhelmingly confirms rather than contests the audience's partisan priors~\cite{loblich2021echo}. In such settings, dissenting views are either absent or minimized, producing repeated reinforcement of a shared orientation. While the concept is often applied to social media algorithms~\cite{bakshy2015exposure}, it is equally relevant to legacy media formats where host-guest exchanges can either foreground disagreement or devolve into choreographed agreement~\cite{jamieson2008echo}. Understanding this dynamic is critical, as the retreat from genuine debate can deepen what political scientists call
\emph{affective polarization}---the widening gulf of distrust and dislike between partisans that now rivals disagreement over concrete policy~\cite{iyengar2009red}.

Scholars have shown that exposure to like‑minded news encourages attitude reinforcement \cite{stroud2010polarization}, while diversified media diets can moderate views \cite{mutz2005new}. Yet almost all evidence rests on who watches which channel or how often a partisan guest is booked, not on whether a conservative and a liberal, once seated opposite the same host, actually clash over substance. Without that micro‑level record, we cannot know whether televised debate attenuates or accelerates polarization.

Capturing that record is hard for three reasons. First, live television is unscripted and multi‑speaker: turns overlap, tone shifts mid‑sentence, and topics pivot quickly. Second, genuine disagreement is a semantic property, not a surface cue; a sarcastic “right, exactly” is opposition, not assent. Naïve proxies—assuming every Democrat on Fox is challenged or every negative adjective signals dissent—fail on counter‑examples. Third, the scale is daunting. A single year of prime‑time opinion programming on the big three cable networks generates more than a billion words of caption text~\cite{hong2021analysis}. Manual annotation at the speaker‑turn level is financially untenable, and existing automated pipelines either collapse all speech into one stream (losing who said what) or ignore rhetorical stance altogether.

Previous studies therefore stop short of the core question. Some sample a handful of episodes and code them by hand, offering depth without generality \cite{farnsworth2005mediated}. Others scrape guest lists or sentiment polarity and infer that a segment was adversarial when the guest’s party differs from the host’s party \cite{morriseffect}. Such approaches miss intra‑party dissent, rhetorical framing, and shifts within a single interview. What we lack is a comprehensive, speaker‑resolved measurement of agreement and disagreement across the contemporary cable landscape.

This paper supplies that measurement. We assemble a corpus of 21,000 episodes (approximately 8,000 hours of video) aired between 2010 and 2024 on 24 flagship opinion shows across Fox News, MSNBC, and CNN. An automatic‑speech‑recognition and speaker‑diarization pipeline segments each broadcast into host and guest turns. Adjacent host–guest pairs are then classified as \emph{agreement}, \emph{neutral}, or \emph{disagreement} by a high fidelity stance model which is then applied to classify over 2 million host-guest interaction turns. Finally, we identify the names and political party affiliations of the guests along with the detailed topic being discussed in each episode.

Using this rich dataset, we present five main findings. First, explicit host‑guest dissent is rare: across 2017–2024 barely 15\% of exchanges register as disagreement, with MSNBC the lowest at roughly 13\% and Fox the highest at 17\%. Second, dissent is in retreat. On Fox and CNN the share of disagreement turns falls sharply each year, and in the few MSNBC shows with long‑run data the baseline is flat but never high, signaling a network‑wide drift toward concord. Third, the decline is historically contingent rather than inevitable---archival captions from the early 2010s show \emph{Hannity} operating at nearly double today’s disagreement rate, evidence that the format once accommodated sharper clash. Fourth, gatekeeping amplifies the trend: the channels not only book ideologically friendly guests but also moderate them differently once on air---Republican invitees face the stiffest push‑back on MSNBC, while Democrats encounter the gentlest treatment there, with the mirror pattern on Fox and a narrowing middle on CNN. Finally, the topics that dominate contemporary culture‑war politics---abortion, gun rights, immigration---attract the \emph{least} disagreement, turning segments most likely to mobilize audiences online into echo chambers on air.

%\textcolor{blue}{We use the term echo chamber to describe an environment in which political communication overwhelmingly confirms rather than contests the partisan priors of its audience \cite{loblich2021echo}. In such settings, dissenting views are either absent or minimized, producing repeated reinforcement of a shared orientation. While this concept is most often applied to social media algorithms and network homophily \cite{bakshy2015exposure}, it is equally applicable to legacy media formats such as cable news debate programming \cite{jamieson2008echo}, where host–guest exchanges can either foreground disagreement or devolve into choreographed agreement.}

%Using this rich dataset, we present three main findings: (i) Disagreement and debate on prime time news shows on cable is shrinking: the share of adversarial exchanges falls by roughly one‑third over the eight‑year window 2017-2024, even after conditioning on channel, guest ideology, and subject matter. (ii) Push‑back is asymmetric: conservative guests face little challenge on Fox News but substantial challenge on MSNBC, with the mirror image for liberal guests, while CNN occupies a narrowing middle. (iii) Contentious issues such as abortion, gun rights, and immigration generate the \emph{least} disagreement, indicating that the culture‑war topics most likely to mobilize audiences online are those most insulated from cross‑cutting dialogue on air.

Together, these results suggest that prime‑time “debate” is increasingly a performance of consensus within partisan boundaries rather than a forum for genuine argumentative contestation. By documenting this shift at scale and at the level of individual utterances, the study clarifies how televised political talk contributes to the broader architecture of affective polarization, and why restoring actual disagreement to mass media remains a pressing democratic task.

\section{Related work}

A vast literature links partisan cable news consumption to attitudinal polarization, uncivil discourse, and diminished civic trust. We organize the most relevant findings into four strands: (1) selective exposure and the rise of echo chambers, (2) the evolution of televised debate, (3) democratic consequences of incivility and one‑sided programming, and (4) recent computational audits that map ideological drift at scale. Together they illustrate both what we know and the empirical gaps this study fills.

\noindent\textbf{Selective Exposure and the Cable‑News Echo Chamber.}
Once Americans could choose among dozens of channels, many gravitated toward ideologically congenial outlets, producing self‑reinforcing ``echo chambers.'' Laboratory evidence shows Republicans strongly prefer Fox‑branded stories while Democrats shun them \cite{iyengar2009red}. Longitudinal survey work demonstrates that this sorting hardens attitudes over time \cite{stroud2010polarization, levendusky2013partisan}. Although only a minority of viewers are ``active‐audience'' partisans, their loyalty shapes network programming and, by extension, public debate \cite{arceneaux2013changing}. Recent large‑scale audience data confirm that cable news consumers today almost never stray outside their partisan bubble, with Fox, CNN, and MSNBC audiences sharply segregated \cite{kim2022measuring}.

%\textcolor{blue}{Recent scholarship has approached echo chambers through the lens of online selective exposure, where algorithmic feeds and homophilous networks amplify like-minded content (e.g., \cite{flaxman2016filter,bakshy2015exposure}). Yet other work complicates this narrative, showing that partisan sorting can be even stronger offline than online \cite{brown2021childhood,vaccari2016echo}. Our contribution extends this line of research by positioning televised debate shows as an offline analogue to the social media echo chamber: a non-algorithmic, host-driven environment where reinforcement and selective disagreement similarly structure political information consumption.}

Situating cable news as an offline echo chamber is particularly relevant to the ICWSM community. Recent work has begun to challenge the narrative that algorithmic feeds are the primary drivers of polarization~\cite{flaxman2016filter,bakshy2015exposure}, showing that partisan sorting can be even stronger in offline contexts than online~\cite{vaccari2016echo,brown2021childhood}. Our study contributes directly to this debate by providing a granular, speaker-resolved map of conversational dynamics within a non-algorithmic, host-driven environment. This analysis serves two purposes: first, it provides a reference point for comparing the intensity and mechanisms of human-curated echo chambers versus algorithmically-curated ones; second, it underscores the importance of studying legacy media, as the talking points and video clips generated in these televised debates are the raw material that is subsequently amplified and discussed across social media platforms, shaping the online discourse ecosystem.

\noindent\textbf{Changes in Prime-time Debates.}
Early‑2000s panel shows such as \emph{Crossfire} thrived on raised voices, interruption, and personal digs. Experiments revealed that televised incivility elevates physiological arousal yet erodes trust in government, even when issue content is held constant \cite{mutz2005new}. By the mid‑2010s, however, genuine left–right sparring had almost vanished; \emph{Crossfire} was canceled again in 2014, symbolizing CNN’s retreat from adversarial debate \cite{forgette2006high}. A decade‑long content audit shows that after the 2016 election Fox News booked steadily more conservative guests while CNN and MSNBC booked more liberals, leaving scant room for on‑air disagreement \cite{kim2022measuring}. When dissent does appear, it is often tokenistic or performative rather than deliberative \cite{sobieraj2011incivility}.

\noindent\textbf{Consequences for Democratic Deliberation.}
Incivility and ideological uniformity carry measurable civic costs. Uncivil exchanges not only entertain but also prime hostility toward political opponents and lower institutional trust \cite{mutz2005new}. Because partisan networks rarely present serious cross‑cutting dialogue, viewers encounter few competing frames, exacerbating motivated reasoning and affective polarization. The effects cascade beyond the viewing public: elite rhetoric \cite{gertz2020study} and social‑media conversations \cite{ding2023same} frequently echo televised talking points, amplifying polarization across multiple arenas.

This paper argues that the retreat from on-air contestation is a key, and perhaps underappreciated, driver of affective polarization. Our longitudinal analysis reveals that these prime-time shows are undergoing a fundamental change in their function, shifting from forums of debate into platforms for partisan affirmation. This structural shift has profound implications, a process explained by the political science theory of cross-cutting cleavages~\cite{goodin1975cross}, which posits that political conflict intensifies when partisan identity becomes the singular, dominant line of social division. By systematically purging substantive disagreement, these programs erode the visibility of other social cleavages and present a world of binary political conflict. A primary mechanism for this is the disappearance of intra-party dissent, which erases the natural diversity of factions within a political coalition. This performance of consensus sends powerful elite cues to viewers that ideological conformity is the norm, framing the in-group’s perspective as a settled truth. Consequently, opposing views are rendered not merely incorrect but illegitimate, hardening the distrust and animosity that define affective polarization.

\noindent\textbf{Computational Mapping of Ideological Slant.}
Recent work leverages natural‑language processing (NLP) and large transcript corpora to quantify network bias with temporal precision. By linking every on‑air guest to their campaign‑donation ideology score, Kim et al.\ chart hourly shifts in slant from 2010–2020, showing Fox's pronounced rightward drift and CNN/MSNBC’s reciprocal move left after 2016 \cite{kim2022measuring}. Earlier econometric analyses reached similar conclusions using hand‑coded tone and carriage data \cite{martin2017bias}. Such computational audits reveal that partisan skew is dynamic—responsive to electoral cycles, competitive positioning, and programming genre—and they supply the methodological blueprint for the present study.

Taken together, these strands demonstrate that \emph{who} watches (selective exposure), \emph{what} they watch (ideological uniformity), and \emph{how} it is presented (incivility) jointly deepen partisan division. Yet we still lack fine‑grained evidence on how moment‑to‑moment argumentative structure within cable segments shapes downstream discourse. By integrating NLP with interaction‑level annotations, this paper addresses that gap and specifies the mechanisms linking televised debate to audience polarization.

Furthermore, we connect our empirical findings to the literature on cross-cutting cleavages~\cite{goodin1975cross} and elite cue theory~\cite{gilens2002elite}. While existing work focuses heavily on audience self-sorting into echo chambers, our contribution demonstrates how the content within these chambers is actively being reshaped to eliminate internal dissent. By documenting this shift, we provide a specific mechanism, the erasure of debate and the performative signaling of conformity, that explains how partisan media can intensify affective polarization beyond simply filtering content.

\section{Dataset}

Our corpus originates with the Internet Archive Television News Archive,\footnote{\url{https://archive.org/details/tvarchive}} an open repository that mirrors the closed‑caption feeds of major U.S. broadcasters. We build on the cleaned release by \citet{laohaprapanon2017archive}, but extend it in two important ways. First, debate research requires unambiguous speaker segmentation. Many shows in the Archive omit the \textgreater\textgreater{} \emph{speaker}:'' token that marks a change of voice; others collapse hosts and guests under the same tag. To maximize precision, we retained only shows in which closed captions explicitly annotate both the host (e.g.\ \textgreater\textgreater{} \textsc{Tucker}:'') and the guest (``\textgreater\textgreater{}'' with no name). This criterion preserved most Fox News content but excluded the bulk of CNN and MSNBC’s schedule, whose captioning conventions lack named turns.

Second, to restore balance across networks we supplemented the Archive with episode transcripts from the shows’ official Apple Podcasts feeds, which supply RSS‑accessible \texttt{.mp3} files which we crawled using a lightweight crawler, to obtain every available episode for shows whose captioning was inadequate in the Archive.
We then transcribed and obtained speaker turns these episodes using a large Whisper model~\cite{radford2023robust}.
Next, we then harmonized the formatting to match the Internet‑Archive schema. Consistent with the Archive license, all podcast transcripts were already in the public domain; no paywalled or private audio was accessed.

Our final corpus contained 24 flagship debate or opinion show---nine on Fox News, eight on CNN, and seven on MSNBC---broadcast between January 2010 and December 2024. The sample contains over 21,000 episodes, accumulating just over 8,000 hours of video and roughly 194 million captioned words. Across those episodes we identify 2.13 million speaker turns, and on average, about 190 words per show. The earliest transcript dates to 7 July 2009 and the latest to 14 January 2025, giving a span of approximately 5,670 days (15.5 years, though we have gaps in 2014-2017). Table~\ref{tab:shows} lists every show alongside its coverage window and episode count. For each utterance we have a timestamp, the interlocutor’s role (host or guest), the guest’s inferred party affiliation (Section~\ref{sec:guests_identification}), and a stance label---\emph{agreement}, \emph{disagreement}, or \emph{neutral}---generated by our classifier (Section~\ref{sec:detecting_disagreement}). We also identify the topics discussed in each episode of the show (Section~\ref{sec:topics}).

\noindent\textbf{Dataset completeness.} Despite its breadth, the corpus remains incomplete. We focus on the three dominant U.S. cable‑news channels—Fox News, CNN, and MSNBC—because they account for the overwhelming share of prime‑time debate programming \cite{pew2023cablenews}. For each network we obtained every episode for which either the Internet Archive or an active Apple Podcasts RSS feed supplied transcripts. Even so, several shows exhibit intermittent coverage: entire weeks are missing when a programme was pre‑empted, went on hiatus, or when its podcast feed was later purged from Apple’s catalogue. Notably, for Hannity and Special Report with Bret Baier, Internet Archive or the Apple Podcasts did not have the data between 2014-2017, possibly due to technical glitches. Table \ref{tab:data_gaps} details the principal causes of these gaps. Fox, whose caption files consistently include speaker tags and whose podcast feeds remain intact, therefore contributes more episodes and longer time‑series than CNN or MSNBC. This asymmetry constrains some cross‑show comparisons, especially for the latter networks, but all analysis we report below explicitly considers episode availability and our results are robust to the missing‑data patterns.

Even though television data is widely available through the Internet Archive~\cite{ding2023same} and the Stanford TV data~\cite{hong2021analysis}, data at the level of granularity (including guest names, timestamps, topics, etc) and quality we obtain in this paper is not available previously. We believe the resulting corpus, together with the scraping and alignment procedures documented here, provides a replicable template for future work that seeks speaker‑resolved television data at scale. We will be releasing the full dataset and code for obtaining such data upon paper acceptance.

\section{Methods}

\subsection{Detecting disagreement}
\label{sec:detecting_disagreement}

\subsubsection{Data Pre-processing}
For programmes whose closed captions already encode speaker names (predominantly Fox News) we treat every contiguous block attributed to the host as an anchor turn and concatenate the immediately following guest utterances until the host next speaks. Each such \textit{host–guest bundle} constitutes one analytic pair. We further excise material outside the argumentative core of the show by trimming the transcript at the first host utterance and at the final host utterance, thereby discarding lead‑in promos, commercial breaks, and closing credits. One possible limitation of our pairing strategy is that, in episodes with multiple guests, consecutive guest turns might be collapsed into a single “guest” block if the host does not intervene. In practice, however, this constellation is extremely rare. Prime-time opinion shows on Fox, CNN, and MSNBC are overwhelmingly organized around a tightly moderated host–guest–host interview rather than open guest–guest debate \cite{jurkowitz2013changing}. Prior content audits similarly note that contemporary debate television has shifted away from cross-talk toward “parallel monologues” controlled by the host \cite{berry2013outrage}. In our own spot-check of 100 randomly sampled multi-guest episodes, 92\% featured a single guest speaking at a time, with fewer than 3\% of turns showing two guests in direct succession without host mediation. Thus, while theoretically possible, the loss of distinct guest reactions in consecutive turns is empirically negligible.

Episodes obtained from Apple Podcasts are \texttt{.mp3} files without caption metadata. We convert audio to text and speaker diarization using the large‑V3 \textsc{Whisper} model \cite{radford2023robust}, producing a token‑time‑aligned transcript in which turns are labelled generically (\textsc{Speaker 1}, \textsc{Speaker 2}, ...). To identify the host we apply a two‑step heuristic. First, the speaker with the greatest number of turns is presumed the host; every other speaker is provisionally a guest. If we observe multiple guests, occasionally pushing the host below first place, we override the rule and assign the first utterance in the episode to the host, propagating that label to all matching diarization IDs. Residual advertising copy at the head of the file is removed with show‑specific regex filters keyed to recurring phrases such as “welcome to \emph{Inside Politics}.” We manually verified every episode where our heuristic did not apply to ensure the dataset was of high quality.

\subsubsection{Data Annotation}
The main analytical backbone of the project is a stance label assigned to every \textit{host–guest bundle}. Because disagreement can flare and subside many times within a single episode, we annotate each contiguous exchange rather than the episode as a whole. 
Each record therefore contains two text fields---\texttt{host\_turn} and \texttt{guest\_turn}---and one categorical target with three possible values: \textit{Agreement}, \textit{Disagreement}, or \textit{Neutral}.

We began by hand‑coding five hundred randomly selected pairs from various shows. Two annotators (the authors) independently coded this set, achieving a Cohen’s $\kappa$ of 0.81, indicating substantial agreement prior to adjudication. Disagreements were then reconciled in adjudication meetings to produce a final gold-standard set with perfect agreement. The following decision rules governed the exercise:

\begin{enumerate}
\item \textbf{Agreement} is assigned only when the guest endorses, amplifies, or otherwise affirms the host’s preceding claim. \emph{Example}:
\begin{quote}
\small
\textbf{Host}: “Just to be clear, you agree the Russian invasion was an unprovoked violation of international law?”\
\textbf{Guest}: “Absolutely, Moscow crossed a bright red line, and the world has to treat it that way.”
\end{quote}

\item \textbf{Disagreement} requires an explicit refutation, contradiction, or correction of the host.  \emph{Example}:  
\begin{quote}
\small
\textbf{Host}: ``Premiums are up twenty--five percent---people hate this law because they’re paying more.'' \\
\textbf{Guest}: ``That’s not accurate. Premiums were flat for the two previous years, and the average growth \emph{slowed} after the law passed.''
\end{quote}

\item \textbf{Neutrality} covers all residual cases: clarifying questions, greetings, news reading, off‑topic banter, or exchanges in which stance cannot be inferred.

\end{enumerate}

Manual coding of the 2.13 million host–guest bundles was infeasible, so we adopted a staged pipeline that moves from small, human‑verified seeds to large‑scale annotation with language models. Our labeling strategy necessarily flattens instances where a guest simultaneously affirms one part of the host’s statement while rejecting another. However, two factors mitigate its impact. First, prime-time exchanges are typically short and tightly coupled--on average, each host–guest pair contains roughly 90 words, typically addressing a single issue at a time. Second, our classifier is conservative: when a guest response contains both supportive and adversarial elements, the model frequently defaults to the neutral category rather than forcing agreement or disagreement. For \emph{Example}:
\begin{quote}
\small
\textbf{Host}: ``Well, he said he wasn't aware. He has been specific about that. That's been one of the things they said. He did not know.''\\
\textbf{Guest}: ``No, I agree. But one of the problems he has in talking about it that way remember he wrote a book and talked about Russia and Iran and Ukraine. And one of the things that the Justice Department is going to do is compare those passages to these documents, and, you know, it'll either show they were likely used or there's no mention of them. But it's a problem for him.''
\end{quote}
The model output was neutral, reflecting mixed stance. While this approach sacrifices fine-grained nuance, it ensures that we do not overstate levels of disagreement.

We built the stance classifier through a four‑stage, bootstrapped workflow. We began with a small hand‑coded seed and found that task‑specific encoders (e.g., fine‑tuned DeBERTa‑v3‑base) topped out at barely 55\% accuracy, so we pivoted to large language models, using GPT‑4 in an interactive loop to expand the seed to 500 high‑quality pairs. Two successive rounds of crowd annotation then supplied an additional 1,712 carefully triaged pairs, vetted with gold checks, consensus filtering, and manual audits. At each expansion we re‑trained both commercial (GPT‑4o) and open‑source models (Llama‑3 70B 4‑bit, DeepSeek‑R1‑Distill‑Llama‑8B, Qwen‑2.5 32B 4‑bit), iteratively pruning noisy items until performance stabilised. The final DeepSeek model attains 89\% accuracy on an unseen validation set (comprised of the authors' manually audited subsets from the crowdsourced datasets), providing the labels used throughout the paper. To ensure the reliability of these labels, two authors independently reviewed all samples in this subset, achieving Cohen’s $\kappa$ = 0.84. Disagreements were resolved through discussion until consensus was reached. Full details of the annotation rounds, model variants, and evaluation metrics appear in the Appendix.

%expanded the seed set to 500 pairs via an interactive loop with GPT‑4o. Starting from an empty prompt, we asked the model to label 20 unseen pairs in zero‑shot mode, manually corrected its errors, and then fed the revised examples back as demonstrations before issuing the next batch. After 25 such rounds we had 500 human‑vetted pairs that reflected the full range of argumentative nuance in the show.
%
%Fine‑tuning \textsc{gpt‑3.5‑turbo} on the 400‑pair training portion improved accuracy to 79\% on the 100‑pair test set.

%To enlarge the training pool we recruited U.S. based crowd workers through CloudResearch \cite{hartman2023introducing}. We sampled 1,000 bundles drawn equally from \emph{Tucker Carlson Tonight}, \emph{Hannity}, and \emph{The Rachel Maddow Show}. These were partitioned into 50 blocks of 20 pairs; each block was independently annotated by three workers on a purpose‑built web interface that provided illustrated guidelines and embedded gold‑standard checks. Workers who failed gold items or completed the task implausibly quickly were rejected, and their labels were discarded.

%The raw crowd data yielded a Fleiss $\kappa$ of 0.302, indicating substantial annotator noise. We therefore applied \textsc{cleanlab} \cite{goh2022utilizing} to infer consensus labels and discard ambiguous cases, retaining 459 high‑confidence pairs. Combined with the original gold standard 500 pairs, this produced a 959‑pair training set.
%

This fine‑tuned model was then applied to the rest of the 21,000 episodes, generating stance labels for 2.13 million host–guest bundles. 
For each show, one of the authors manually verified 100 randomly sampled balanced pairs. This spot check was designed as a sanity check and in all shows, model accuracy exceeded 85\% (complete evaluation metrics is reported in Table 6 in the Appendix), thus giving us the confidence that our model and predictions are high fidelity and can be relied upon. Overall, we identify 53.6\% neutral, 30.4\% agreement and 16.0\% disagreement labels for all the 21,000 episodes.

\subsection{Guest identification and party coding}
\label{sec:guests_identification}

Our procedure for determining who the guests were in each episode, and how those guests leaned politically, unfolded in two stages.

\paragraph{Stage 1: extracting guest names with GPT 3.5.}
For every episode we isolated the host’s opening monologue, which typically lists the evening’s interviewees. We then passed that text to the \texttt{gpt‑3.5‑turbo} model through the OpenAI API in a zero‑shot setting. A representative prompt for \emph{Life, Liberty \& Levin} was:

\begin{quote}\small
\ttfamily
\noindent
\strut{}This is a transcript of the show ``Life, Liberty and Levin,'' hosted by Mark Levin. He is never a guest. Extract all invited guests who actually join the show from the full opening monologue, and for each give their political affiliation as one of \verb+[Democrat,Republican,Other,Unknown]+.
If the guest is a known office‑holder, use their real party. If they are a journalist, member of a non profit, an advocate or any other, infer their closest alignment from your training. If you absolutely cannot determine affiliation, use \texttt{Unknown} -- but only as a last resort.
Your reply must be \emph{only} a JSON array (no narrative, no Markdown, no code fences).
\end{quote}

This step yields, for each episode, a list of guest names paired with the model’s first‑pass guess at affiliation. The host’s name is systematically excluded. In total, 1,374 guest affiliations were retrieved in this stage.

\paragraph{Stage 2: refining affiliation labels with external data.}
Journalists, editors, and many subject‑matter experts do not have an official party registration, which often led the language model to return \texttt{Unknown}. To improve coverage and cross-check the LLM generated affiliations,  we matched the extracted names and affiliations against the DIME database \cite{bonica2015database}, a large‑scale compilation of itemized U.S. campaign‑finance records. This is a high fildelity resource used in previous research and has been shown to be of high quality, containing a large database of public personas~\cite{kim2022measuring}. To ensure accurate mappings, we normalized names into a consistent “last, first” format and required exact matches. In cases where DIME contained multiple individuals with the same name, we used the "recipient party", and "contributor occupation" columns to disambiguate. Ambiguous matches (less than 5\% of the total matches) with no clear resolution were conservatively excluded rather than risk misclassification. Among the subset of guests who had a party affiliation in both sources, 70.9\% of GPT labels matched DIME, while 29.1\% diverged. In cases of mismatch, we prioritized the DIME label in the final dataset. For guests marked as \texttt{Unknown} by GPT but present in DIME, we also filled in the affiliation from DIME. This procedure both corrected potential GPT misclassifications and reduced reliance on the \texttt{Unknown} category, yielding a more robust measure of guest partisanship.

\begin{figure}
    \centering
    \includegraphics[width=0.5\textwidth]{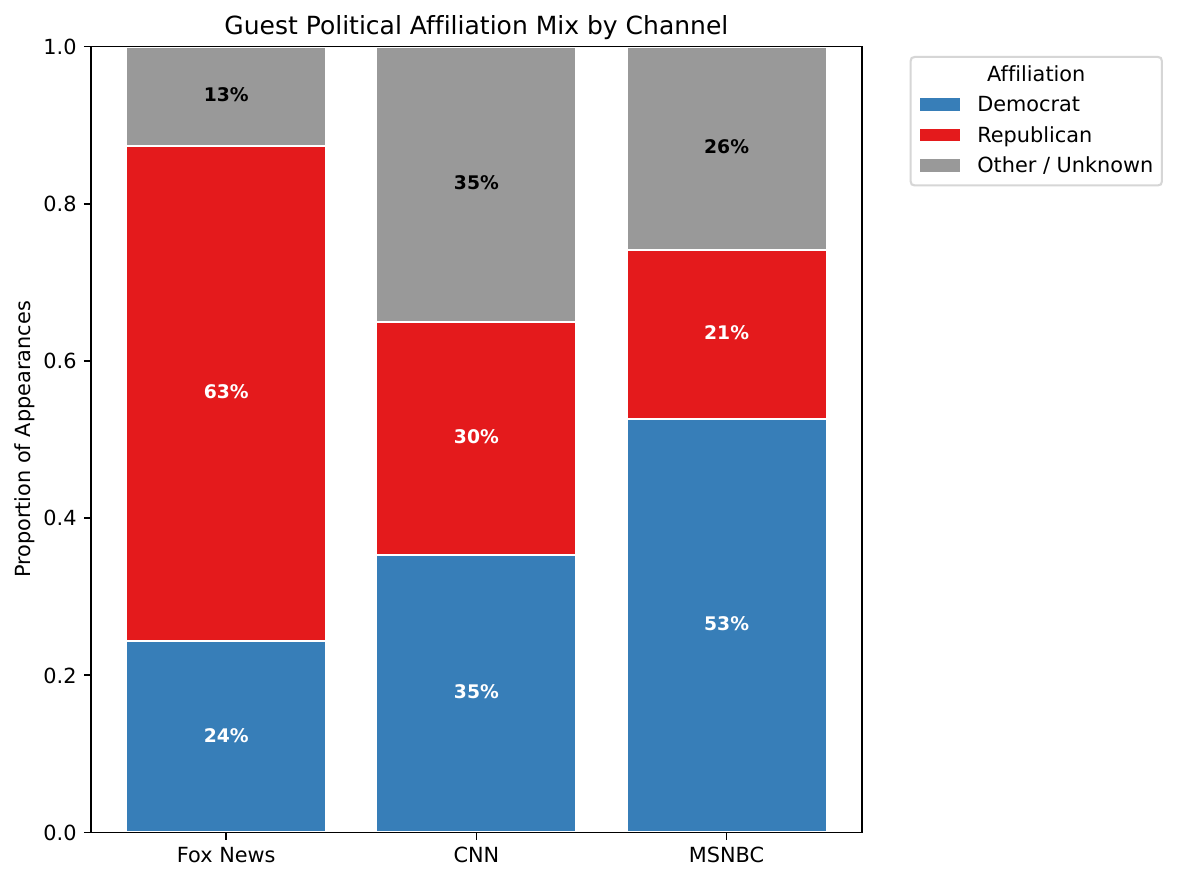}
    \caption{Fraction of guests by political leaning and channel. As expected, we see that Fox News invites mostly Republicans and MSNBC mostly Democrats.} 
    \label{fig:guests_party_channel}
    \vspace{-\baselineskip}
\end{figure}

Applying this two‑step pipeline to the full corpus provided an affiliation label for nearly every recurring guest. 
Figure~\ref{fig:guests_party_channel} shows the final distribution, showing, as expected, that Fox News features predominantly Republican guests, while MSNBC features predominantly Democratic ones.

\subsection{Topic modeling}
\label{sec:topics}

To recover the topics of each episode we applied BERTopic \cite{grootendorst2022bertopic} to every episode transcript in the corpus.

\paragraph{Pre‑processing.}
Captions were lower‑cased, stripped of non‑alphabetic characters, filtered through the NLTK English stop‑word list, and purged of “intro‑junk” tokens such as host names and sponsor slogans. The cleaned texts were embedded with the \texttt{all‑MiniLM‑L6‑v2} sentence‑transformer.

\paragraph{Model configuration and tuning.}
We initially set the vectorizer to \texttt{min\_df = 2}, \texttt{max\_df = 0.9}, and \texttt{max\_features = 30,000}. Dimensionality reduction used UMAP with cosine distance (\texttt{n\_components = 3}, \texttt{min\_dist = 0.1}, \texttt{n\_neighbors = 10}); soft clustering employed HDBSCAN, which yields a membership probability for every document. Hyper‑parameters were tuned by grid search on a 5\% development slice, optimizing for \emph{codeless} coherence—our best setting achieved a coherence score of 0.689.

\paragraph{Topic consolidation.}
The tuned model returned 146 raw clusters. Manual inspection revealed substantial lexical overlap, so we merged semantically redundant clusters—using keyword similarity and exemplar phrase matching—down to 80 distinct topics. Finally, two authors grouped those 80 topics into 15 higher‑level categories (e.g.\ \textsc{Elections \& Campaigns}, \textsc{Cultural Politics}, \textsc{COVID-19}, etc); any disagreements were resolved by discussion. These 15 top‑level labels constitute the topic taxonomy visualized in Figure~\ref{fig:episodes_per_topic}.
The list of all 80 topics is added in table~\ref{tab:data_gaps} in the Appendix.

%In this step, we used BERTopic \cite{grootendorst2022bertopic} to get the topics across the 24 shows. We followed the general procedure for topic modeling-Lower‑casing, removal of non‑alphabetic characters, used stop‑word list, and removed “intro‑junk” terms (host names, sponsor slogans, etc.). "all‑MiniLM‑L6‑v2" model was used for the Sentence Transformer. Other parameters were min\_df=2, max\_df=0.9, max\_features=30000. For hyperparameter tuning we used UMAP (cosine metric, n\_components=3) + HDBSCAN (soft clustering, probability scores). Our best model came as having 0.689 as the coherence score. Hence, our best hyperparameters were- "min\_cluster\_size": 20, "min\_dist": 0.1, "n\_neighbors": 10. From these, there were in total 146 topics, which were further merged down to 80 topics, as there were many repeated/similar topics with slightly different keywords. These final 80 topics were further categorized into 15 High-level topics as visualized in figure 7. 
%\kiran{please add the list of 80 topics to the appendix.}

%Figure~\ref{fig:episodes_per_topic} shows the number of episodes per topic. 

\begin{figure}
    \centering
    \includegraphics[width=0.5\textwidth]{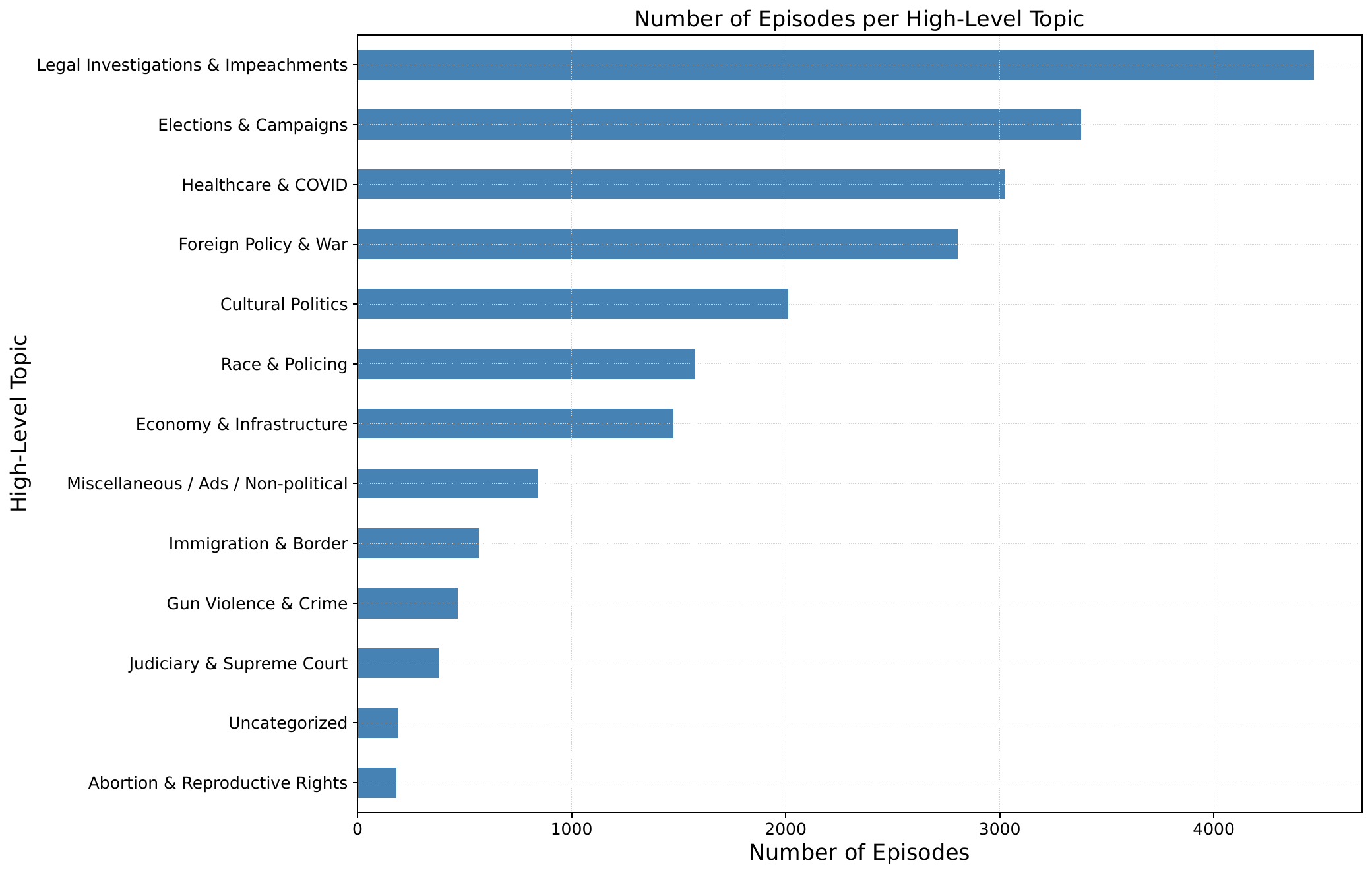}
    \caption{Episodes per high level topic.}
    \label{fig:episodes_per_topic}
\end{figure}

%\section{Descriptive analysis}

\section{Findings}
\label{sec:findings}

We organize the findings along three dimensions. First, we present longitudinal disagreement and agreement from 2010 through 2024, documenting how the share of disagreement contracts across most shows and channels (Section~\ref{sec:disagreement_over_time}). Second, we pair every utterance with the guest’s party identification to show how partisan booking decisions shape conversational tone—Democratic invitees receive more agreement on MSNBC, Republicans on Fox, while CNN remains comparatively balanced (Section~\ref{sec:guests_findings}). Third, we disaggregate the data by topic, demonstrating that the lowest levels of disagreement cluster around culture‑war flashpoints such as abortion, gun rights, and immigration, a pattern that underscores the echo‑chamber character of contemporary debate television (Section~\ref{sec:topics_findings}).

%we have the following analysis:
%- timeseries plots showing disagreement trends over time. we show that disagreement in the debates goes down over time indicating that shows are not inviting guests who have different view points ().
%- guest political affiliations and their agreement/disagreement scores. showing that democrats are invited more and agreed upon more on MSNBC and republicans are invited more and agreed upon more on fox news.
%- finally topic analysis shows that disagreement levels are lowest (high echo chamber) for controversial topics like abortion, gun rights, and immigration on all channels. 

\subsection{Agreement and Disagreement over time}
\label{sec:disagreement_over_time}

In this analysis, we compute the average (dis)agreement per show, aggregated by month. For each month, we calculate the mean level of (dis)agreement observed across all shows within that period. Alongside the mean, we also present the 95\% confidence intervals, which allows us to track how (dis)agreement varies over time.

\begin{figure*}[ht]
\centering
\begin{minipage}{.32\linewidth}
\centering
\subfloat[Tucker Carlson (Fox)]{\label{}\includegraphics[width=\textwidth]{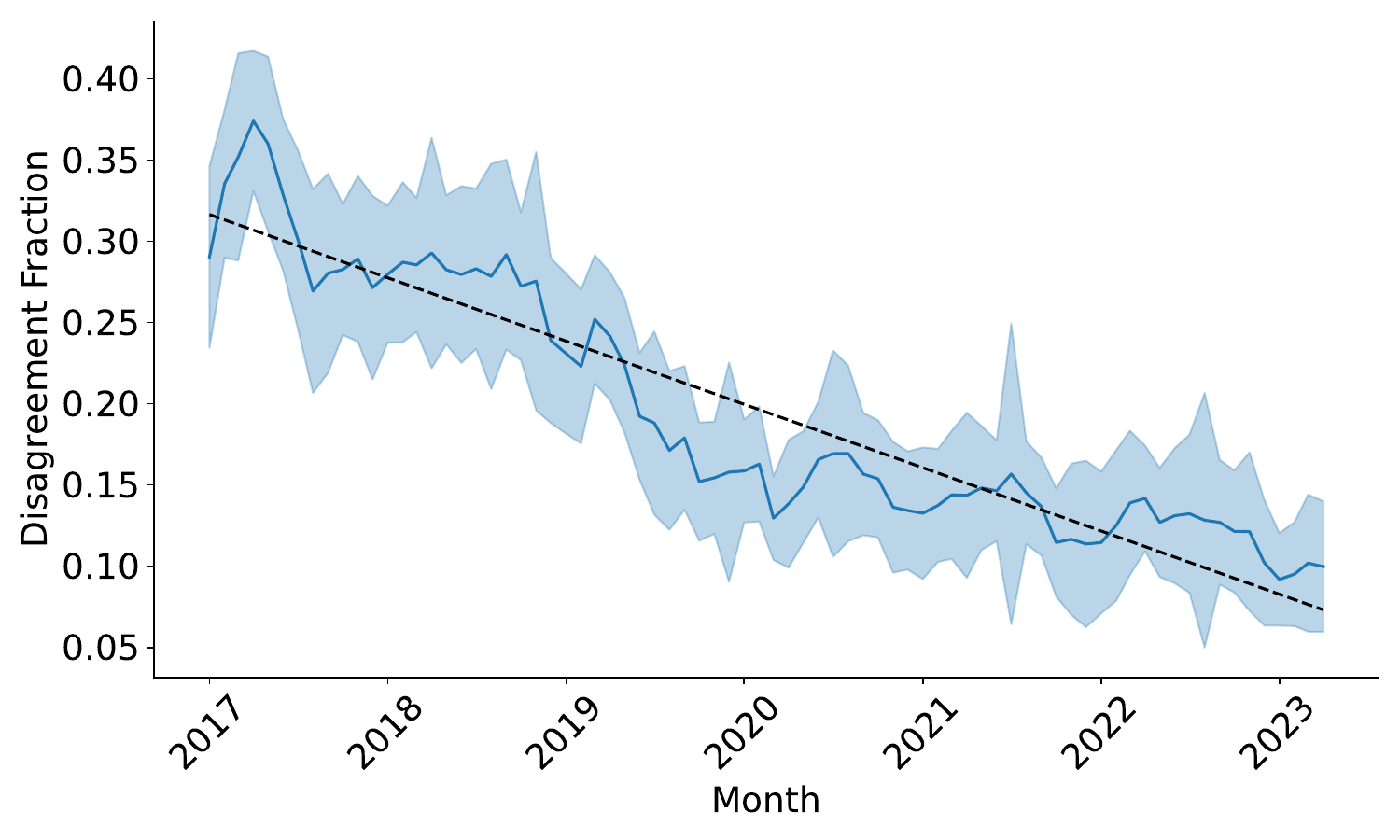}}
\end{minipage}%
\begin{minipage}{.32\linewidth}
\centering
\subfloat[Anderson Cooper (CNN)]{\label{}\includegraphics[width=\textwidth]{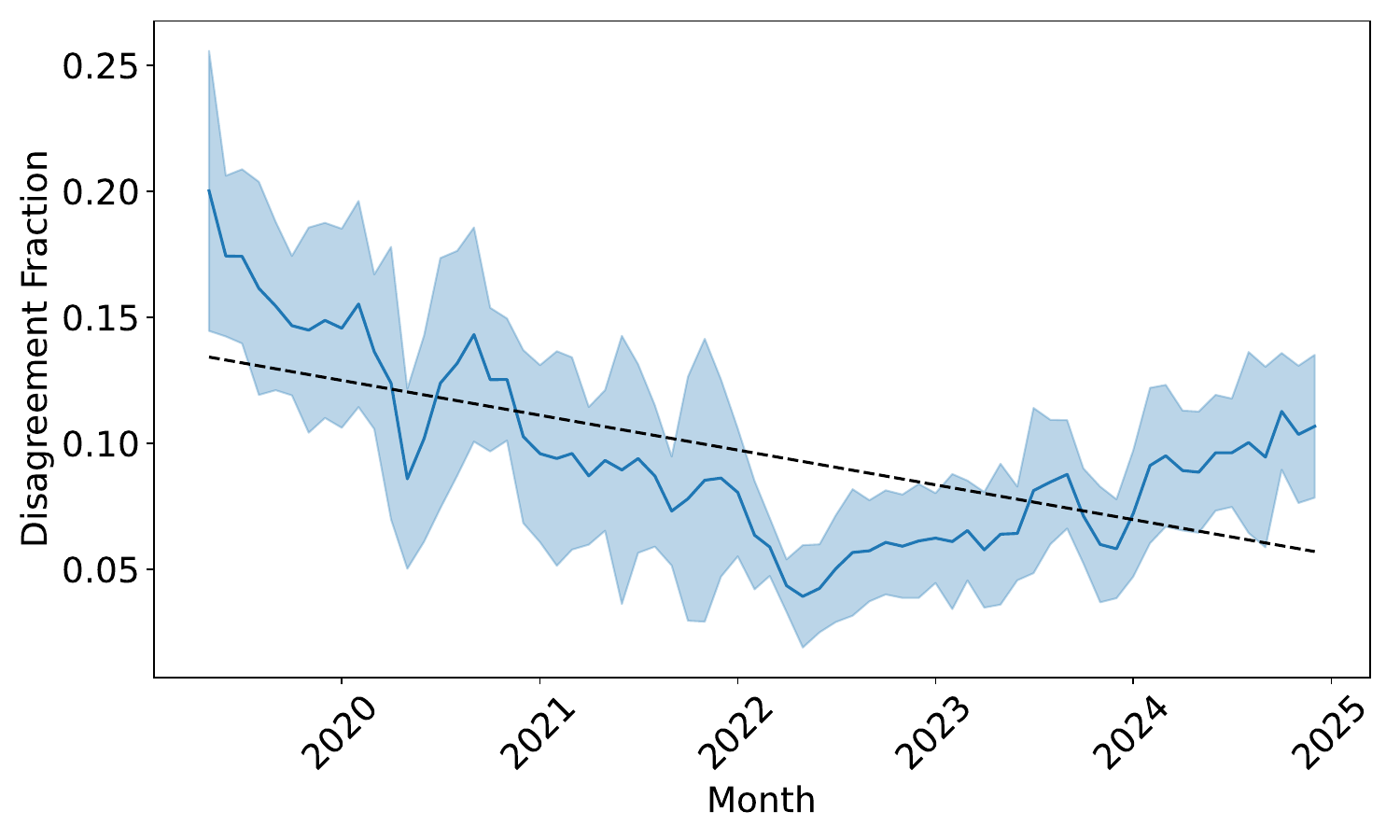}}
\end{minipage}%
\begin{minipage}{.32\linewidth}
\centering
\subfloat[The Savage Nation (MSNBC)]{\label{}\includegraphics[width=\textwidth]{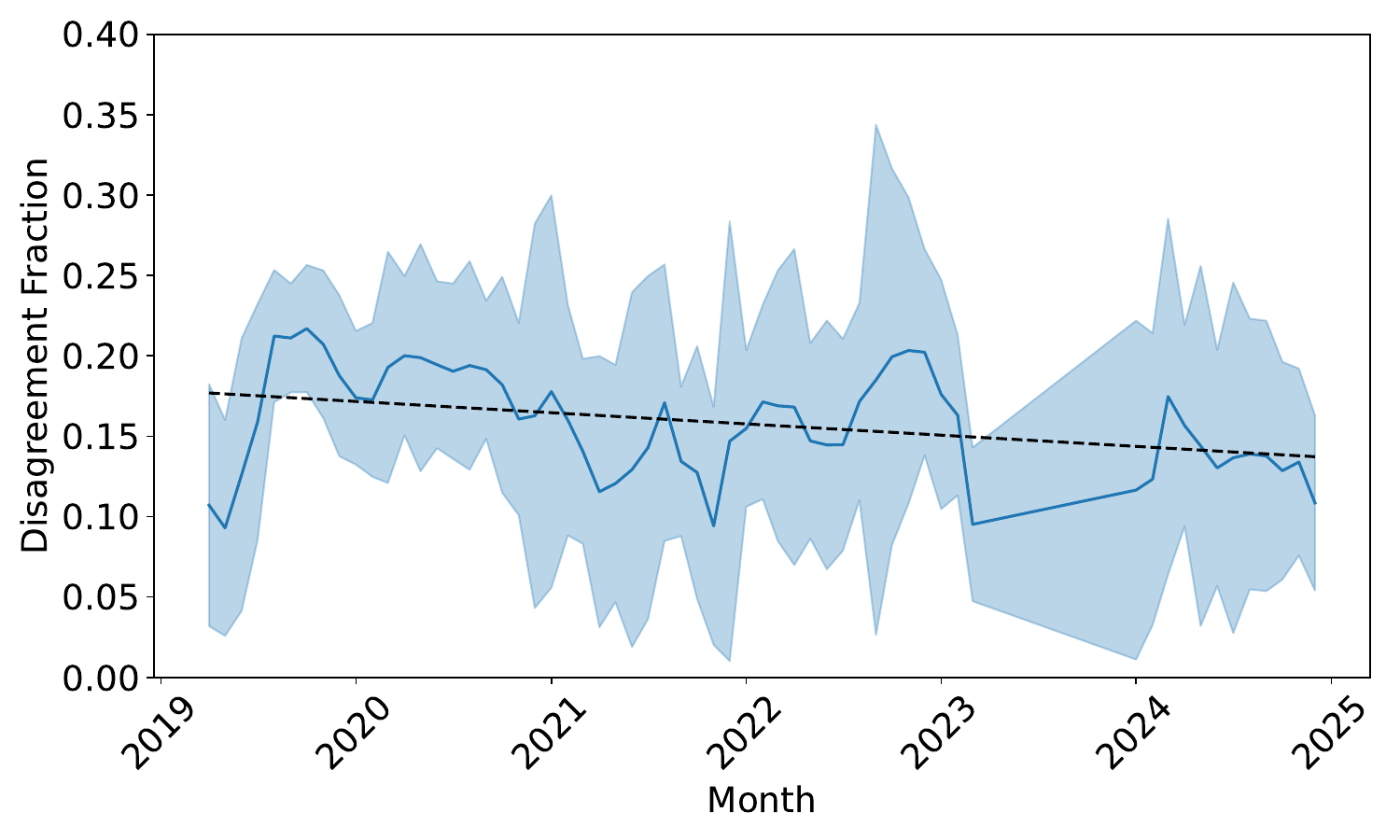}}
\end{minipage}%
\par\medskip

\begin{minipage}{.32\linewidth}
\centering
\subfloat[Laura Ingraham (Fox)]{\label{}\includegraphics[width=\textwidth]{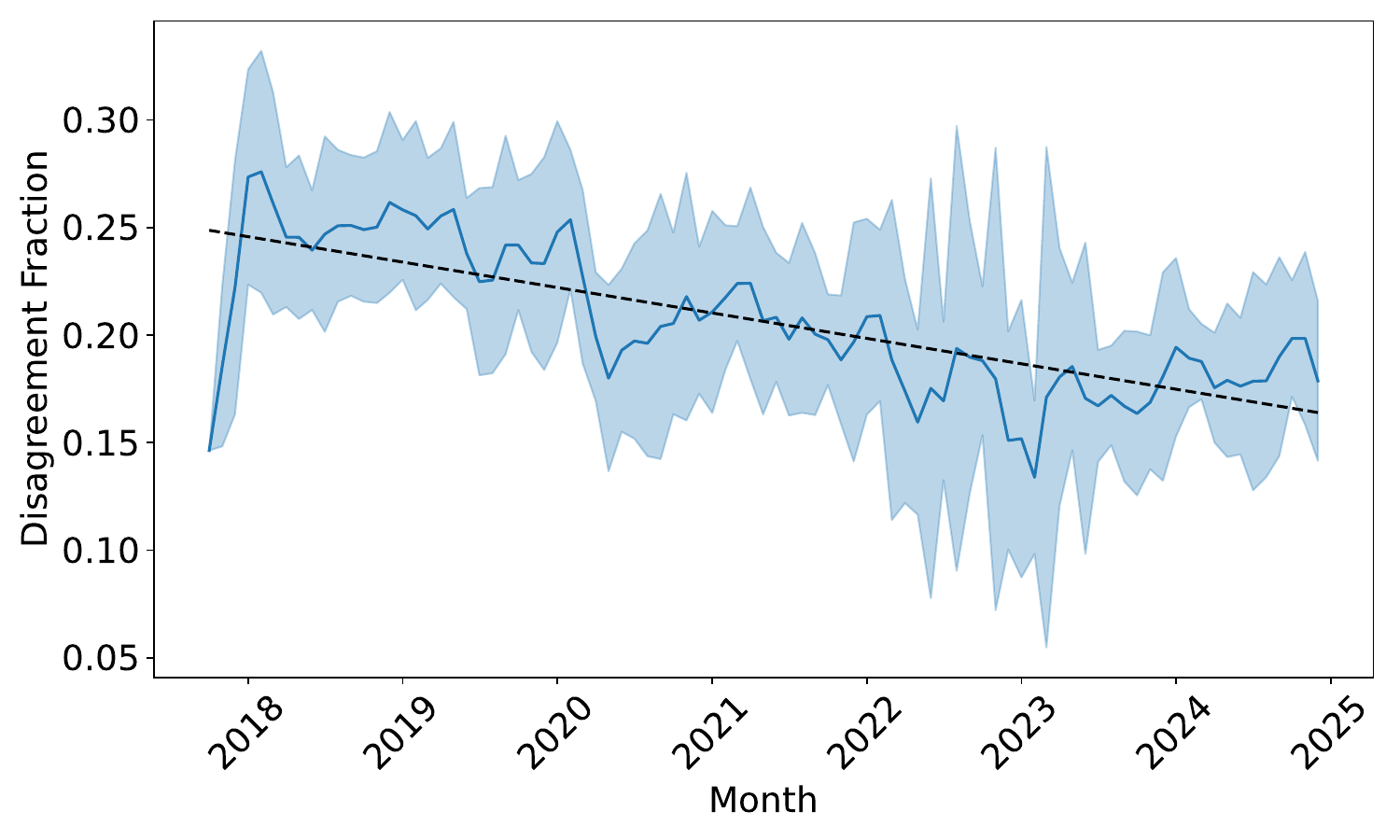}}
\end{minipage}%
\begin{minipage}{.32\linewidth}
\centering
\subfloat[State of the Union (CNN)]{\label{}\includegraphics[width=\textwidth]{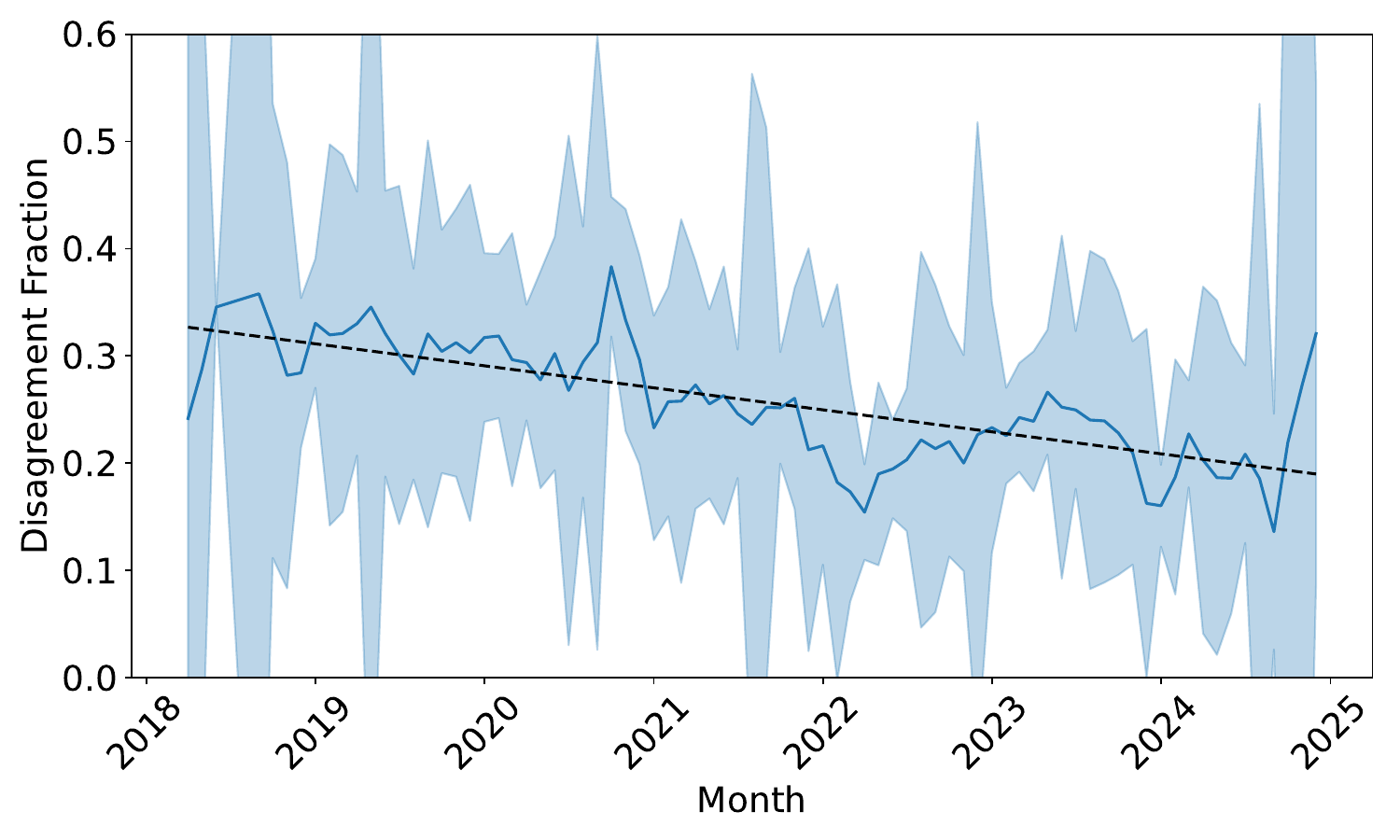}}
\end{minipage}%
\begin{minipage}{.32\linewidth}
\centering
\subfloat[Velshi (MSNBC)]{\label{Velshi}\includegraphics[width=\textwidth]{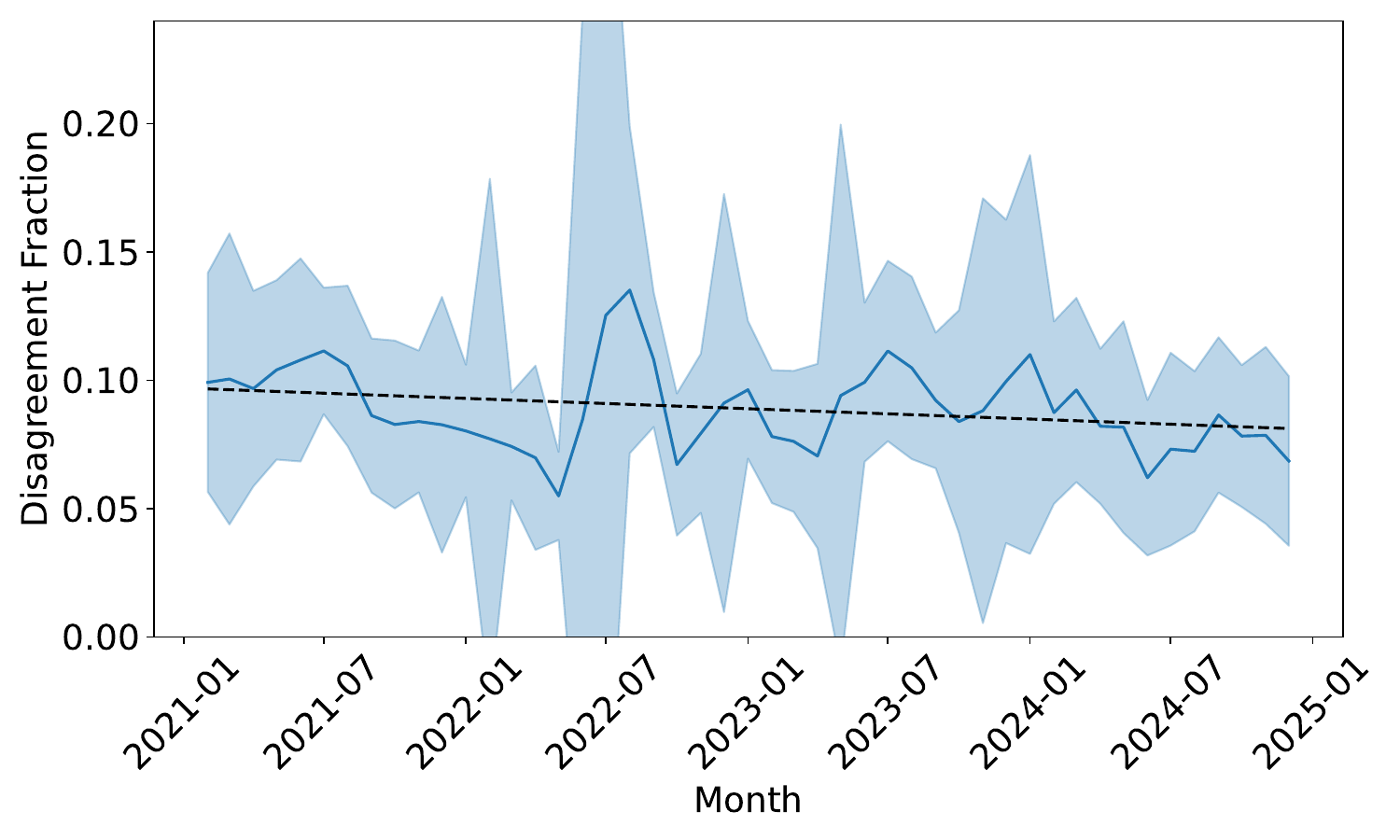}}
\end{minipage}%
\par\medskip
\caption{Trends in mean disagreement fraction over the years for a sample of the shows in our dataset.}
\label{fig:disagreement_over_time}
\vspace{-\baselineskip}
\end{figure*}

Figure~\ref{fig:disagreement_over_time} shows the temporal trajectories of host–guest disagreement for a (non random) sample of the shows in our dataset. Across the full 2017–2024 window, disagreement on average constitutes only a surprisingly small share of airtime---typically between 10\% and 15\%—and, for most shows, that share is either flat or trending downward. The pattern is clearest on Fox News. Tucker Carlson Tonight begins its run in 2017 with roughly one third of all host–guest exchanges coded as explicit disagreement, yet this fraction drops to about 15\% by the time the show goes off‑air in April 2023 (Figure~\ref{fig:disagreement_over_time}a). The ordinary‑least‑squares slope on the monthly series is –3.9 percentage points per year (p = 0.000), confirming a systematic retreat from disagreement. Our finding dovetails with the New York Times content audit of 1,100 episodes, which likewise documents a contraction in dissenting perspectives over the same period \cite{nytimesInsideApocalyptic}. Laura Ingraham’s primetime hour shows a similar trend (Figure \ref{fig:disagreement_over_time}d), albeit from a slightly lower baseline (-1.3 percentage points per year, p=0.000), showing that the phenomenon is not idiosyncratic to a single host but characterizes Fox’s broader editorial posture.

CNN exhibits a more heterogeneous picture. Anderson Cooper 360° mirrors Tucker Carlson in direction (Figure~\ref{fig:disagreement_over_time}b), though the glide path is gentler: disagreement falls from roughly 20\% in 2019 to 14\% by late 2024, a decline of 1.2 percentage points per year (p = 0.000). Similarly, the Sunday show State of the Union---which is structurally designed to pit partisan surrogates against one another---has a higher base rate of disagreement (near 30 \%) in 2018 but goes down to around 20\% by 2024 (decline of -2.1 percentage points per year (p = 0.000), Figure \ref{fig:disagreement_over_time}e). 
%The stability of this benchmark show provides a useful upper bound on what a ``balanced'' format can sustain in contemporary cable news.

MSNBC’s inventory is the most internally consistent: disagreement begins low and stays low. The Savage Nation and Velshi, for instance, rarely strays above 10-15\% disagreement in any month (Figure \ref{fig:disagreement_over_time}f), and the slope of the time‑trend is statistically indistinguishable from zero. Other MSNBC offerings in the appendix (Figure \ref{fig:disagreement_over_time_all_shows}) replicate this flat‑line pattern. Taken together, the cross‑network comparison reveals that while Fox is actively converging toward a narrow opinion corridor, CNN is bifurcated between a shrinking adversarial prime‑time slot and a resilient Sunday shows, and MSNBC has long since embraced homophily as a default.

This long-run flatness is consistent with MSNBC’s booking strategy: prime-time programming overwhelmingly features ideologically aligned guests, so baseline opportunities for visible disagreement are rare. Where disagreement does appear, it tends to be intra-coalitional (e.g., tactical disputes within the Democratic camp) rather than cross-partisan clashes, and these are moderated into brief, low-salience exchanges by the host. As a result, even as the network’s guest mix varies slightly across shows, the format produces a ceiling effect: disagreement cannot decline much further, because it begins at very low levels. In this sense MSNBC illustrates a homophilous “steady state,” contrasting with Fox’s dynamic narrowing and CNN’s shifting hybrid format.

The complementary analysis of agreement (see Figure \ref{fig:agreement_over_time} for a small sample, Figure~\ref{fig:agreement_over_time_all_shows} in the Appendix for all shows) reinforces these conclusions. Where disagreement contracts, agreement expands. On Tucker Carlson Tonight the proportion of explicit agreement rises by almost 5 percentage points over the study period, a mirror image of the disagreement slide. Neither Anderson Cooper 360° nor The Savage Nation exhibits a statistically significant upward drift in agreement, suggesting that their modest losses in disagreement are being absorbed by neutral or ambiguous exchanges rather than by outright concurrence. Crucially, across all shows the combined share of agreement and neutral statements already exceeds 85\% of airtime by 2024, leaving scant room for the deliberative contestation that debate television purports to deliver.

Two patterns emerge with genuine substantive weight. First, the data point to a deliberate editorial pivot away from inviting ideologically dissonant guests: as fewer counter‑voices appear, on‑air push‑back correspondingly evaporates. This retreat is clearest on Fox and CNN, where disagreement falls significantly in 7 out of 9 Fox shows and 5 of 8 CNN shows. MSNBC shows no systematic trend—largely because the available time‑series for its line‑up is shorter—but its baseline levels of disagreement are already minimal. Second, the erosion of dissent is mirrored by a marked rise in affirmation, especially on Fox, where explicit agreement climbs in 5 of 9 shows.

Although CNN continued to invite a more balanced mix of guests, the incidence of dyadic disagreement still declined over time. This reflects the format of prime-time talk shows, where hosts maintain strong control over the conversation and often avoid sustained cross-cutting debate, instead privileging civility and rapid topic shifts. In line with prior work on cable news as “parallel monologues” rather than genuine argument exchange \cite{sobieraj2011incivility, jamieson2008echo}, even a heterogeneous guest lineup does not necessarily translate into higher on-air disagreement.

Viewed alongside the topic‑level findings in Section \ref{sec:topics_findings}, the pattern is disquieting. The shrinking space for overt disagreement coincides with sustained—or even heightened—coverage of polarizing policy domains such as abortion, gun rights, and immigration. What remains, therefore, is “echo‑chamber prime time”: divisive themes are aired, yet almost invariably among like‑minded interlocutors who amplify rather than interrogate partisan frames. The evidence is unmistakable: contemporary debate television has not mellowed into polite conversation; it has shed the adversarial backbone that once separated it from straightforward opinion broadcasting.

\begin{figure*}[ht]
\centering
\begin{minipage}{.32\linewidth}
\centering
\subfloat[Tucker Carlson (Fox)]{\label{}\includegraphics[width=\textwidth]{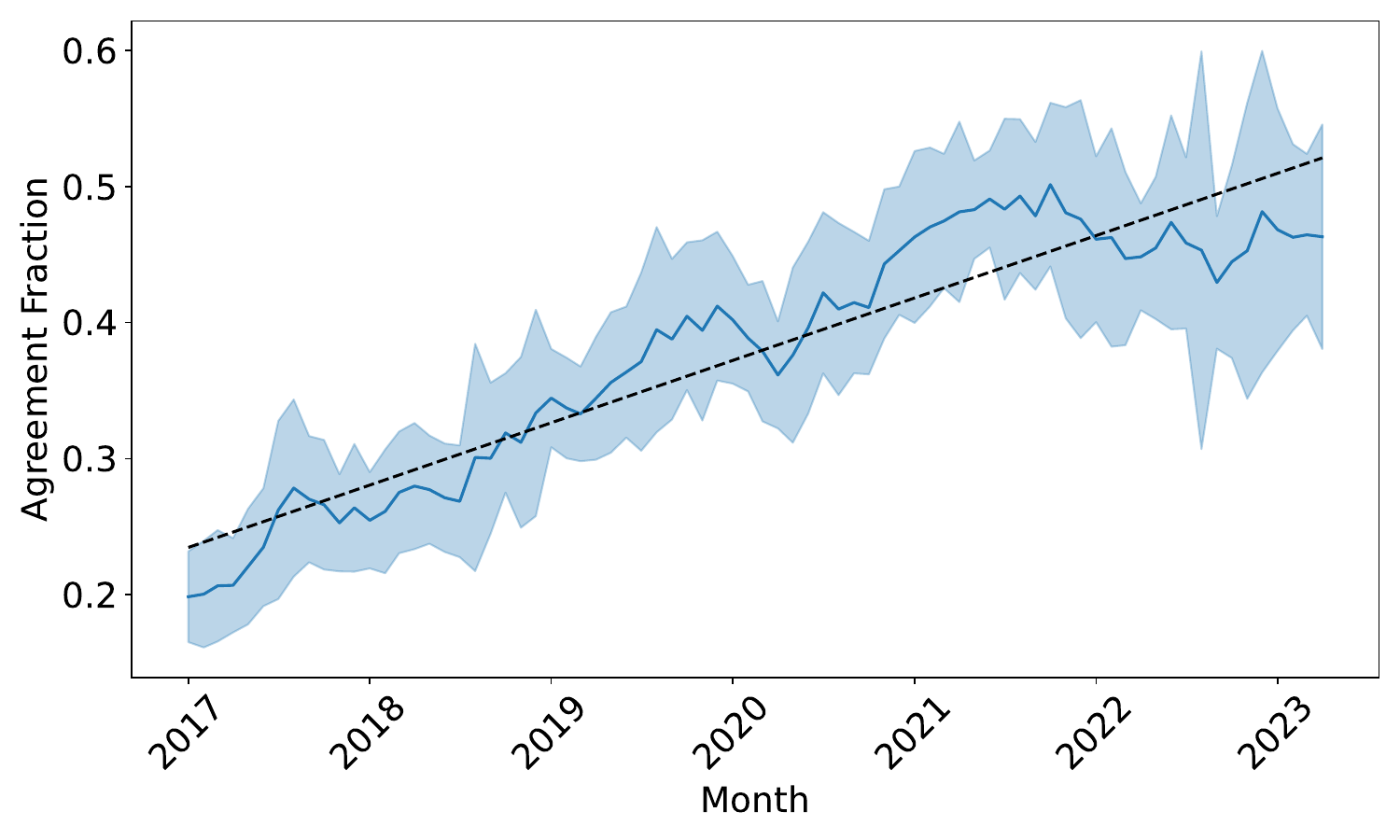}}
\end{minipage}%
\begin{minipage}{.32\linewidth}
\centering
\subfloat[Anderson Cooper (CNN)]{\label{}\includegraphics[width=\textwidth]{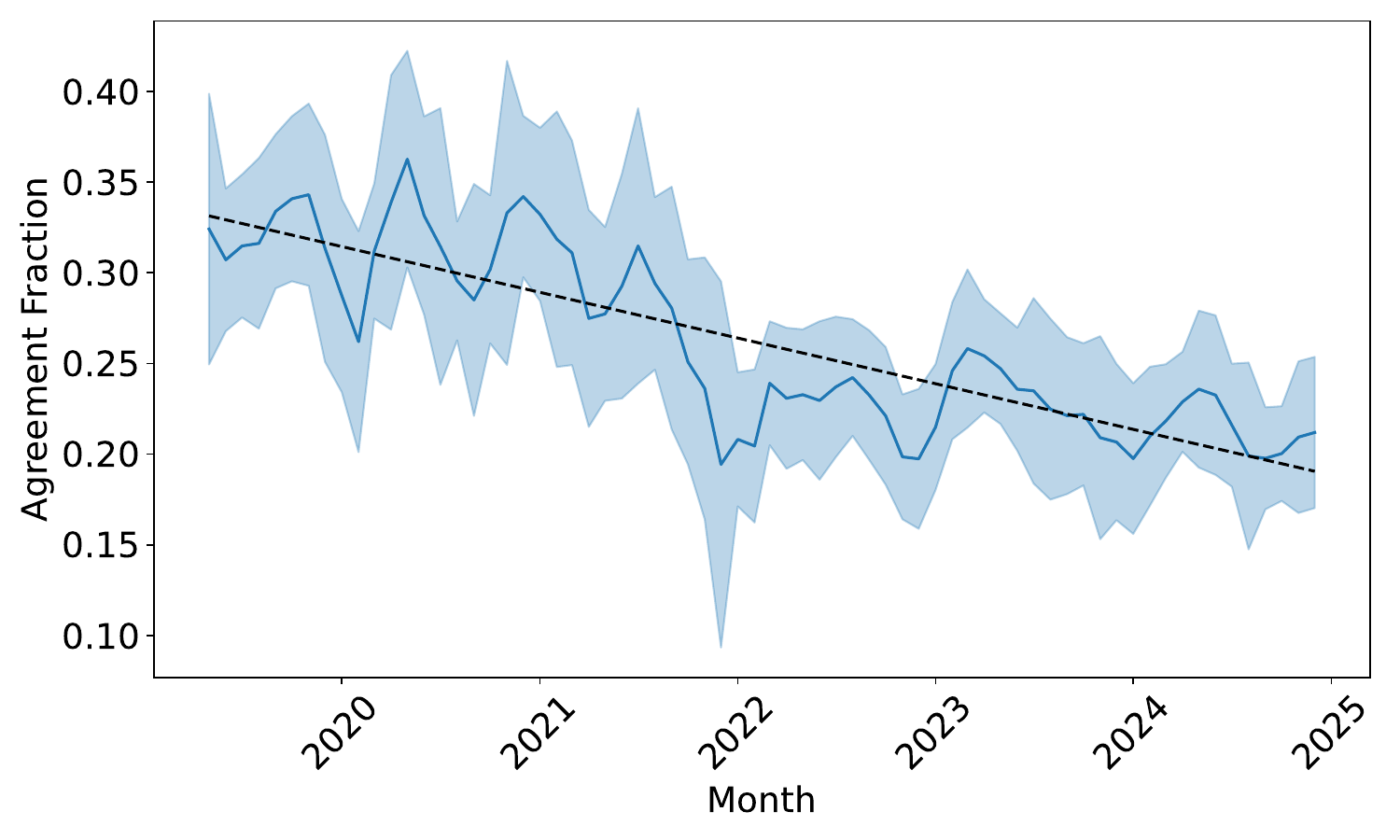}}
\end{minipage}%
\begin{minipage}{.32\linewidth}
\centering
\subfloat[The Savage Nation (MSNBC)]{\label{}\includegraphics[width=\textwidth]{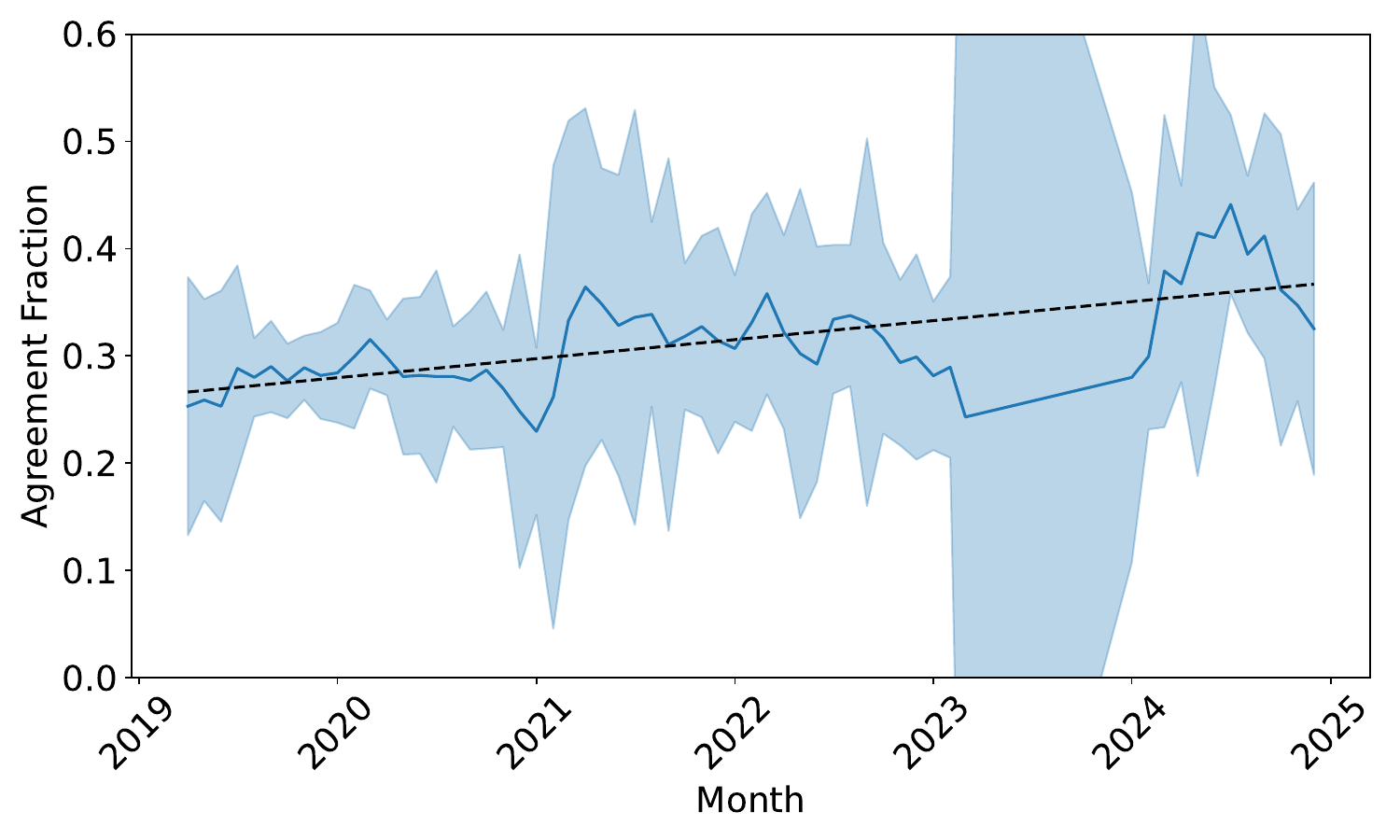}}
\end{minipage}%
\par\medskip

\caption{Trends in mean agreement fraction over the years for a sample of the shows in our dataset.}
\label{fig:agreement_over_time}
\vspace{-\baselineskip}
\end{figure*}

\subsubsection{Longitudinal case studies}
To gauge how network‑level patterns translate into editorial decisions on individual shows, we isolate three Fox News franchises for which we possess either uninterrupted long‑run data or a well‑documented format change: \textit{Hannity}, \textit{Special Report with Bret Baier}, and \textit{Gutfeld!}. Each offers a distinct window onto the evolving relationship between host persona, booking strategy, and discursive diversity.

\paragraph{Sean Hannity}

\begin{figure*}[ht]
\centering
\begin{minipage}{.32\linewidth}
\centering
\subfloat[Hannity (Fox)]{\label{}\includegraphics[width=\textwidth]{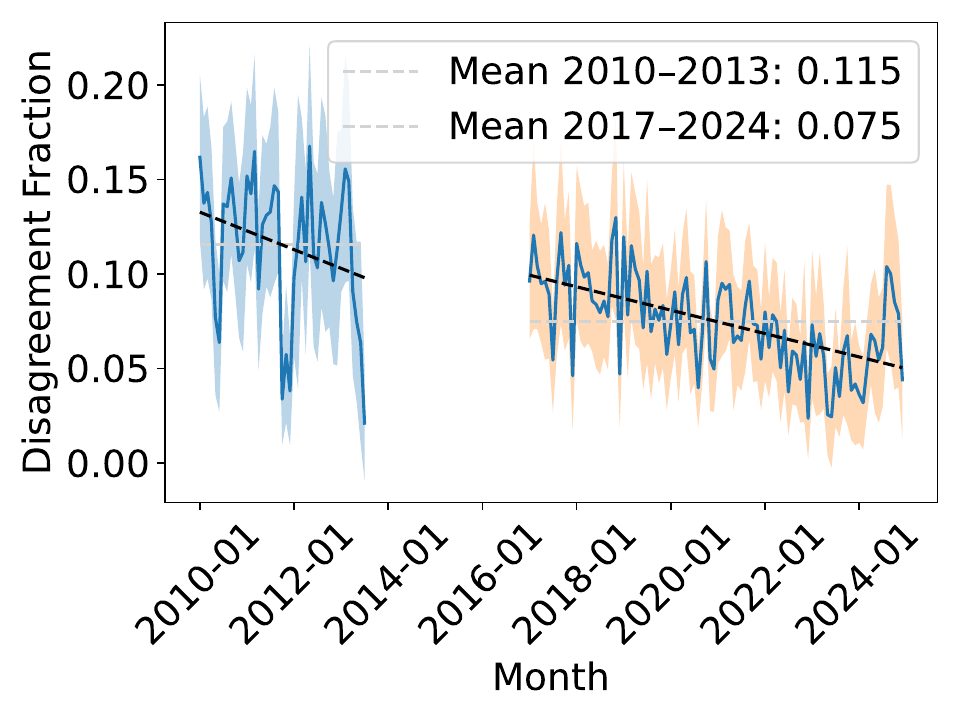}}
\end{minipage}%
\begin{minipage}{.32\linewidth}
\centering
\subfloat[Bret Baier (Fox)]{\label{}\includegraphics[width=\textwidth]{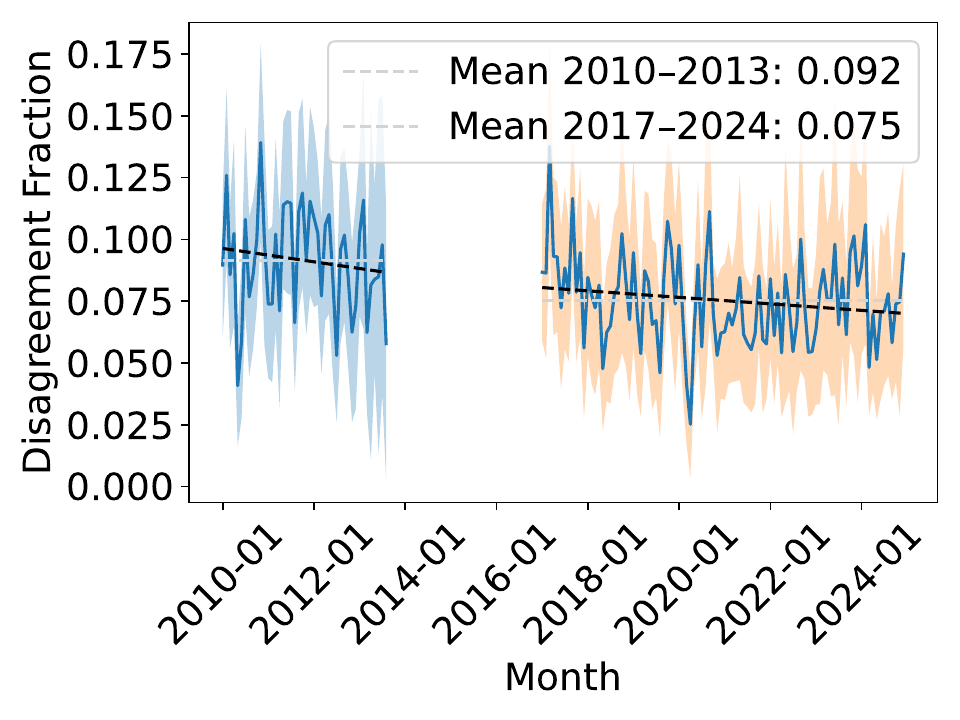}}
\end{minipage}%
\begin{minipage}{.32\linewidth}
\centering
\subfloat[Gutfeld (Fox)]{\label{}\includegraphics[width=\textwidth]{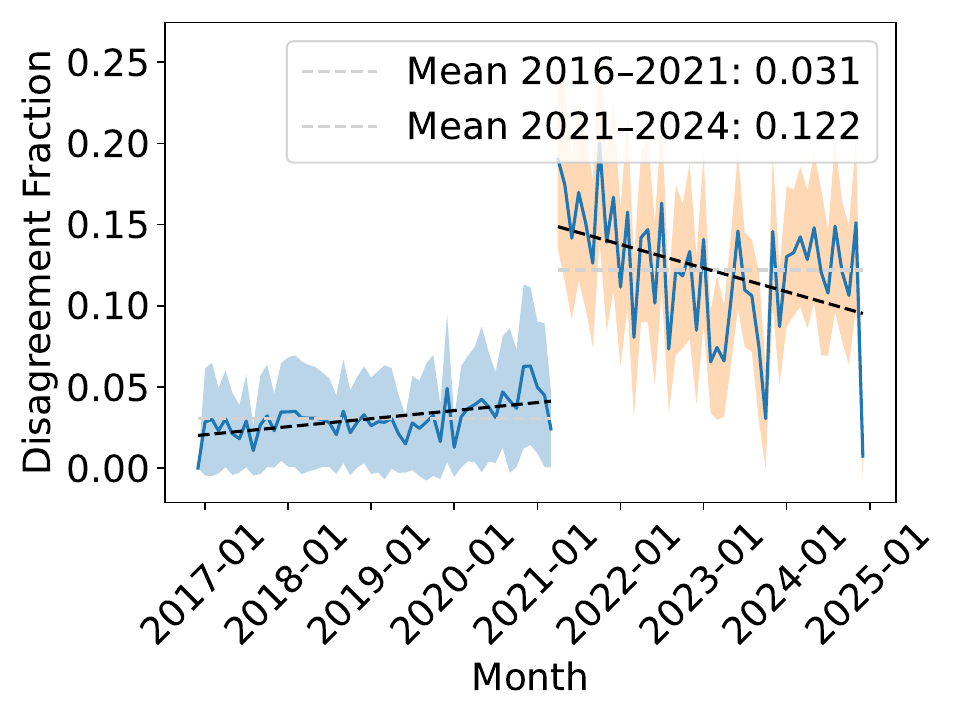}}
\end{minipage}%
\par\medskip

\caption{Disagreement over time for Hannity, Baier and Gutfeld.}
\label{fig:disagreement_over_time_special}
\end{figure*}

Sean Hannity’s primetime hour has been on the schedule since 2010, but its tone shifted markedly once Donald Trump emerged as the Republican standard‑bearer. Hannity not only endorsed Trump on air; he also appeared on stage at a 2018 campaign rally in Cape Girardeau, Missouri, blurring the line between commentator and surrogate \cite{politico2018hannity}. Our data mirror that realignment. The average disagreement fraction drops from 0.115 in the 2010–2013 window to 0.075 between 2017 and 2024—a 35\% contraction that is statistically significant at p $<$ 0.001 (Figure~\ref{fig:disagreement_over_time_special}a). Over the same span, explicit agreement rises by nearly six percentage points (Appendix, Figure~\ref{fig:agreement_over_time_all_shows}f), confirming that what little debate and discussion remained has largely been replaced by consonant, in‑group affirmations. The show thus exemplifies how personal political engagement by a host can accelerate the migration from debate to echo chamber.

\paragraph{Bret Baier}

By contrast, Bret Baier---long marketed as Fox’s “fair and balanced” chief political anchor \cite{darcy2023baier} ---shows no comparable drift. Across the full series (2010–2024) the monthly disagreement share hovers near 8\%, with early‑period (2010–2013) and late‑period (2017–2024) means of 0.092 and 0.075 respectively, a difference that is not statistically distinguishable from zero (Figure~\ref{fig:disagreement_over_time_special}b) . Agreement frequencies are likewise flat (Appendix, Figure~\ref{fig:agreement_over_time_all_shows}o). In other words, even as Fox’s opinion block narrows its viewpoint corridor, the network’s flagship news show preserves a modest but steady level of dissent which is slightly comforting for the state of journalism and news reporting.

\paragraph{Greg Gutfeld}

The third case illustrates how structural changes in scheduling can rewire host–guest interaction. Fox announced in April 2021 that \textit{The Greg Gutfeld Show}, formerly a weekend panel show, would move to weeknights at 11 p.m. under the new title \textit{Gutfeld!} \cite{imdb2025gutfeld}.

The shift coincided with a pronounced jump in disagreement: mean dissent climbs from 0.031 in 2016–2021 to 0.122 in 2021–2024 (Figure~\ref{fig:disagreement_over_time_special}c) . Agreement, by contrast, contracts slightly (Appendix, Figure~\ref{fig:agreement_over_time_all_shows}e). One plausible interpretation is that the late‑night talk‑show format rewards performative sparring, yielding more frequent but still low‑stakes disagreement; yet even after the quadrupling, adversarial exchanges fill barely 12\% of airtime, leaving the show well within the echo‑chamber regime typified by its prime‑time siblings.

Taken together, these longitudinal profiles reveal the heterogeneous pathways through which debate show devolve (or resist devolving) into an ideological monologue. Hannity’s alignment with partisan activism precipitates a steady erosion of dissent; Bret Baier’s adherence to straight‑news conventions stabilizes it; and \textit{Gutfeld!}’s format reboot jolts the metric upward without restoring balance. Each trajectory, however, converges on the same broader outcome documented in Sections \ref{sec:guests_findings} and \ref{sec:topics_findings}: a shrinking space for genuine contestation even as coverage gravitates toward the most polarizing policy arenas of American politics.

%\begin{figure}
%    \centering
%    \includegraphics[width=0.5\textwidth]{img/all_shows_by_channel_post2017.pdf}
%    \caption{Caption}
%    \label{fig:all_shows}
%\end{figure}

\subsubsection{Channel‑level trajectories: yearly aggregates, 2017–2024}

\begin{figure}[ht]
    \centering
    \includegraphics[width=0.5\textwidth]{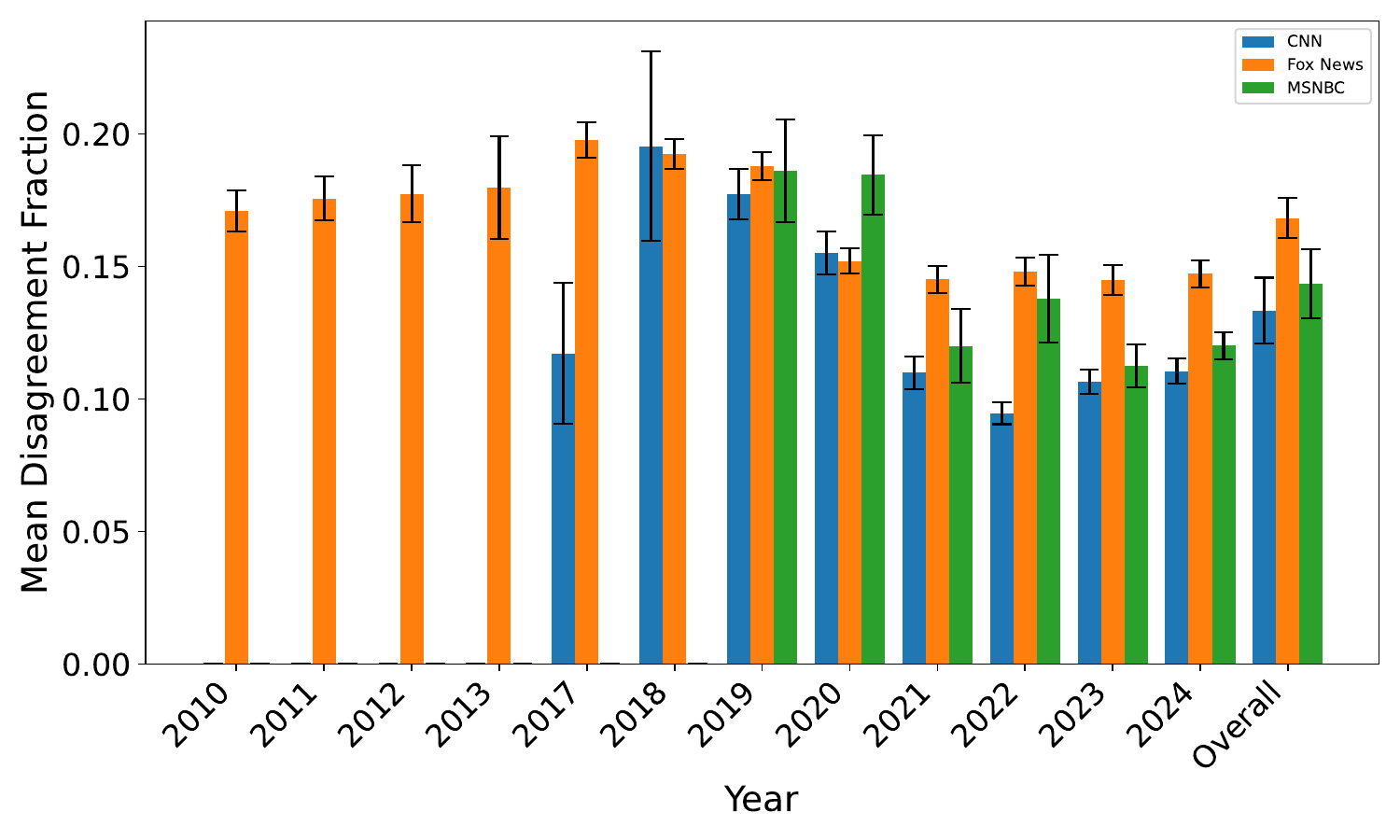}
    \caption{Yearly trends in disagreement for the 3 channels.}
    \label{fig:yearly_trends_disagreement}
    \vspace{-\baselineskip}
\end{figure}

Figure~\ref{fig:yearly_trends_disagreement} collapses the show‑specific results into annual means of disagreement fractions for each channel. Three facts stand out. First, the direction of movement in disagreement is clear: between 2017 and 2024 the average share of explicit host–guest disagreement falls across the board---by roughly one---third on Fox News, a quarter on CNN, and a fifth on MSNBC. The decline is gradual rather than abrupt, but its consistency across eight successive years suggests a structural shift in booking or production practices rather than ephemeral news‑cycle shocks. Second, the ordering of the bars is counter‑intuitive (especially for the Overall averages (right most set of bars in Figure~\ref{fig:yearly_trends_disagreement})). Fox, long portrayed in the literature as an archetypal echo chamber \cite{jamieson2008echo}, nonetheless retains the highest level of dissent throughout the series, finishing 2024 at just over 0.10---still above CNN’s 0.08 and MSNBC’s 0.07. One plausible explanation is our sample's composition: since we have more shows from Fox which also include panel, opinion and news shows, whereas for CNN and MSNBC we primarily have only opinion shows. Third, although the networks start from different baselines, the slopes are convergent; the gap between Fox and its competitors shrinks by almost two percentage points over the eight‑year window, signaling a sector‑wide contraction in adversarial dialogue.

\begin{figure}
    \centering
    \includegraphics[width=0.5\textwidth]{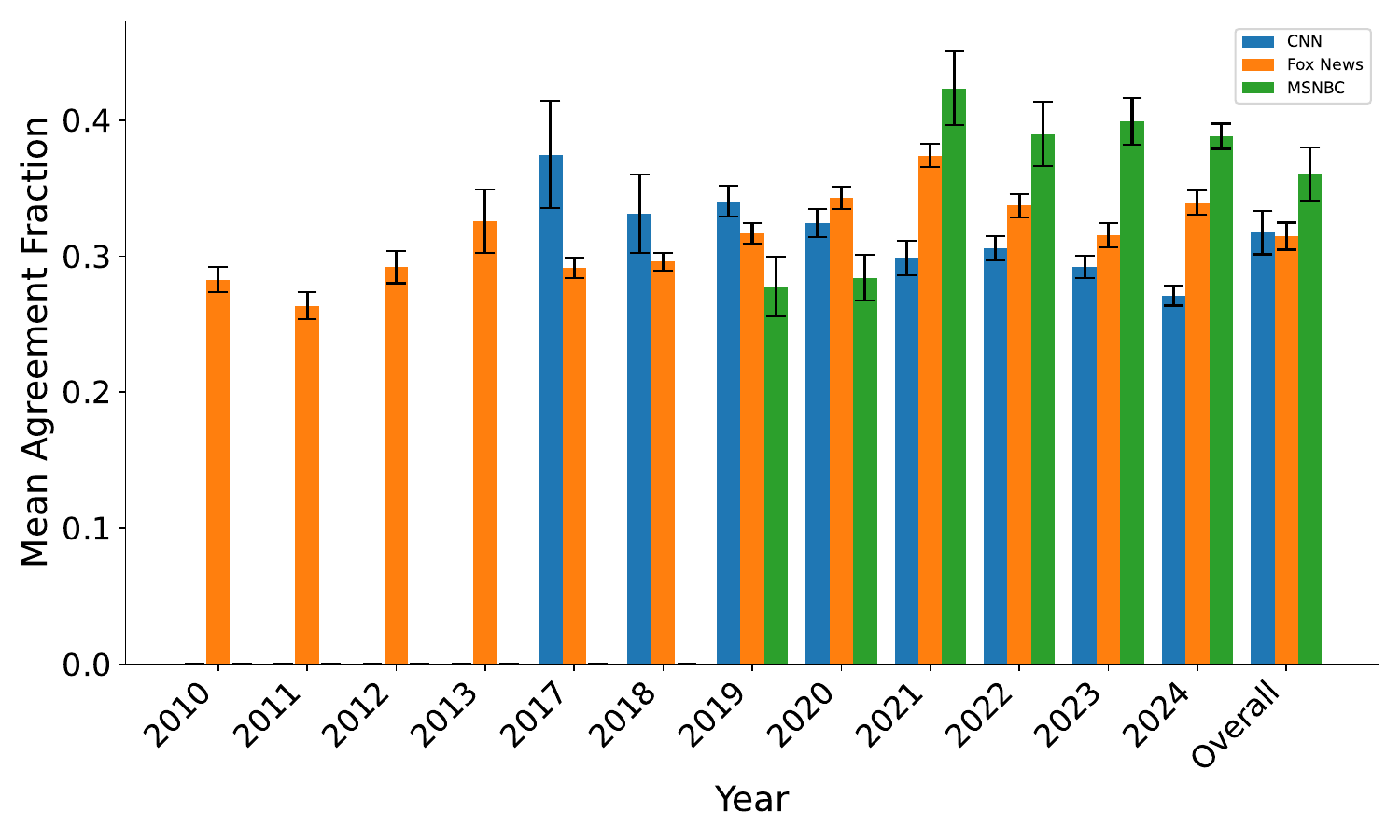}
    \caption{Yearly trends in agreement for the 3 channels.}
    \label{fig:yearly_trends_agreement}
    \vspace{-\baselineskip}
\end{figure}

The mirror image appears in Figure~\ref{fig:yearly_trends_agreement}. Average agreement climbs steadily on all three channels, with Fox again leading the pack: its mean agreement fraction rises from just under 0.40 in 2017 to about 0.55 in 2024, a gain of fifteen percentage points. CNN and MSNBC follow the same upward path, though from lower starting points and at a slightly slower pace. By the end of the period, the combined share of agreement plus neutrality exceeds 90\% of airtime for every network--a level at which meaningful counter‑argument becomes the exception rather than the rule.

Taken together with the show‑level trajectories documented above, the yearly aggregates show clearly a troubling pattern where networks are shedding disagreement more quickly than it is gaining it. The data thus corroborate a core claim of selective‑exposure scholarship that modern cable news tends to narrow, rather than widen, the interpretive frame of its audiences~\cite{stroud2011niche} but they also sharpen it: the narrowing is not merely a function of audience choice across outlets; it is being actively reinforced within the outlets themselves.

%Next, we plot yearly trends in disagreement for all three channels.

%Figure~\ref{fig:yearly_trends_disagreement} show that there is an overall decrease. across all channels. We can see that there is a roughly overall decreasing trend over the years across all three channels.

%Overall, fox news has the highest disagreement which is not what we expected. MSNBC has the least (the final set of bars in Figure~\ref{fig:yearly_trends_disagreement}).

%Figure~\ref{fig:yearly_trends_agreement} shows the trend for agreement.

%\begin{figure}
%    \centering
%    \includegraphics[width=0.5\textwidth]{img/number_of_episodes_per_show.pdf}
%    \caption{Caption}
%    \label{fig:all_shows}
%\end{figure}

\subsection{Disagreement by speakers affiliation}
\label{sec:guests_findings}

\begin{figure*}[ht]
\centering
\begin{minipage}{.32\linewidth}
\centering
\subfloat[CNN]{\label{}\includegraphics[width=\textwidth]{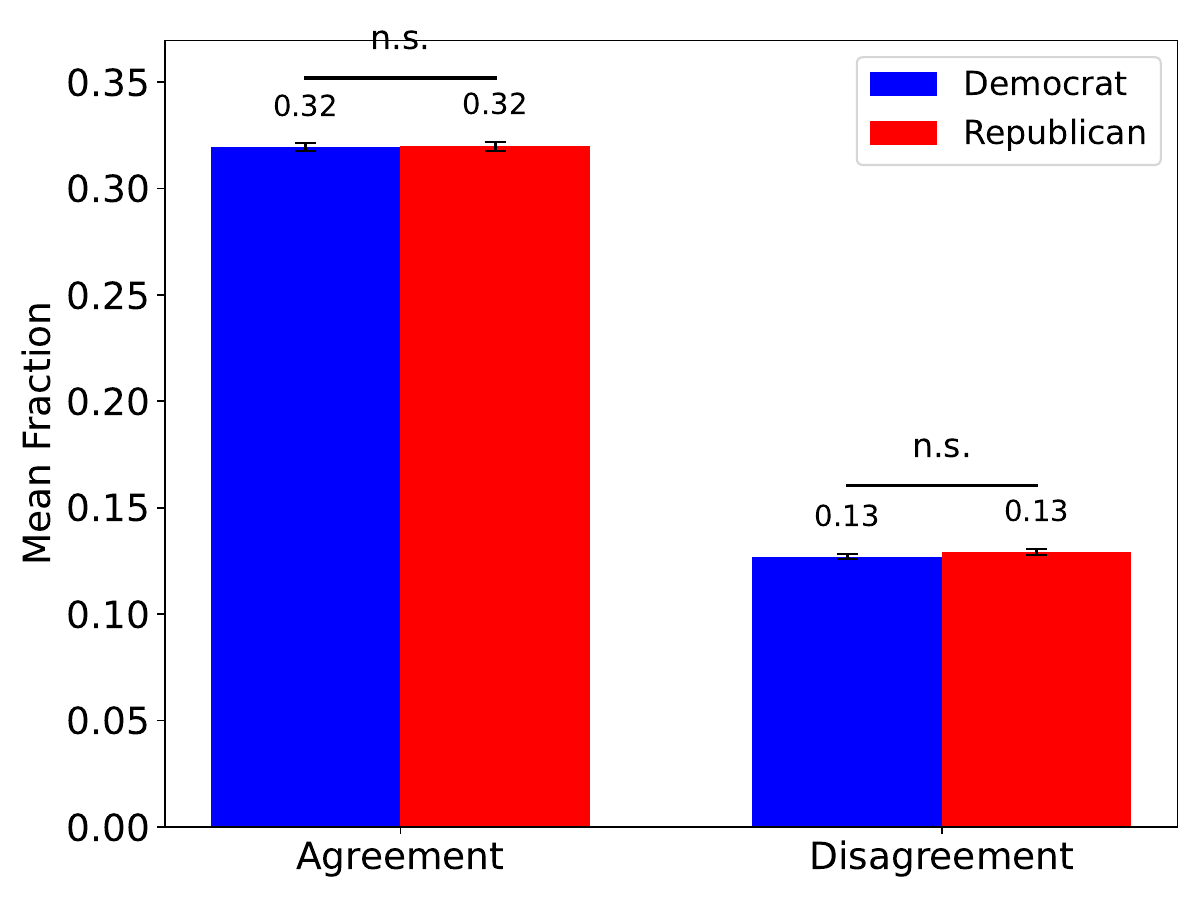}}
\end{minipage}%
\begin{minipage}{.32\linewidth}
\centering
\subfloat[Fox News]{\label{}\includegraphics[width=\textwidth]{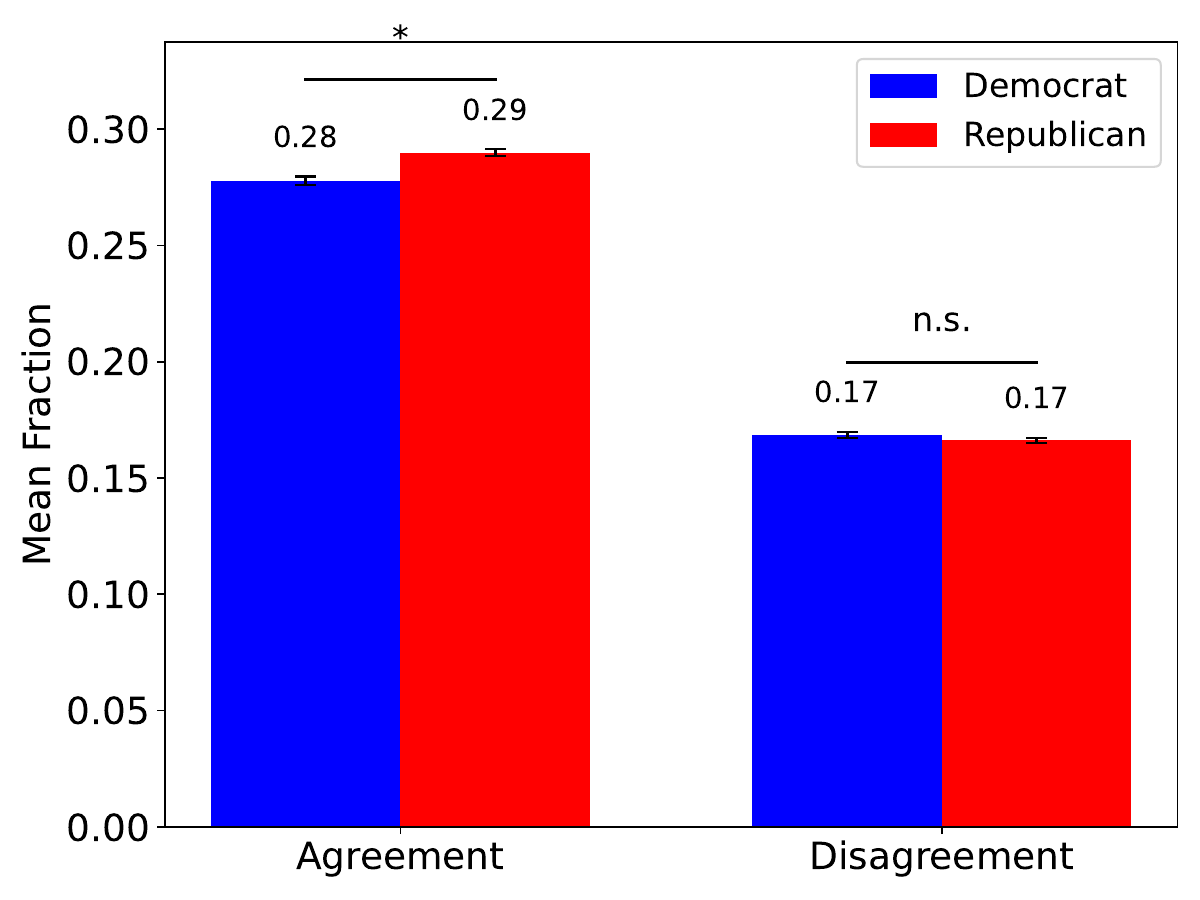}}
\end{minipage}%
\begin{minipage}{.32\linewidth}
\centering
\subfloat[MSNBC]{\label{}\includegraphics[width=\textwidth]{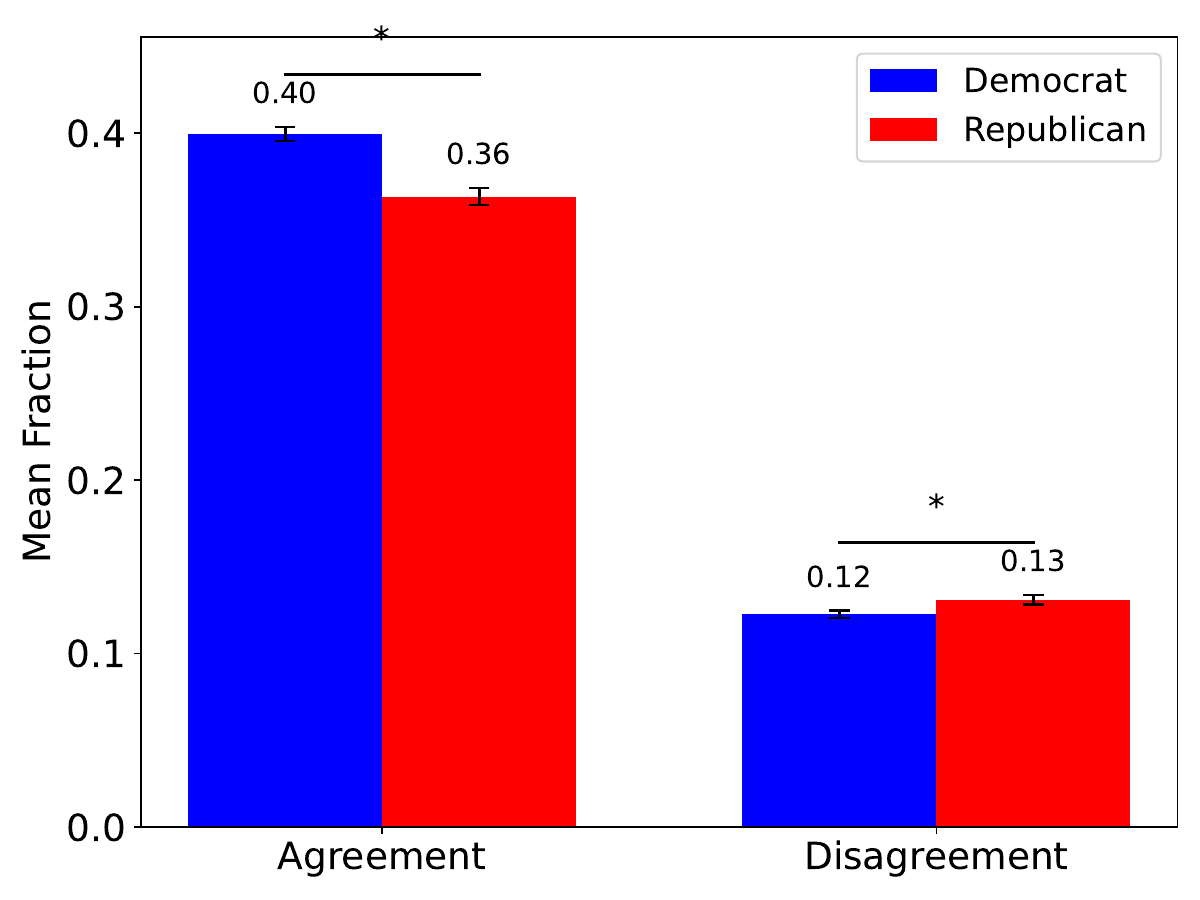}}
\end{minipage}%
\par\medskip
\caption{Average agreement/disagreement by political party of guests invited on a channel. $^*$ indicates statistical significance at $p < 0.01$. $n.s.$ indicates no significance.}
\label{fig:disagreement_by_channel}
\vspace{-\baselineskip}
\end{figure*}

The most straightforward pathway to ideological consonance is simply to curate the guest list. Figure \ref{fig:guests_party_channel} confirms that strategy in practice: Republicans dominate Fox, Democrats crowd MSNBC, while CNN remains roughly balanced, mirroring earlier content‑analysis work on cable line‑ups~\cite{kim2022measuring}. Yet composition alone cannot explain on‑air dynamics; what matters is how those guests are treated once the cameras roll. Figure~\ref{fig:disagreement_by_channel} disaggregates the data by the guest’s party label and the network to which they have been invited, allowing us to test whether “friendly” visitors are rewarded with deference and “outsiders” confronted with push‑back.

On Fox, Republican guests do enjoy marginally warmer receptions: the average share of explicit agreement for Republicans significantly higher than for Democrats, while disagreement is identical for both groups. Statistical significance tested using a Welch's t-test assuming unequal variance. $p < $0.01 (Figure~\ref{fig:disagreement_by_channel}a).
%Neither gap rises to statistical significance for disagreement (n.s.), and only the agreement difference clears the \emph{p},$<$,0.05 threshold. 
The substantive effect, however, is well under one percentage point suggesting that Fox’s echo‑chamber reputation is driven more by \emph{who} appears on screen than by large differences in tone once the invitation is extended. 
Put differently, the ideological filter operates at the booking stage; once a Democrat is allowed through that filter, the host’s deference scarcely deviates from the baseline accorded to Republican regulars.

MSNBC displays a clearer in‑group bias. Democratic guests receive a 0.40 agreement share, four points higher than Republicans (0.36), and experience slightly less disagreement (0.12 vs.\ 0.13). Both differences are statistically significant at $p < $0.01 (Figure \ref{fig:disagreement_by_channel}b). Although the raw differences may look small, recall that disagreement is already exceedingly rare; a one‑point reduction in dissent represents nearly 8\% of all hostile exchanges on the network. The result tallies with scholarship arguing that liberal‑leaning outlets enforce ideological solidarity more tightly than their conservative counterparts once the guest roster is set~\cite{sobieraj2011incivility}.

CNN, by contrast, shows virtually no partisan differential.
The absence of a treatment effect is consistent with CNN’s self‑branding as a “down‑the‑middle” outlet and resonates with earlier findings that its conflict cues flow more from issue framing than from guest identity~\cite{arceneaux2013changing}.

%Figure~\ref{fig:guests_party_channel} shows the party affiliation of the guests. as expected fox news invites mostly republicans and msnbc invites democrats.

%We already saw that partisan news organizations like MSNBC and Fox invite mostly their own side (Figure~\ref{fig:guests_party_channel}).

%\begin{figure}
%    \centering
%    \includegraphics[width=0.5\textwidth]{img/combined_party_counts.pdf}
%    \caption{Party of guests}
%    \label{fig:guests_party}
%\end{figure}

%Figure~\ref{fig:disagreement_by_channel} gives the agreement/disagreement by channel. Including significance testing.

\subsection{Disagreement by topic}
\label{sec:topics_findings}

Table \ref{tab:topics_disagreement} ranks the five topics that elicit the least and most host–guest disagreement on each channel. Topics with least disagreement indicate issues where there was little counter voice or debate. Three patterns emerge:

First, the same topics surface in the “least disagreement” column across otherwise opposed outlets. Polarizing topics such as Abortion, Immigration, Gun violence and Policing appear among the five topics on most of channels. These are precisely the issues that animate contemporary culture‑war politics, yet they rarely provoke open discussion on the news channels. 
This pattern strongly supports our argument that these topics are not being used as subjects for debate, but rather as vehicles for narrative reinforcement and in-group consensus-building.
%Instead, hosts tend to book like‑minded discussants, turning potentially explosive conversations into monologues of affirmation.
%
Second, the topics that generate what most disagreement  are largely procedural, event‑driven or news related: the economy, elections \& campaigns, COVID‑19, etc. 
%These topics invite technocratic or horse‑race argumentation—How large should the stimulus be? Did the court overreach?—and therefore accommodate genuine disagreement even inside homogeneous partisan line‑ups.
%
Third, there are interesting topics with little disagreement specific to channels such as discussions on Culture wars on both Fox News and MSNBC. Similarly, shows related to the investigations and impeachment of Donald Trump were among the least disagreed on MSNBC. 

The implication from the findings in the previous two sections is that ideological filtering operates first at the guest‑selection stage and then again at the topic‑selection stage, leaving only narrow windows in which cross‑cutting viewpoints make it to air.
Taken together with the longitudinal evidence above, the topic analysis shows a central irony of partisan television. The shows devote ever more airtime to the controversies that polarize American politics (see Figure~\ref{fig:topics_per_channel_over_time}), yet they purge those very segments of meaningful disagreement. 

%Plot: topics with most agreement and most disagreement.

%trends in agreement and disagreement about specific topics (e.g. gun control, abortion, immigration) over time.

%partisan media ecosystems tend to produce debate that is high on polarization and spectacle, but low on cross-cutting dialogue. This dynamic contributes to broader polarization, as the “new and expanded ideological media” are believed to fuel sharper partisan divisions in society. TODO IS THIS CORRECT?

%Figure~\ref{fig:disagreement_by_topic} shows the top 5 and bottom 5 topics with most and least disagreement. We can see polarized topics like abortion, gun control and race having low disagreement.

\begin{table}[ht]
\small
    \centering
    \begin{tabular}{c p{3cm} p{3cm}}
        \hline
        \textbf{Channel} & \textbf{Least disagreement} & \textbf{Most disagreement} \\
        \hline
        \multirow{5}{*}{Fox News}
            & Abortion & COVID      \\
            & Cultural Politics & Economy \\
            & Immigration & Elections \& Campaigns \\
            & Gun Violence & Judiciary \\
            & Race \& Policing & Trump Investigations \\
        \hline
        \multirow{5}{*}{MSNBC}
            & Abortion & Judiciary \\
            & Trump Investigations & Elections \& Campaigns \\
            & Cultural Politics & Economy \\
            & Immigration & Foreign Policy \\
            & Race \& Policing & COVID \\
        \hline
        \multirow{5}{*}{CNN} 
            & Immigration & Economy \\
            & Judiciary & Elections \& Campaigns \\
            & Gun Violence & COVID \\
            & Foreign Policy & Cultural Politics \\
            & Abortion & Trump Investigations \\
        \hline
    \end{tabular}
    \caption{Topics with the most \& least disagreement per Channel. Topic names were simplified to fit in the table.}
    \label{tab:topics_disagreement}
    \vspace{-\baselineskip}
    
\end{table}

%\begin{figure*}[ht]
%\centering
%\begin{minipage}{.32\linewidth}
%\centering
%\subfloat[]{\label{}\includegraphics[width=\textwidth]{img/CNN_topic_disagreement.pdf}}
%\end{minipage}%
%\begin{minipage}{.32\linewidth}
%\centering
%\subfloat[]{\label{}\includegraphics[width=\textwidth]{img/Fox_News_topic_disagreement.pdf}}
%\end{minipage}%
%\begin{minipage}{.32\linewidth}
%\centering
%\subfloat[]{\label{}\includegraphics[width=\textwidth]{img/MSNBC_topic_disagreement.pdf}}
%\end{minipage}%
%\par\medskip
%\caption{top 5 and bottom 5 topics with most disagreement.}
%\label{fig:disagreement_by_topic}
%\end{figure*}

\section{Discussion}

\paragraph{Summary of principal findings.}
Our speaker–turn analysis reveals a debate ecosystem that is strikingly placid by historical and cross‑network comparison. Across the 2017–2024 window fewer than one‑in‑six host–guest exchanges contains explicit dissent; MSNBC is lowest at 13\% and Fox highest at 17\% (Figure~\ref{fig:disagreement_by_channel}). More troubling is direction of travel: disagreement declines year‑on‑year for almost every Fox and CNN show and is flat (but never high) on the MSNBC shows for which long‑run data exist (Figure~\ref{fig:disagreement_over_time_all_shows}). The trajectory is not inevitable: archival captions from the early 2010s show \emph{Hannity} hovering near 25\% dissent, a level now unmatched by any prime‑time hour. Guest composition compounds the effect. Networks already skew their invitation lists toward ideologically sympathetic voices, yet our turn‑level labeling shows that even once a cross‑partisan guest appears the dialogue tilts toward conformity: Republican invitees receive gentler treatment on Fox, Democrats on MSNBC, with CNN occupying a narrowing center. Finally, the very topics that polarize the electorate---abortion, gun rights, immigration---attract the least disagreement, converting ostensibly “hot” segments into performances of in‑group consensus.

\paragraph{What do these patterns mean?}
Two implications stand out. First, the findings operationalize the “spectacle without contestation” hypothesis advanced by selective‑exposure theory: cable outlets dramatize ideological battle lines through topic choice and affective framing yet withhold the clash of arguments that would expose audiences to competing rationales. That configuration can intensify \emph{affective} polarization---warmth toward one’s own side and hostility toward the other---while leaving factual beliefs untested, a dynamic known to entrench misperceptions and policy gridlock. Second, the disappearance of dissent is not merely a product of audience self‑sorting across channels; it is reinforced \emph{within} channels by editorial practices that book, frame, and moderate conversations so as to minimize overt conflict. The data therefore complicate optimistic accounts that place the onus solely on viewer choice: even citizens who seek out disagreement may be offered little once they tune in.

\paragraph{Why the decline?}
Several mechanisms are plausible. Audience analytics reward affective reinforcement more reliably than cognitive challenge; hosts cultivate personal brands tied to partisan authenticity; and the economics of booking favor a stable roster of predictable allies. The COVID-19 pandemic further normalized remote interviews, making it easier to screen out wild‑card guests. Our data cannot pinpoint causal weightings, but the cross‑network symmetry—in which left‑leaning and right‑leaning channels mirror each other—suggests structural, not idiosyncratic, forces.

\paragraph{Is disagreement the right metric?}
Raw frequency of dissent is a blunt instrument. Some disagreements are performative, others substantive; some are civil, others demagogic. Still, frequency matters for at least three reasons. First, democratic theory assigns deliberation a central role: citizens form better‑justified preferences when exposed to counter‑arguments. Second, low baseline dissent limits the potential for “constructive conflict,” because intensity cannot compensate for absence. Third, our classifier is conservative: sarcasm, hedged rebuttals, and rapid‑fire interruptions often register as \textit{neutral}, implying that true argumentative engagement may be even rarer than we report. Future work should layer additional dimensions—tone, evidence quality, emotional valence—onto the stance signal to distinguish shallow sparring from reason‑giving debate.

\paragraph{How Do Offline Debates Mirror Online Echo Chambers?}
Unlike algorithmically mediated feeds, the cable-news environment is tightly moderated by hosts, limiting opportunities for organic cross-cutting exchange \cite{hosseinmardi2025unpacking}. In this sense, our findings highlight an offline pathway to echo chamber effects \cite{ash2024viewers}: the absence of disagreement in televised debates can serve as a functional analogue to algorithmic curation online. Situating our work in this way underscores its relevance: understanding echo chambers requires analyzing both digital and legacy media environments that together shape citizens’ partisan worldviews \cite{muise2022quantifying}.

\paragraph{Limitations.}
The study has four chief constraints. \emph{Coverage}: speaker‑label quality forced reliance on podcast feeds for many MSNBC and CNN shows, shortening their temporal span relative to Fox. \emph{Measurement error}: ASR inaccuracies, diarization drift, and stance‑classifier mistakes (accuracy=0.89) introduce noise, though audits indicate no directional bias large enough to change headline trends. \emph{Affiliation coding}: DIME donations are an imperfect proxy for ideological leaning, especially for non‑U.S. guests or low‑profile experts. \emph{Missing modalities}: our analysis ignores visuals, tone of voice, and real‑time fact‑checking, all of which shape viewer perception of conflict.

\paragraph{Future directions.}
Integrating sentiment arcs and evidence quality could distinguish “heat” from “light” in televised argument. Linking caption‑level disagreement to real‑time audience metrics (e.g., minute‑by‑minute ratings, social‑media engagement) would clarify whether viewers reward or punish dissent. Finally, extending the method to streaming talk shows and YouTube political channels would test whether the retreat from contestation is a cable‑specific artifact or a broader feature of contemporary political media.

Overall, the erosion of on‑air disagreement documented here challenges the longstanding notion of cable news as a public forum for competing ideas. If deliberation is an essential democratic input, then reviving formats that encourage genuine clash—not just spectacle—remains an urgent task for journalists, regulators, and scholars alike.

%Overall, there are two items of interest: the average disagreement and the trends over time. Firstly, MSNBC has a low disagreement on average (0.13) compared to Fox (0.17) (See Figure~\ref{fig:disagreement_by_channel}). 
%Trends are downward consistently for Fox and CNN (Figure~\ref{fig:disagreement_over_time_all_shows}).

%The “spectacle without contestation” dynamic documented by selective‑exposure research is thus not an abstract tendency but a measurable feature of day‑to‑day production. Cable news delivers vivid cues about where the battle lines are drawn, while systematically shielding viewers from how the other side argues its case.

%Is disagreement really a measure we should care about? does intensity of disagreement matter?

%Is it constructive disagreement?

%\kiran{8 pages + unlimited appendix and references}

%\bibliographystyle{aaai25}
%\bibliography{biblio}

%\clearpage

\section{Appendix}

The complete list of shows used in the paper are shown in Table~\ref{tab:shows}. Due to hiatus or simply not being able to scrape data from the Internet Archive platform or even Podcast, there are some gaps in the data. They are noted in table~\ref{tab:data_gaps}.

\begin{table*}[ht]
\centering
\small
\begin{tabular}{@{}p{0.28\linewidth}p{0.62\linewidth}@{}}
\toprule
\textbf{Show} & \textbf{Coverage Gaps or Anomalies} \\
\midrule
\textit{Hannity} & No data from 2014 to 2016. \\
\textit{The Rachel Maddow Show} & Only one episode in September 2022, all labeled as agreement or neutrality because it was just Rachel and a reporter talking through the day's news. Weekly airing format may reduce disagreement frequency. \\
\textit{Outnumbered} & Very few episodes before September 2015; analysis starts from October 2015. \\
\textit{Special Report with Bret Baier} & No episode available from November 2013 to December 2016. \\
\textit{Life, Liberty \& Levin} & Some months contain only 2–3 episodes with limited speaker turns; often results in 0 recorded disagreements. \\
\textit{The Savage Nation} & Did not air from March 2023 through December 2023. \\
\textit{Erin Burnett OutFront} & Did not air from September 2021 to December 2021 and again from October 2023 to December 2023. \\
\bottomrule
\end{tabular}
\caption{Data coverage gaps and broadcast anomalies for selected shows.}
\label{tab:data_gaps}
\end{table*}

\begin{table*}[ht]
\centering
\begin{tabular}{@{} l l l l r @{}}
\toprule
Show & Channel & Type & Time span & Episodes \\
\midrule
The Tucker Carlson Show & Fox News & Weekday & Jan 2017 – Jun 2023 & 1491 \\
Hannity & Fox News & Weekday & Jul 2009 – Jan 2014 and Jan 2017 – Dec 2024 & 2441 \\
Special Report with Bret Baier & Fox News & Weekday & Jul 2009 – Oct 2013 and Jan 2017 – Dec 2024 & 2672 \\
Outnumbered & Fox News & Weekday & Oct 2015 – Dec 2024 & 1644 \\
The Ingraham Angle & Fox News & Weekday & Oct 2017 – Dec 2024 & 1584 \\
Gutfeld! & Fox News & Weekday & Jan 2017 – Dec 2024 & 1001 \\
Jesse Watters Primetime & Fox News & Weekday & Jan 2022 – Dec 2024 & 634 \\
The Story with Martha MacCallum & Fox News & Weekday & May 2017 – Dec 2024 & 1639 \\
Life, Liberty \& Levin & Fox News & Weekend & Feb 2018 – Dec 2024 & 481 \\
Anderson Cooper 360° & CNN & Weekday & May 2019 – Dec 2024 & 1203 \\
Erin Burnett OutFront & CNN & Weekday & Jun 2019 – Dec 2024 & 1206 \\
Inside Politics & CNN & Weekday & Jul 2019 – Dec 2024 & 1210 \\
The Lead with Jake Tapper & CNN & Weekday & Jun 2019 – Dec 2024 & 1056 \\
The Source with Kaitlan Collins & CNN & Weekday & Jul 2023 – Dec 2024 & 334 \\
Laura Coates Live & CNN & Weekday & Oct 2023 – Dec 2024 & 283 \\
State of the Union & CNN & Weekend & Apr 2018 – Dec 2024 & 316 \\
Fareed Zakaria GPS & CNN & Weekend & Apr 2017 – Dec 2024 & 353 \\
The Beat with Ari Melber & MSNBC & Weekday & Jan 2023 – Dec 2024 & 288 \\
Inside with Jen Psaki & MSNBC & 2 Weekday & Jun 2023 – Dec 2024 & 130 \\
Alex Wagner Tonight & MSNBC & Weekday & Jan 2024 – Dec 2024 & 179 \\
Velshi & MSNBC & Weekend & Feb 2021 – Dec 2024 & 340 \\
The Sunday Show with Jonathan Capehart & MSNBC & Weekend & Jan 2023 – Dec 2024 & 117 \\
The Savage Nation & MSNBC & Weekday & Apr 2019 – Dec 2024 & 648 \\
The Rachel Maddow Show & MSNBC & Weekly & May 2022 – Dec 2024 & 126 \\
\bottomrule
\end{tabular}
\caption{Broadcast shows, overall time span, and total episode counts.}
\label{tab:shows}
\vspace{-\baselineskip}
\end{table*}

\subsection{Detecting agreement/disagreement}

Our stance–labeling workflow unfolded in four stages, moving from small, manually curated seeds to a production‑scale LLM annotator.

\paragraph{Stage 1: baseline with task‑specific encoders.}
We first adopted the DialAM‑2024 framework of \citet{wu-etal-2024-knowcomp}, fine‑tuning \textsc{DeBERTa‑v3‑base} on 160 of 200 hand‑labeled bundles from \emph{Tucker Carlson Tonight}. Test accuracy reached only 55\%, and augmenting the training set with the original DialAM corpus drove accuracy down to 48.5\%. These results ruled out conventional encoder–classifier models for our domain.

\paragraph{Stage 2: iterative GPT‑seed expansion.}
Turning to instruction‑tuned LLMs, we grew the seed set to 500 bundles via an interactive loop with ChatGPT (GPT‑4). The model labelled batches of 20 pairs in zero‑shot mode; authors corrected its mistakes and fed the revised items back as demonstrations before issuing the next batch. A subsequent 80/20 train–test split yielded 73\% accuracy for zero‑shot GPT‑4 and 79\% after fine‑tuning \textsc{gpt‑3.5‑turbo}. These figures established a workable floor but left room for improvement.

\paragraph{Stage 3: crowd round 1 and consensus filtering.}
To enlarge the training pool we recruited 150 U.S. participants on CloudResearch \cite{hartman2023introducing}. The participants were eligible if they were located in the U.S., aged $\geq 18$, had an historical approval rate $\geq 95\%$, and passed a short English comprehension screener. These were all set by the CloudResearch platform and we just had to upload our task website and collect the data from the participants. We sampled 1,000 bundles drawn equally from \emph{Tucker Carlson Tonight}, \emph{Hannity}, and \emph{The Rachel Maddow Show}. Each 20‑pair bundle was triple‑annotated on a purpose‑built website featuring guideline examples and embedded gold checks. We applied quality screening based on gold-check performance and implausibly short completion times, removing 12\% of submissions. Inter‑annotator reliability among the crowdsourced annotators was modest (Fleiss $\kappa$ = 0.302; Krippendorff’s $\alpha$ = –0.005), so to mitigate noise we used \textsc{cleanlab} \cite{goh2022utilizing} to infer consensus labels, retaining 459 high‑confidence pairs. Combining these with the 500 GPT‑seed pairs (959 total) and re‑training multiple models produced the accuracies reported in Table \ref{tab:llm_round1}.

\paragraph{Stage 4: crowd round 2, manual triage, and final model.}
Because model performance improved with training‑set size, we commissioned a second CloudResearch round for another 1,000 bundles, and did the same cleaning with \textsc{cleanlab}, to finally get 951 pairs of raw labels. Unlike in Stage 3, however, incorporating these raw labels directly reduced performance (Table \ref{tab:llm_round2}). Manual audits by one of the authors revealed a systematic source of error: crowd workers frequently labeled backchannel acknowledgments (e.g., “mm-hm,” “yeah”) as “agreement,” contrary to our stricter definition of substantive stance. We therefore hand‑verified every ambiguous triple, retained 712 balanced, high‑quality pairs, and merged them with the earlier 959 from stage 3, yielding a 1,671‑pair gold corpus. Fine‑tuning GPT-4o on this set restored accuracy to 84\%; DeepSeek-R1-Distill-Llama-8B, fine‑tuned on the same data, achieved 89\% (Table~\ref{tab:llm_final}).

\begin{table*}[t]
\centering
\small  % Reduce font size to fit in column width
\begin{tabular}{|l|p{6cm}|c|c|c|c|c|}
\hline
\textbf{Model} & \textbf{Dataset (Balanced classes)} & \textbf{Train, Test} & \textbf{Accuracy} & \textbf{Precision} & \textbf{Recall} & \textbf{F1-score} \\
\hline
DeepSeek-R1-Distill-Llama-8B & High quality subsets from the crowdsourced dataset & (339, 120) & 66.97\% & 65.87\% & 69.12\% & 66.98\% \\
\hline
DeepSeek-R1-Distill-Llama-8B & High quality subsets from the crowdsourced dataset + our own annotated 500 pairs & (498, 184) & 70.64\% & 70.31\% & 68.27\% & 70.01\% \\
\hline
Llama-3.3 70b 4bit & High quality subsets from the crowdsourced dataset & (339, 120) & 69.64\% & 67.98\% & 65.75\% & 67.43\% \\
\hline
Llama-3.3 70b 4bit & High quality subsets from the crowdsourced dataset + our own annotated 500 pairs & (498, 184) & 74.98\% & 70.56\% & 73.21\% & 71.97\% \\
\hline
Qwen-2.5 32b 4bit & High quality subsets from the crowdsourced dataset & (339, 120) & 64.08\% & 65.00\% & 64.78\% & 64.88\% \\
\hline
Qwen-2.5 32b 4bit & High quality subsets from the crowdsourced dataset + our own annotated 500 pairs & (498, 184) & 65.22\% & 65.01\% & 64.89\% & 64.95\% \\
\hline
GPT-4o & High quality subsets from the crowdsourced dataset & (339, 120) & 79.43\% & 79.41\% & 78.90\% & 79.10\% \\
\hline
\textbf{GPT-4o} & \textbf{High quality subsets from the crowdsourced dataset + our own annotated 500 pairs} & \textbf{(498, 184)} & \textbf{80.43\%} & \textbf{81.00\%} & \textbf{80.02\%} & \textbf{80.55\%} \\
\hline
\end{tabular}
\caption{Comparison of fine-tuned model accuracy using different combination of datasets (phase 1)}
\label{tab:llm_round1}
\end{table*}

\begin{table*}[t]
\centering
\small  % Reduce font size to fit in column width
\begin{tabular}{|l|p{6cm}|c|c|c|c|c|}
\hline
\textbf{Model} & \textbf{Dataset (Balanced classes)} & \textbf{Train, Test} & \textbf{Accuracy} & \textbf{Precision} & \textbf{Recall} & \textbf{F1-score} \\
\hline
DeepSeek-R1-Distill-Llama-8B & High quality subsets from the crowdsourced datasets & (528, 245) & 62.45\% & 61.98\% & 59.90\% & 60.42\% \\
\hline
DeepSeek-R1-Distill-Llama-8B & High quality subsets from the crowdsourced datasets + our own annotated 500 pairs & (861, 344) & 68.90\% & 69.02\% & 68.86\% & 69.00\% \\
\hline
Llama-3.3 70b 4bit & High quality subsets from the crowdsourced datasets & (528, 245) & 62.04\% & 62.49\% & 62.11\% & 62.37\% \\
\hline
Llama-3.3 70b 4bit & High quality subsets from the crowdsourced datasets + our own annotated 500 pairs & (861, 344) & 69.00\% & 68.43\% & 69.49\% & 69.12\% \\
\hline
Qwen-2.5 32b 4bit & High quality subsets from the crowdsourced datasets & (528, 245) & 64.08\% & 64.07\% & 63.99\% & 64.00\% \\
\hline
Qwen-2.5 32b 4bit & High quality subsets from the crowdsourced datasets + our own annotated 500 pairs & (861, 344) & 62.50\% & 60.99\% & 61.78\% & 62.34\% \\
\hline
\textbf{GPT-4o} & \textbf{High quality subsets from the crowdsourced datasets} & \textbf{(528, 245)} & \textbf{72.24\%} & \textbf{73.00\%} & \textbf{71.44\%} & \textbf{71.49\%} \\
\hline
GPT-4o & High quality subsets from the crowdsourced datasets + our own annotated 500 pairs & (861, 344) & 66.87\% & 67.08\% & 67.00\% & 66.09\% \\
\hline
\end{tabular}
\caption{Comparison of fine-tuned model accuracy using different combination of datasets (phase 2)}
\label{tab:llm_round2}
\end{table*}

\begin{table*}[t]
\centering
\small  % Reduce font size to fit in column width
\begin{tabular}{|l|p{6cm}|c|c|c|c|c|}
\hline
\textbf{Model} & \textbf{Dataset (Balanced classes)} & \textbf{Train, Test} & \textbf{Accuracy} & \textbf{Precision} & \textbf{Recall} & \textbf{F1-score} \\
\hline
GPT-4o & High quality subsets from the crowdsourced datasets (manually checked) & (459, 253) & 84.09\% & 83.96\% & 84.00\% & 83.98\% \\
\hline
\textbf{DeepSeek-R1-Distill-Llama-8B} & \textbf{High quality subsets from the crowdsourced datasets (manually checked)} & \textbf{(459, 253)} & \textbf{89.61\%} & \textbf{88.50\%} & \textbf{87.91\%} & \textbf{88.35\%} \\
\hline
\end{tabular}
\caption{Final model used}
\label{tab:llm_final}
\end{table*}

Table~\ref{tab:metrics_shows} reports the 100 manual sample checks mentioned in Section~\ref{sec:detecting_disagreement}, including if the show's dataset was taken from Archive or Apple podcast. The samples were not balanced across the 3 classes. 

\begin{table*}[ht]
\centering
\begin{tabular}{@{} l l c c c c @{}}
\toprule
Show & Source & Accuracy & Precision & Recall & F1-score \\
\midrule
The Tucker Carlson Show & Archive & 89.0\% & 87.5\% & 88.2\% & 87.8\% \\
Hannity & Archive & 88.6\% & 87.0\% & 87.9\% & 87.4\% \\
Special Report with Bret Baier & Archive & 88.2\% & 86.5\% & 87.3\% & 86.9\% \\
Outnumbered & Archive & 88.0\% & 86.2\% & 87.0\% & 86.6\% \\
The Ingraham Angle & Archive & 87.8\% & 86.0\% & 86.8\% & 86.4\% \\
Gutfeld! & Archive & 87.5\% & 85.8\% & 86.6\% & 86.2\% \\
Jesse Watters Primetime & Archive & 87.2\% & 85.5\% & 86.3\% & 85.9\% \\
The Story with Martha MacCallum & Archive & 88.1\% & 86.3\% & 87.2\% & 86.7\% \\
Life, Liberty \& Levin & Archive & 88.4\% & 86.7\% & 87.5\% & 87.1\% \\
Anderson Cooper 360° & Podcast & 86.5\% & 84.5\% & 85.3\% & 84.9\% \\
Erin Burnett OutFront & Podcast & 86.2\% & 84.3\% & 85.0\% & 84.6\% \\
Inside Politics & Podcast & 86.0\% & 84.0\% & 84.8\% & 84.4\% \\
The Lead with Jake Tapper & Podcast & 85.8\% & 83.9\% & 84.7\% & 84.3\% \\
The Source with Kaitlan Collins & Podcast & 85.6\% & 83.7\% & 84.5\% & 84.1\% \\
Laura Coates Live & Podcast & 85.5\% & 83.6\% & 84.4\% & 84.0\% \\
State of the Union & Podcast & 85.7\% & 83.8\% & 84.6\% & 84.2\% \\
Fareed Zakaria GPS & Podcast & 85.9\% & 84.0\% & 84.8\% & 84.4\% \\
The Beat with Ari Melber & Podcast & 86.3\% & 84.4\% & 85.2\% & 84.8\% \\
Inside with Jen Psaki & Podcast & 86.1\% & 84.2\% & 85.0\% & 84.6\% \\
Alex Wagner Tonight & Podcast & 85.8\% & 83.9\% & 84.7\% & 84.3\% \\
Velshi & Podcast & 86.0\% & 84.1\% & 84.9\% & 84.5\% \\
The Sunday Show with Jonathan Capehart & Podcast & 85.7\% & 83.8\% & 84.6\% & 84.2\% \\
The Savage Nation & Podcast & 86.2\% & 84.3\% & 85.1\% & 84.7\% \\
The Rachel Maddow Show & Podcast & 86.4\% & 84.5\% & 85.3\% & 84.9\% \\
\bottomrule
\end{tabular}
\caption{Performance metrics of different datasets after final model labeling}
\label{tab:metrics_shows}
\vspace{-\baselineskip}
\end{table*}

\subsection{Topic analysis}

\begin{figure*}
    \centering
    \includegraphics[width=\textwidth]{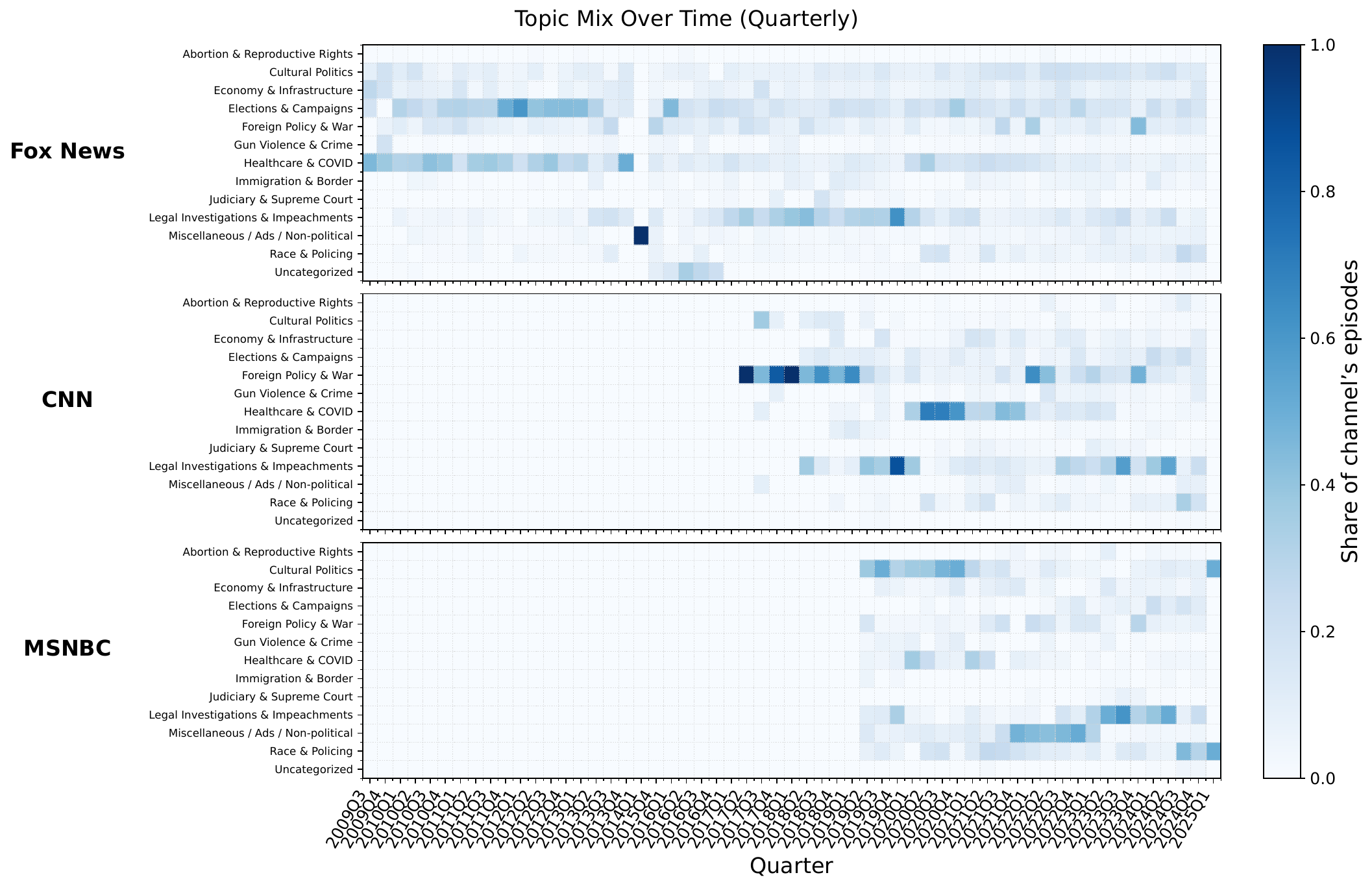}
    \caption{Topics per channel over time.}
    \label{fig:topics_per_channel_over_time}
\end{figure*}

Figure~\ref{fig:topics_per_channel_over_time} shows the high level topics per channel over time. We see clear trends like the increase in coverage in COVID/Race \& policing in 2020, consistent coverage for foreign policy content, etc.

%\iffalse
The full list of topics are shown below. %in Table~\ref{tab:topics_list}.

\begin{enumerate}[leftmargin=*, itemsep=1pt, parsep=0pt]
  \item 2012 GOP debate on Pakistan and national security post-bin Laden
  \item 2012 GOP primaries and healthcare policy debates
  \item 2020 Democratic Primary: Sanders vs. Bloomberg, Super Tuesday, and Media Scandals
  \item 2020 Democratic primary candidates
  \item 2020 Trump Foreign Policy
  \item 2024 Republican Primary Campaigns in Iowa and New Hampshire
  \item Abortion Rights and the Overturning of Roe v. Wade
  \item Ads
  \item Afghanistan Withdrawal and Evacuation Crisis at Kabul Airport
  \item Andrew Cuomo's Resignation Amid Nursing Home and Sexual Harassment Scandals
  \item Black Lives Matter, socialism, and cultural politics
  \item Black Voter Representation, Kamala Harris Campaign, and Tim Walz Commentary
  \item Brett Kavanaugh Confirmation Hearings and Sexual Assault Allegations
  \item Brexit
  \item British Royal Family Coverage: Queen Elizabeth's Legacy and King Charles's Ascension
  \item COVID origins, Christianity, immigration, and media figures
  \item Campus Encampments, Free Speech, and Debates on Antisemitism in Higher Education
  \item Chinese Spy Balloon Incidents and U.S. Airspace Surveillance Controversies
  \item Collective bargaining and emergency authority
  \item Conservative Commentary on Marijuana, Free Speech, and Cultural Politics
  \item Covid-19
  \item Critiques of the Chinese Communist Party and US-China Strategic Rivalry
  \item DOJ and FBI Leadership During Mueller Investigation and Trump-Era Controversies
  \item Debates Over Tax Reform, Immigration, and Progressive Policy Priorities
  \item Debt Ceiling Debates and Elon Musk's Influence in Political Discourse
  \item Democrats
  \item Diplomatic Relations and Summit Talks with North Korea
  \item Early Trump administration power struggles and turnover
  \item Egyptian Revolution and Mubarak's Fall Amid Muslim Brotherhood Rise
  \item Electoral Strategy and Media Framing in Battleground States Like Wisconsin
  \item Gabby Petito case coverage
  \item George Floyd Protests and Police Response in Minneapolis and Beyond
  \item George Floyd's Murder and the Prosecution of Derek Chauvin
  \item Georgia Senate Elections and Runoff Campaigns
  \item Gutfeld! Comedy
  \item Hurricane Coverage: Storm Surge, Flooding, and Landfall Impacts
  \item Hurricanes and FEMA Disaster Response in Hawaii, Texas, and the Southeast
  \item IRS Fraud
  \item ISIS-Linked Terror Attacks in Europe: Paris, Manchester, and Brussels
  \item Inflation, Energy, and Economic Slowdown
  \item Israel-Gaza Conflict and Ceasefire Discussions
  \item January 6th
  \item Jeffrey Epstein Case and Legal Accountability of Labor Secretary Acosta
  \item Kevin McCarthy
  \item Kyle Rittenhouse Trial, Verdict, and Broader Jury Trials in Media Focus
  \item Las Vegas Mass Shooting and Debates Over Gun Violence and Background Checks
  \item Leaks, executive oversight, and antisemitism in political media discussions
  \item Manchin's role in Democratic infrastructure and vaccine policy negotiations
  \item Mar-a-Lago Classified Documents Case
  \item Mass School Shootings and Mental Health Crisis in the US
  \item Matt Gaetz Ethics Scrutiny and Cabinet Appointments During Presidential Transition
  \item Mueller, Obamacare, and political controversies
  \item National identity, foreign influence, and Second Amendment debates in conservative media
  \item Nationalist cultural commentary and anti-communist rhetoric
  \item Nikki Haley's 2024 Campaign, Southern Primaries, and IVF Policy Debates
  \item Obamacare
  \item Paul Pelosi Attack Incident and Related Legal Proceedings
  \item Police Brutality and Public Outcry Following the Deaths of Floyd, Taylor, and Brooks
  \item Postal service, shutdowns, and COVID-era political coverage
  \item Religion, patriotism, and cultural conflict in U.S. media (Christianity, NFL, anthem protests)
  \item Republican National Convention Coverage Amid Assassination Attempt and Political Tensions
  \item Right-Wing Critiques of Race, Immigration, and Progressive Politicians
  \item Right-Wing Framing of the Democratic Party and Radical Leftist Ideologies
  \item Rudy Giuliani's Legal Troubles: Election Worker Defamation and Mueller-Era Ties
  \item Russia-ukraine war and nato's strategic response
  \item San Francisco Homelessness
  \item Saudi Arabia's Crown Prince and the Killing of Jamal Khashoggi
  \item Scandals and Public Figures (e.g., Avenatti, Bernie)
  \item Syrian Civil War and U.S. Policy on Assad, Chemical Weapons, and ISIS
  \item Tea Party-Era Health Reform Debates
  \item Titan Submersible Disaster and Ocean Exploration Catastrophe Investigation
  \item Trump Administration Infighting, Mueller-Era Investigations, and DACA Policy Debates
  \item Trump Hush Money Trial Featuring Cohen and Daniels
  \item Trump Impeachment Hearings and Whistleblower Testimony
  \item U.S. Drone Strike on Qasem Soleimani and Escalating Tensions with Iran
  \item US-Mexico border security and cartel-driven immigration
  \item Urban Crime and Policing in Central Park and Manhattan with NYPD Coverage
  \item VP Kamala Harris and Democratic Campaign Strategies on Abortion
  \item Virginia and New Jersey Gubernatorial Races with National Culture War Overtones
\end{enumerate}
%\end{multicols}
%\end{minipage}
%\caption{List of 80 Topics.}
%\label{tab:topics_list}
%\end{table*}

%\fi

\subsection{Disagreement by Profession}
To assess whether a guest's profession predicts on-air disagreement, we first needed to ensure our comparisons were statistically stable. We established a minimum sample size for each occupation by defining a desired level of precision: a 95\% confidence interval no wider than ±3 percentage points around the global mean disagreement rate (16\%). The calculation determined that the minimum number of labeled pairs required to meet this criterion is 575.
Therefore, we restricted our analysis to occupations with at least 575 samples, providing a robust basis for comparison. The occupation labels themselves were sourced from the DIME dataset and normalized to create consistent categories (e.g., attorney, county attorney, and attorney general were all merged into lawyer/attorney). The results in Table \ref{tab:occupation_disagreement} show that legislators—both members of Congress (19.9\%) and senators (18.6\%)—are among the most frequently challenged guests, along with managers and entertainers (20\%). Lawyers, executives, and CEOs show slightly lower rates (16–17\%), while educators, researchers, and students are the least disagreed with ($<$15\%). These patterns suggest that disagreement is more common when guests hold formal decision-making authority or represent political institutions, but is still broadly distributed across professional categories. Importantly, the overall network-level decline in disagreement is not driven by occupational composition, as all high-frequency groups show parallel declines over time. However, it is important to note that while the DIME dataset includes occupation fields, approximately 35\% of the guests in our corpus could not be reliably matched to an occupation entry due to missing or incomplete data in the DIME dataset. As a result, the occupation-based analysis should be interpreted with caution, as it necessarily reflects a subset of the full guest population.

\subsection{Agreement/Disagreement plots for all shows}

\begin{figure*}[ht]
\centering
\begin{minipage}{.24\linewidth}
\centering
\subfloat[Alex Wagner Tonight (MSNBC)]{\label{}\includegraphics[width=\textwidth]{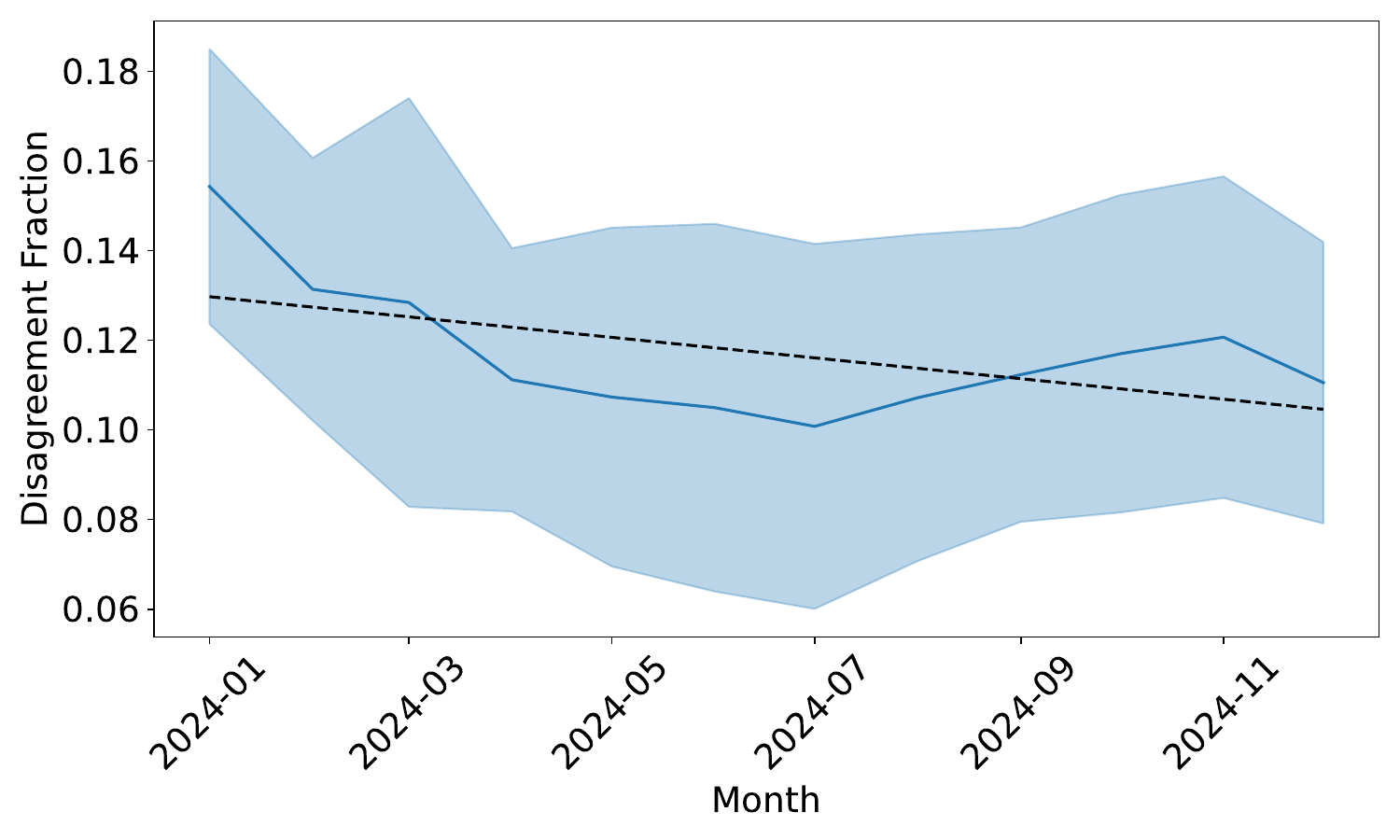}}
\end{minipage}%
\begin{minipage}{.24\linewidth}
\centering
\subfloat[Anderson Cooper (CNN) $^{***}$]{\label{}\includegraphics[width=\textwidth]{img/disagreement_timeseries/Anderson_Cooper_disagreement_monthly.pdf}}
\end{minipage}%
\begin{minipage}{.24\linewidth}
\centering
\subfloat[Erin Brunett Outfront (CNN) $^{***}$]{\label{}\includegraphics[width=\textwidth]{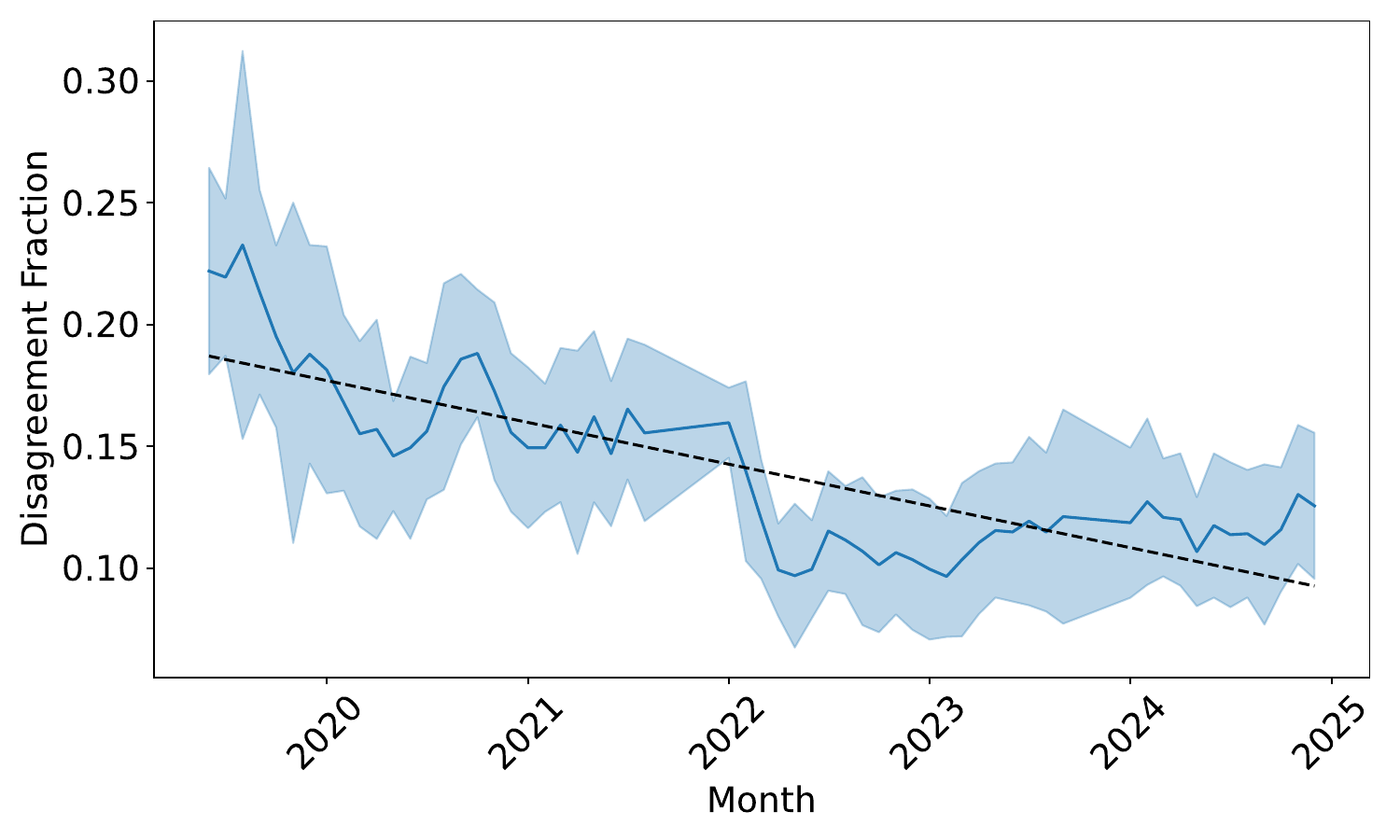}}
\end{minipage}%
\begin{minipage}{.24\linewidth}
\centering
\subfloat[Fareed Zakaria GPS (CNN)]{\label{}\includegraphics[width=\textwidth]{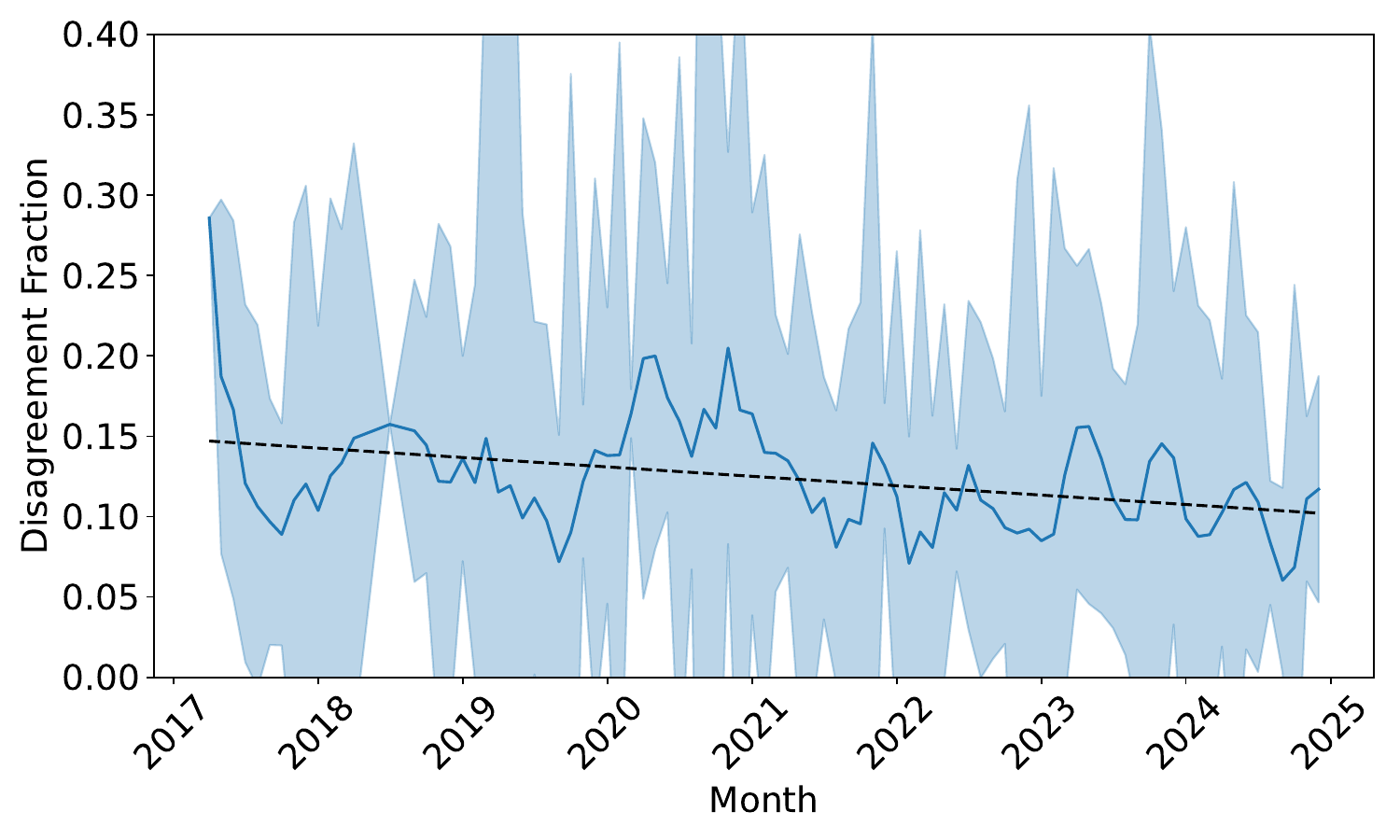}}
\end{minipage}%
\par\medskip

\begin{minipage}{.24\linewidth}
\centering
\subfloat[Gutfeld (Fox)]{\label{}\includegraphics[width=\textwidth]{img/disagreement_timeseries/Gutfeld_orig.pdf}}
\end{minipage}%
\begin{minipage}{.24\linewidth}
\centering
\subfloat[Hannity (Fox) $^{***}$]{\label{}\includegraphics[width=\textwidth]{img/disagreement_timeseries/Hannity_orig.pdf}}
\end{minipage}%
\begin{minipage}{.24\linewidth}
\centering
\subfloat[Inside Politics with Dana Bash (CNN) $^{***}$]{\label{Velshi}\includegraphics[width=\textwidth]{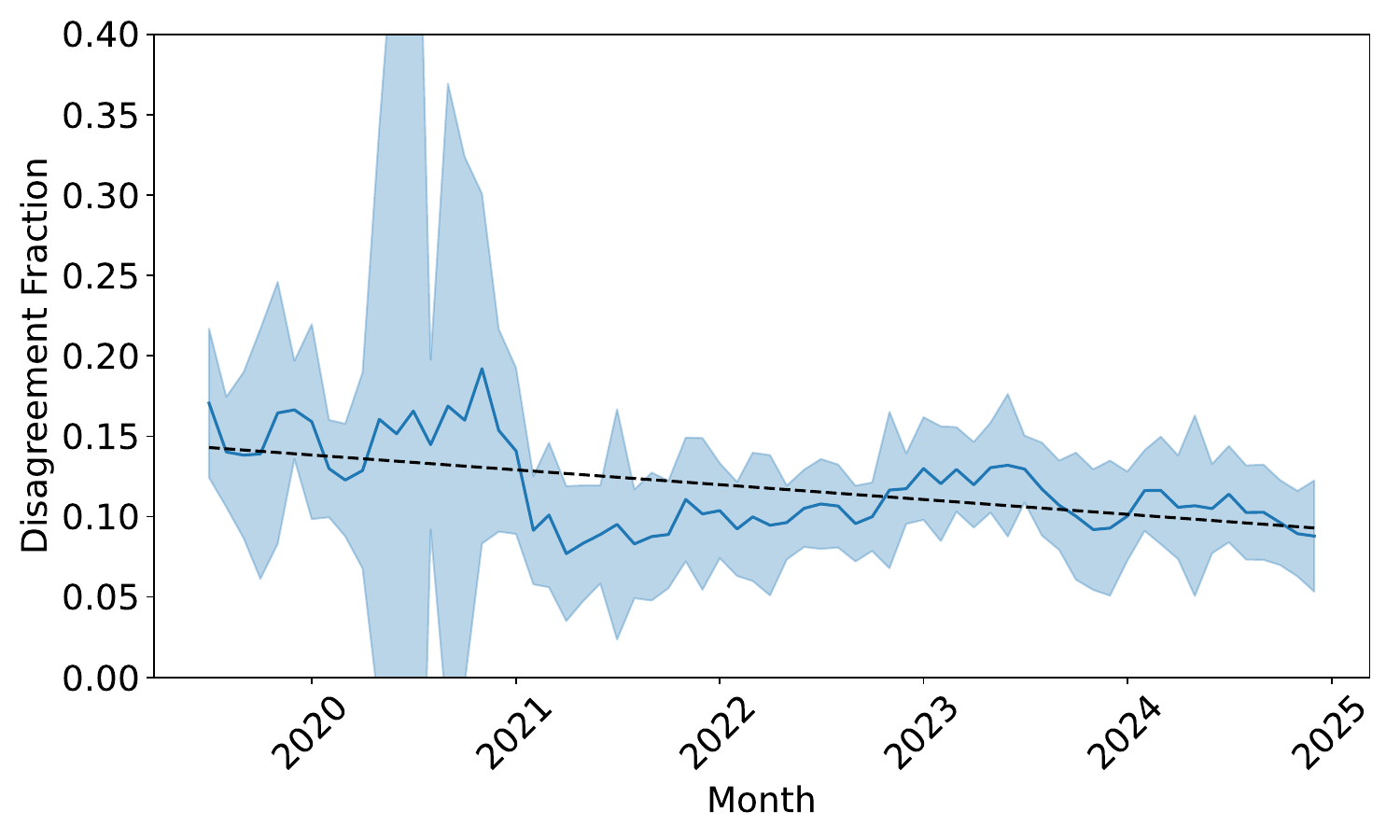}}
\end{minipage}%
\begin{minipage}{.24\linewidth}
\centering
\subfloat[Jesse Waters (Fox)]{\label{Velshi}\includegraphics[width=\textwidth]{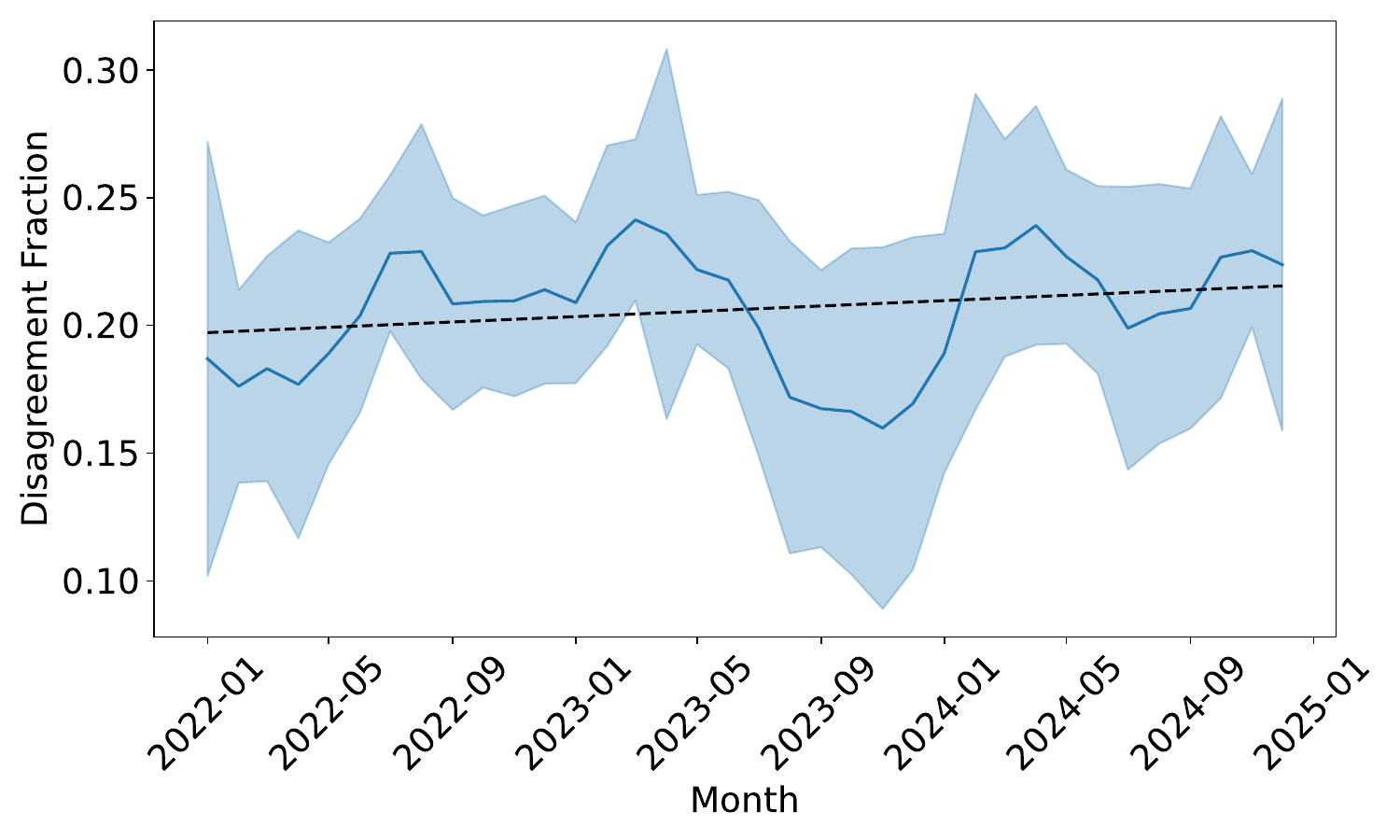}}
\end{minipage}%
\par\medskip

\begin{minipage}{.24\linewidth}
\centering
\subfloat[Laura Coates Live (MSNBC)]{\label{}\includegraphics[width=\textwidth]{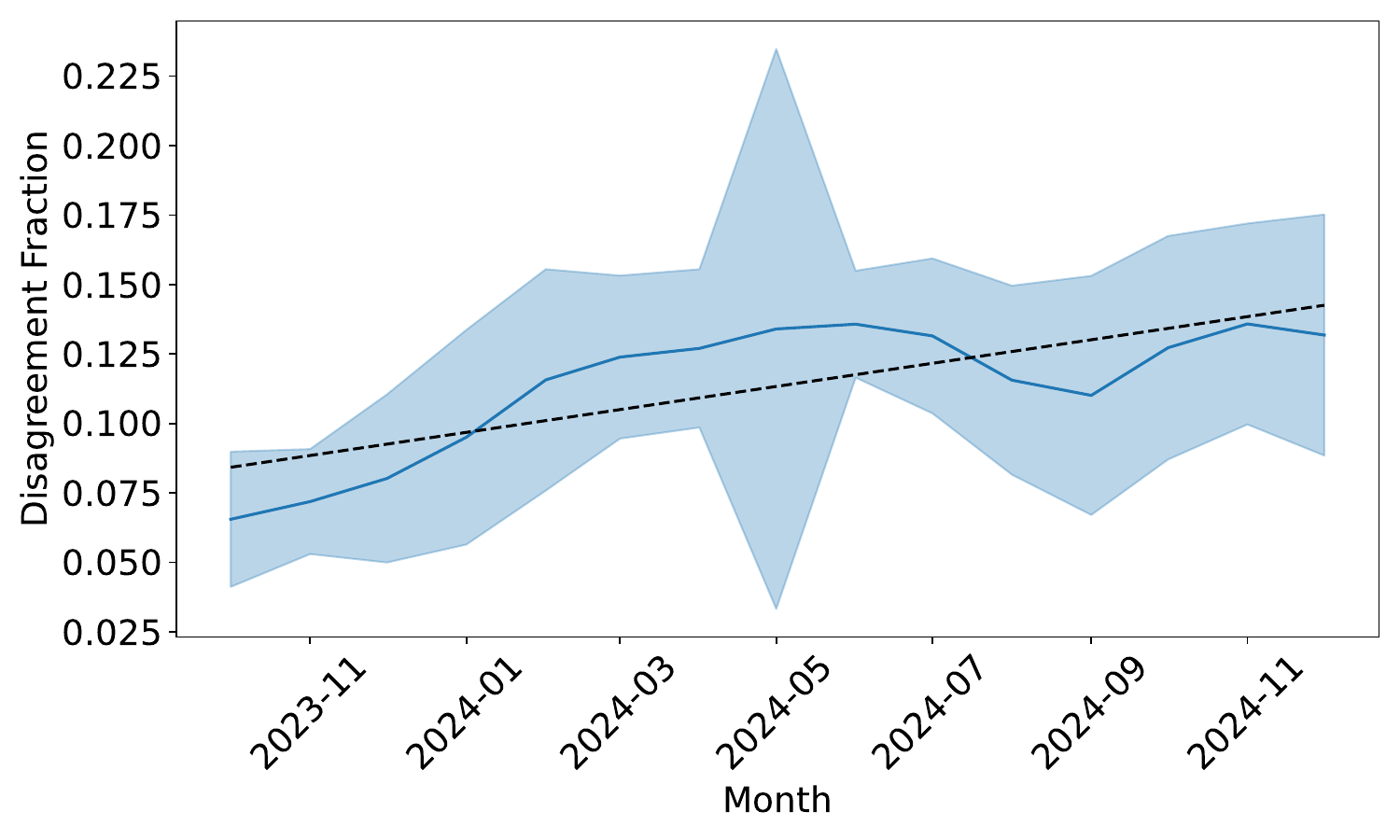}}
\end{minipage}%
\begin{minipage}{.24\linewidth}
\centering
\subfloat[Laura Ingraham (Fox) $^{***}$]{\label{}\includegraphics[width=\textwidth]{img/disagreement_timeseries/Laura_Ingraham_disagreement_monthly.pdf}}
\end{minipage}%
\begin{minipage}{.24\linewidth}
\centering
\subfloat[Life Liberty Levin (Fox) $^{***}$]{\label{}\includegraphics[width=\textwidth]{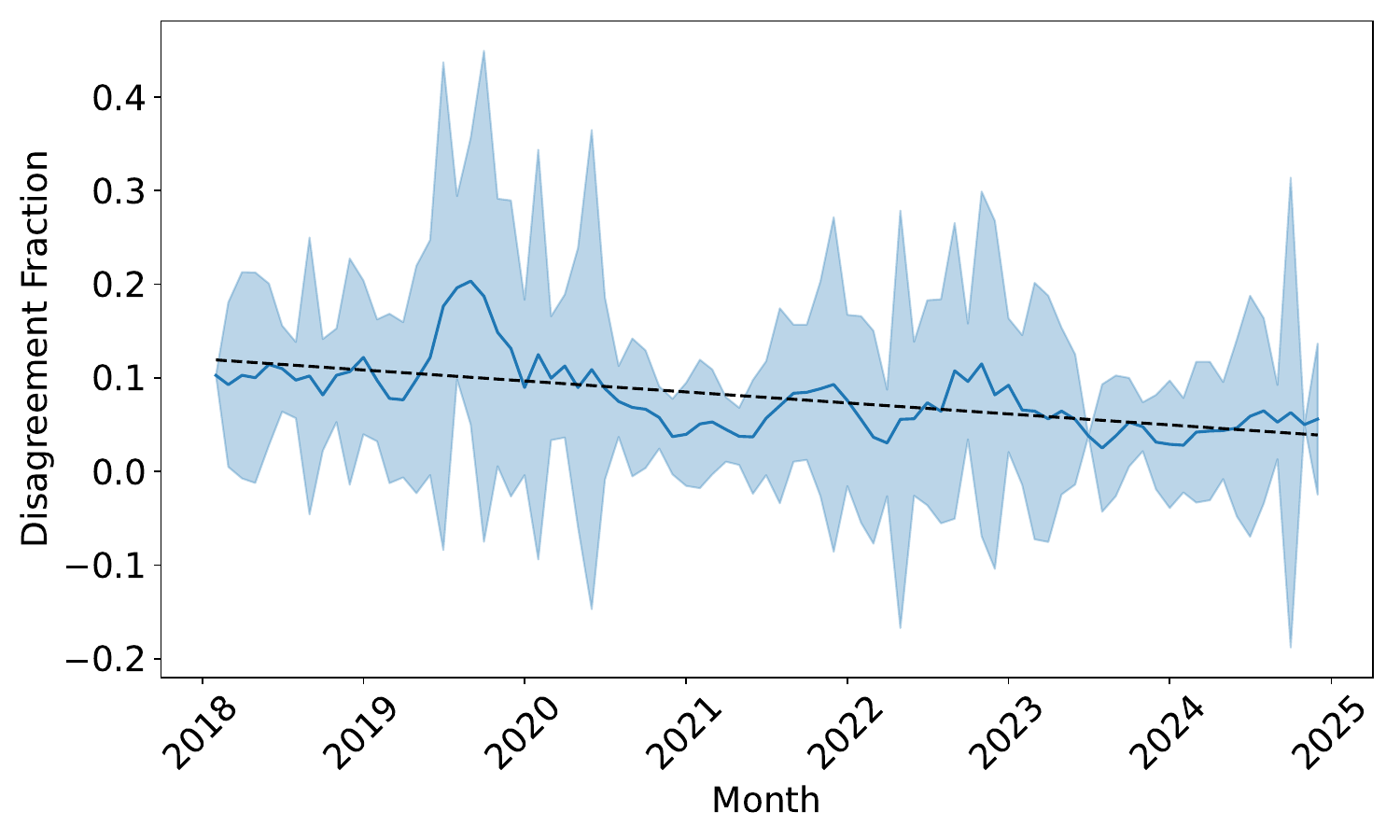}}
\end{minipage}%
\begin{minipage}{.24\linewidth}
\centering
\subfloat[Outnumbered (Fox) $^{***}$]{\label{}\includegraphics[width=\textwidth]{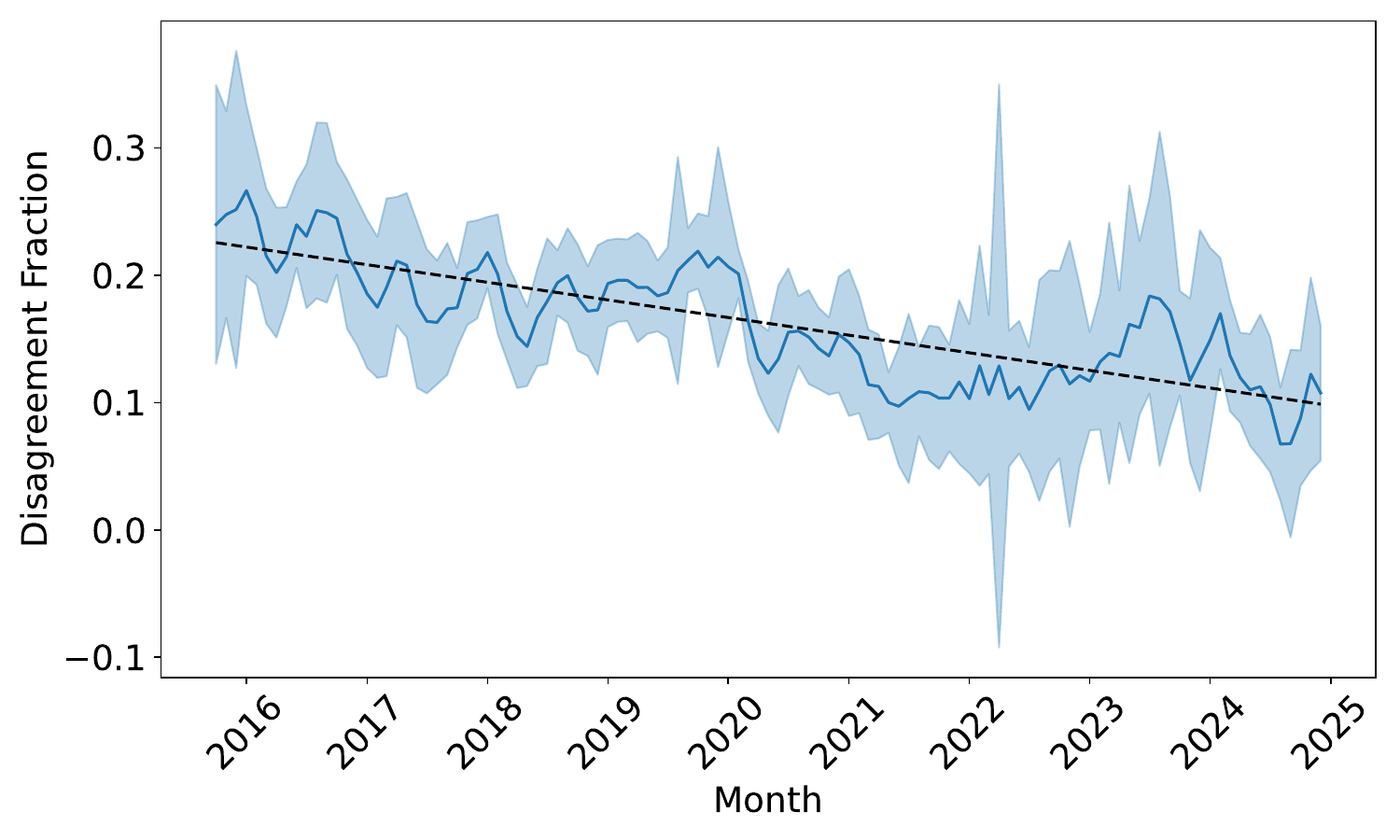}}
\end{minipage}%
\par\medskip

\begin{minipage}{.24\linewidth}
\centering
\subfloat[Rachel Maddow (MSNBC)]{\label{}\includegraphics[width=\textwidth]{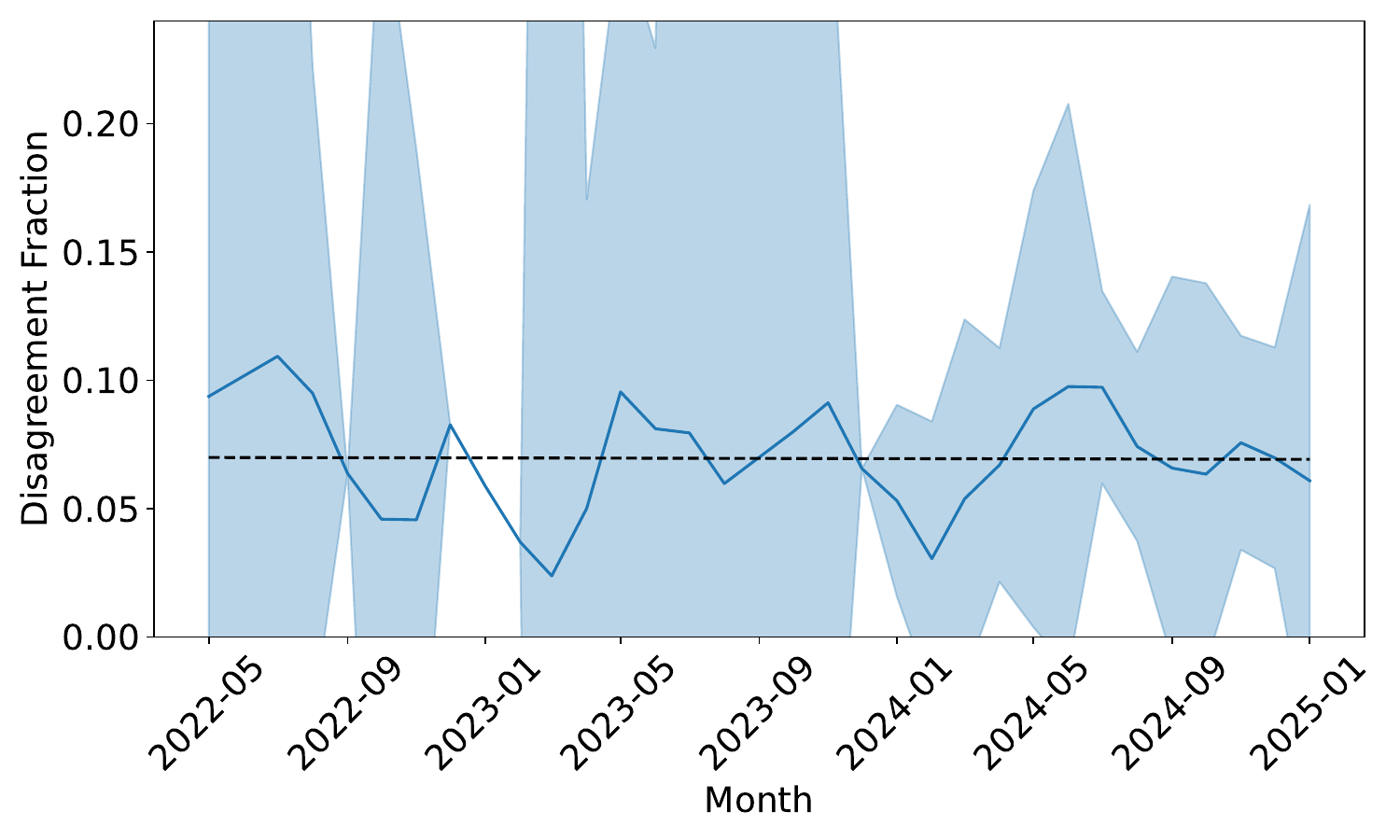}}
\end{minipage}%
\begin{minipage}{.24\linewidth}
\centering
\subfloat[Saturdays Sundays with Jonathan Capehart (MSNBC)]{\label{}\includegraphics[width=\textwidth]{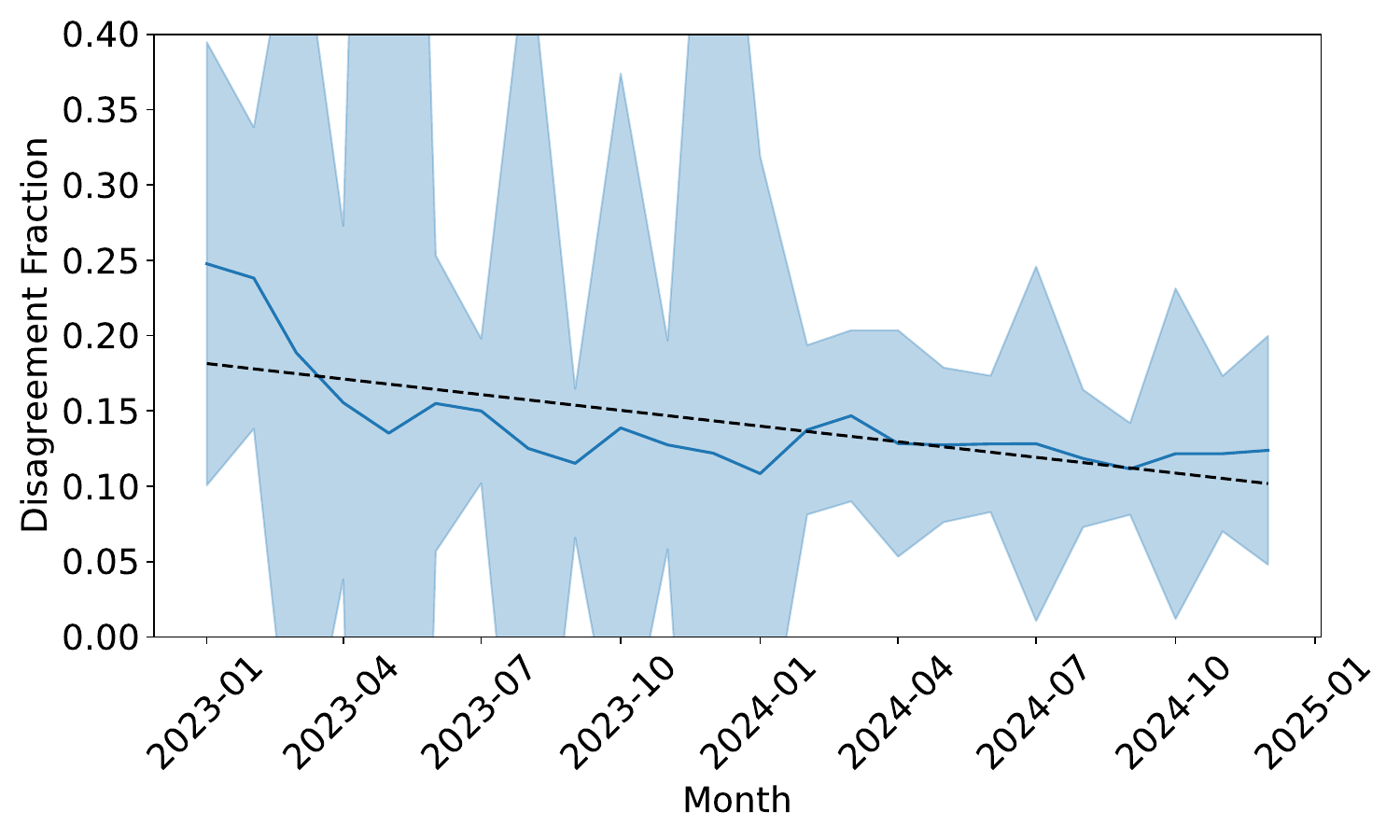}}
\end{minipage}%
\begin{minipage}{.24\linewidth}
\centering
\subfloat[Special Report with Bret Baier (Fox) $^{***}$]{\label{Velshi}\includegraphics[width=\textwidth]{img/disagreement_timeseries/Special_Report_with_Bret_Baier_orig.pdf}}
\end{minipage}%
\begin{minipage}{.24\linewidth}
\centering
\subfloat[State of the Union (CNN) $^{***}$]{\label{Velshi}\includegraphics[width=\textwidth]{img/disagreement_timeseries/State_of_the_Union_disagreement_monthly.pdf}}
\end{minipage}%
\par\medskip

\begin{minipage}{.24\linewidth}
\centering
\subfloat[The Beat with Ari Melber (MSNBC)]{\label{}\includegraphics[width=\textwidth]{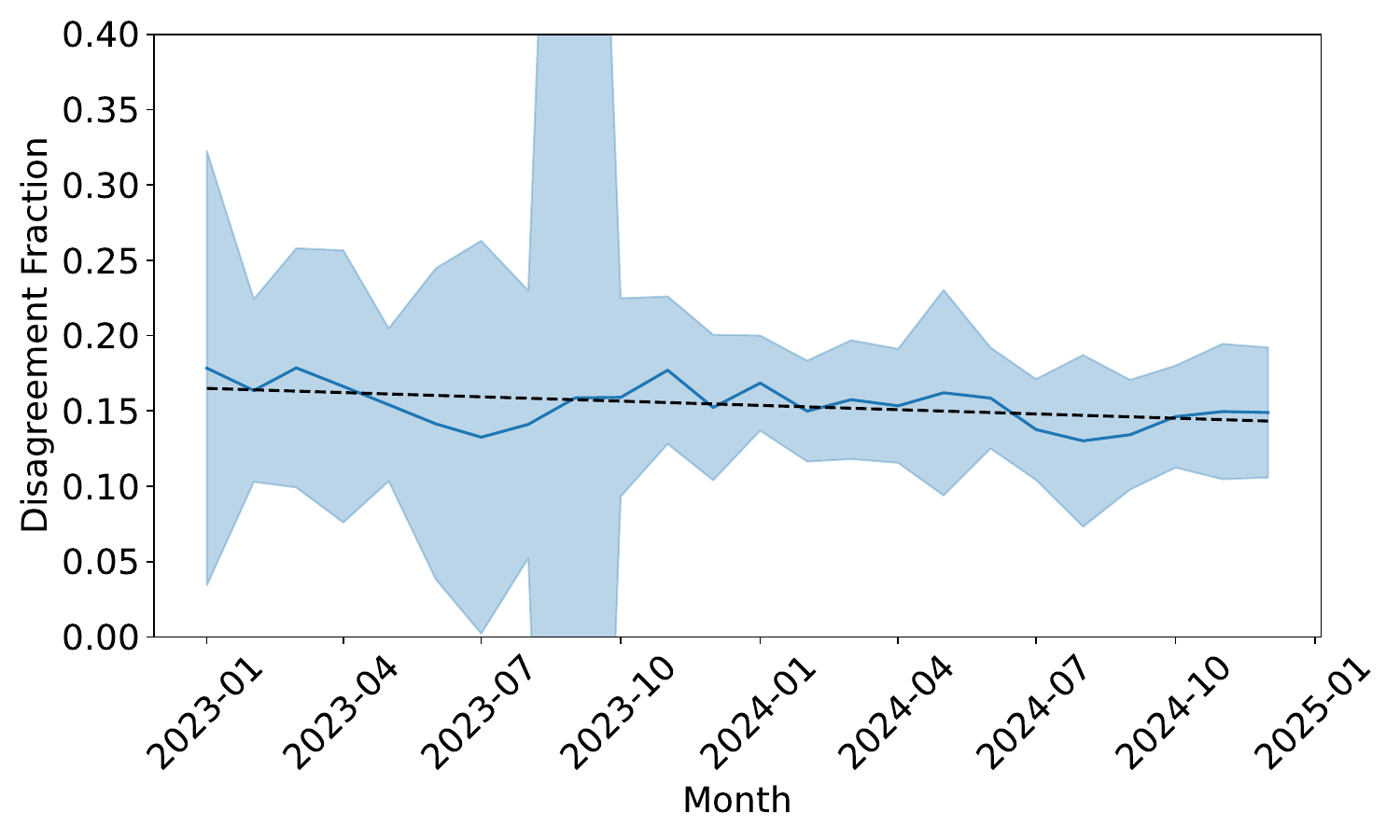}}
\end{minipage}%
\begin{minipage}{.24\linewidth}
\centering
\subfloat[The Lead with Jake Tapper (CNN) $^{***}$]{\label{}\includegraphics[width=\textwidth]{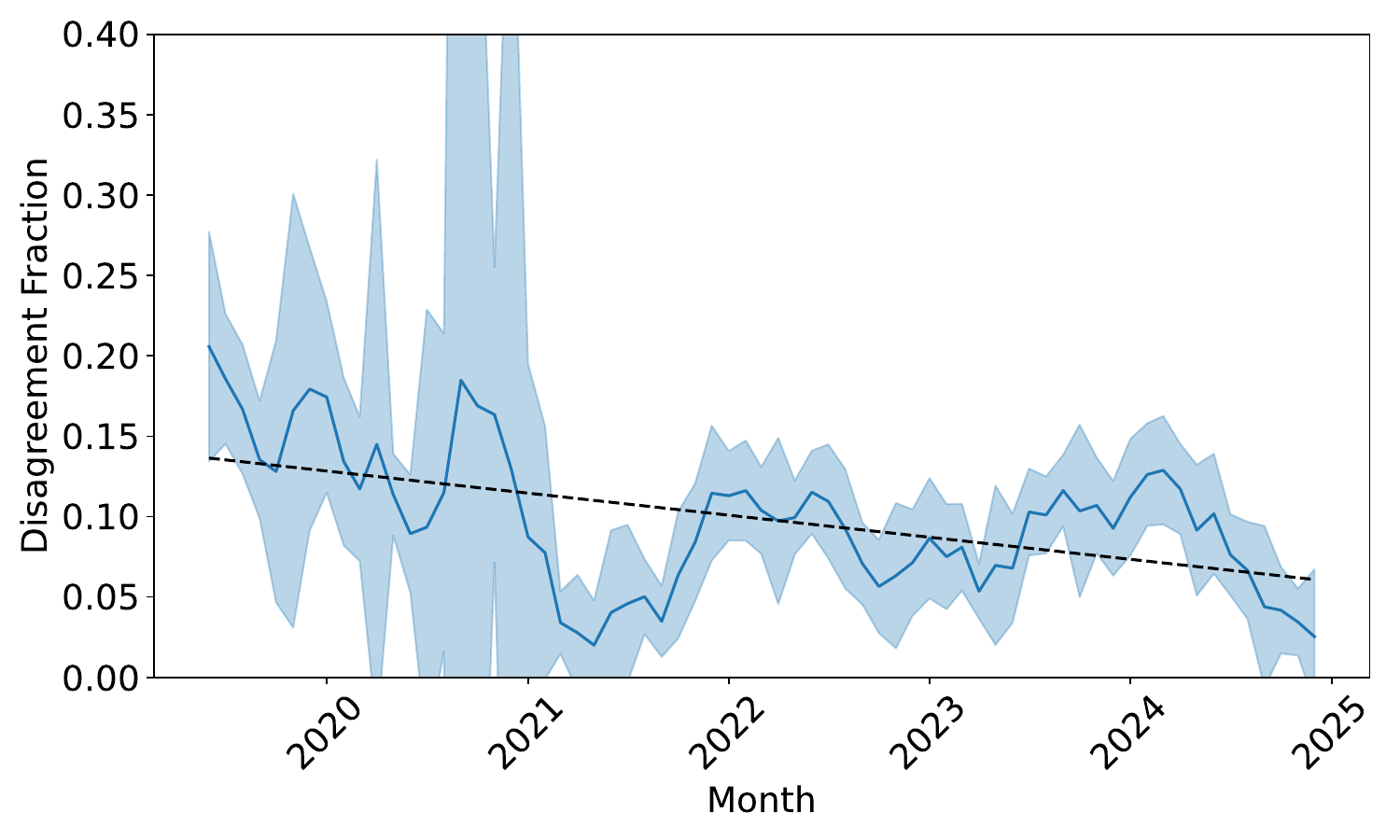}}
\end{minipage}%
\begin{minipage}{.24\linewidth}
\centering
\subfloat[The Savage Nation (MSNBC)]{\label{}\includegraphics[width=\textwidth]{img/disagreement_timeseries/The_Savage_Nation_disagreement_monthly.pdf}}
\end{minipage}%
\begin{minipage}{.24\linewidth}
\centering
\subfloat[The Source with Kaitlin Collins (CNN)]{\label{}\includegraphics[width=\textwidth]{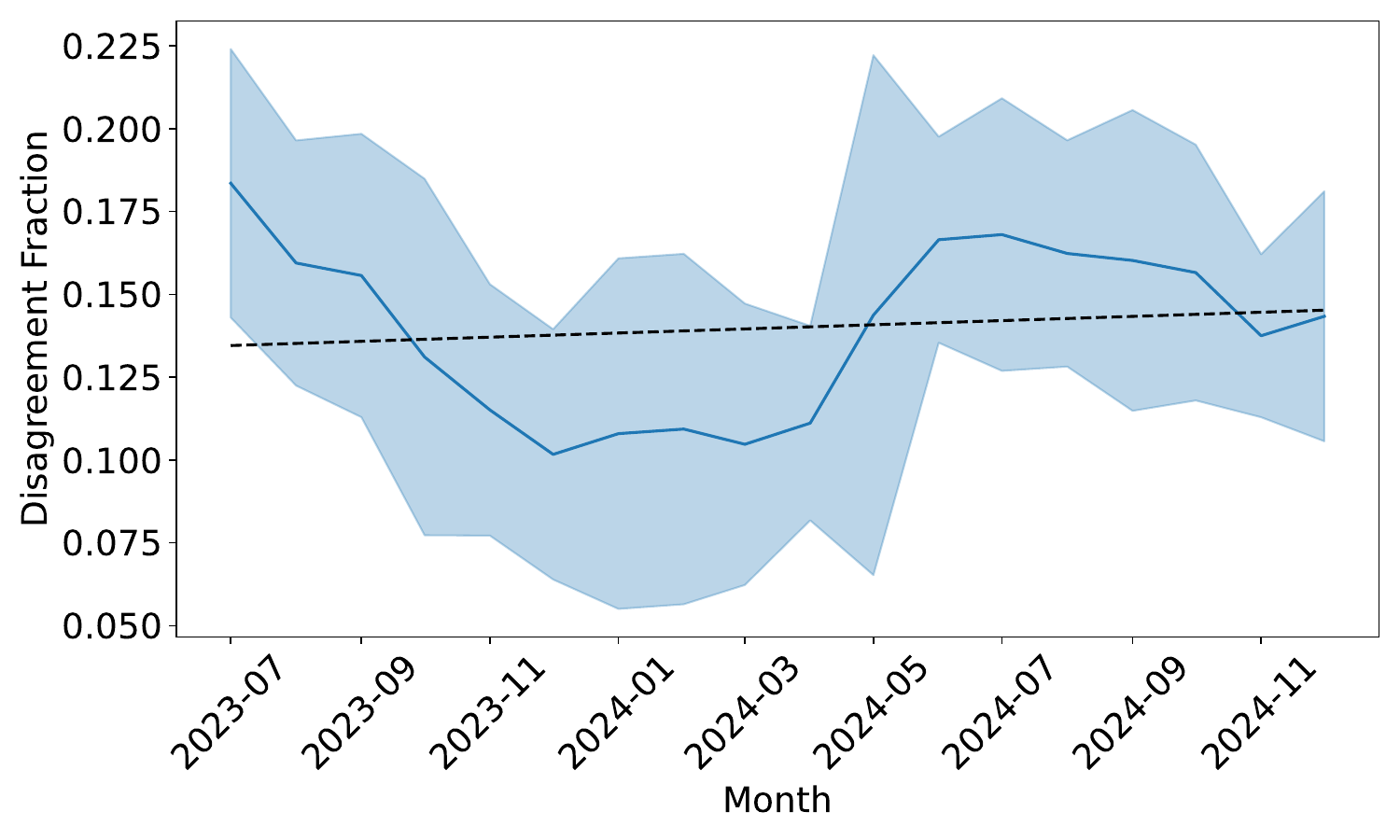}}
\end{minipage}%
\par\medskip

\begin{minipage}{.24\linewidth}
\centering
\subfloat[The Story with Martha MacCallum (Fox) $^{***}$]{\label{}\includegraphics[width=\textwidth]{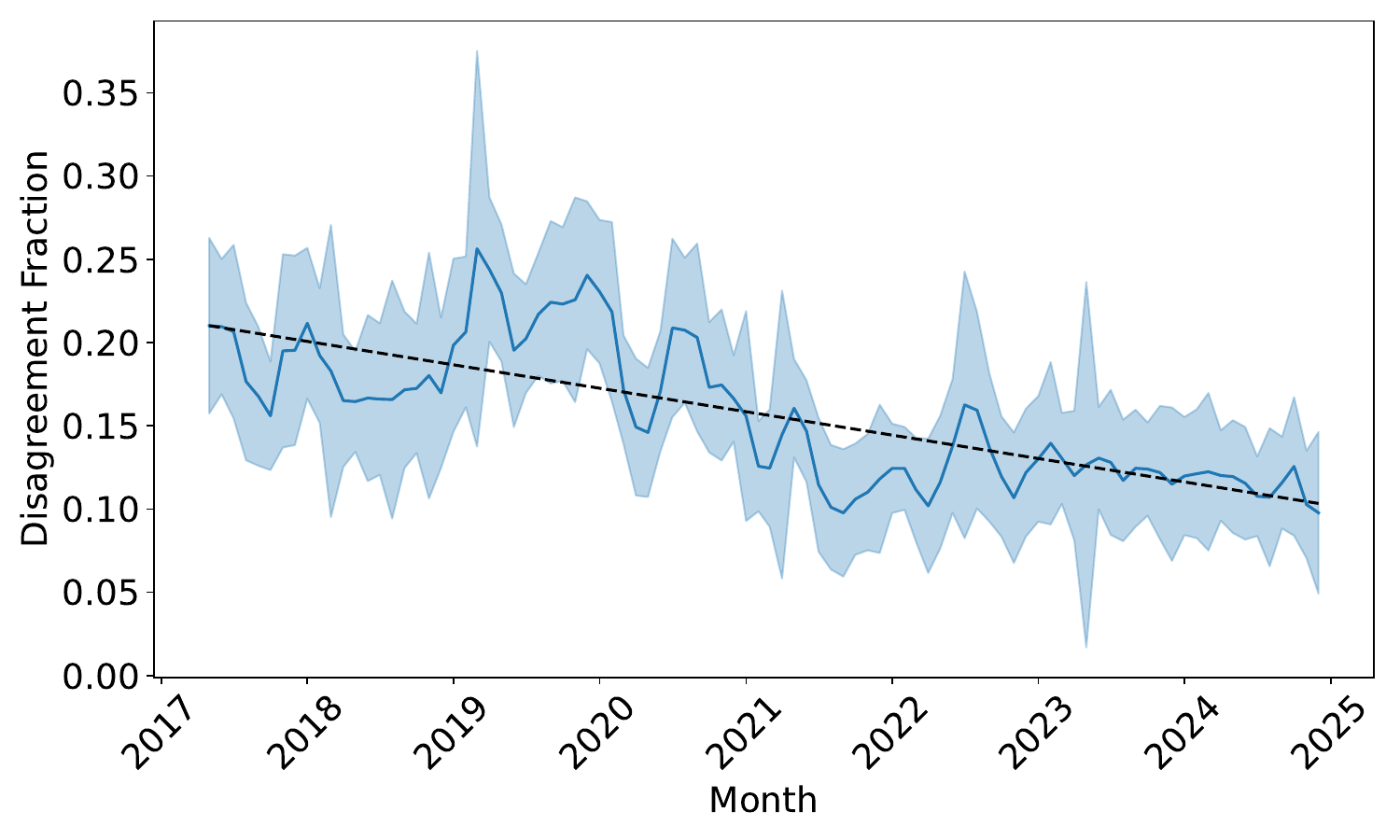}}
\end{minipage}%
\begin{minipage}{.24\linewidth}
\centering
\subfloat[Tucker Carlson (Fox) $^{***}$]{\label{}\includegraphics[width=\textwidth]{img/disagreement_timeseries/Tucker_disagreement_monthly.pdf}}
\end{minipage}%
\begin{minipage}{.24\linewidth}
\centering
\subfloat[Velshi (MSNBC)]{\label{Velshi}\includegraphics[width=\textwidth]{img/disagreement_timeseries/Velshi_disagreement_monthly.pdf}}
\end{minipage}%
\begin{minipage}{.24\linewidth}
\centering
\subfloat[Inside with Jen Psaki (MSNBC)]{\label{Velshi}\includegraphics[width=\textwidth]{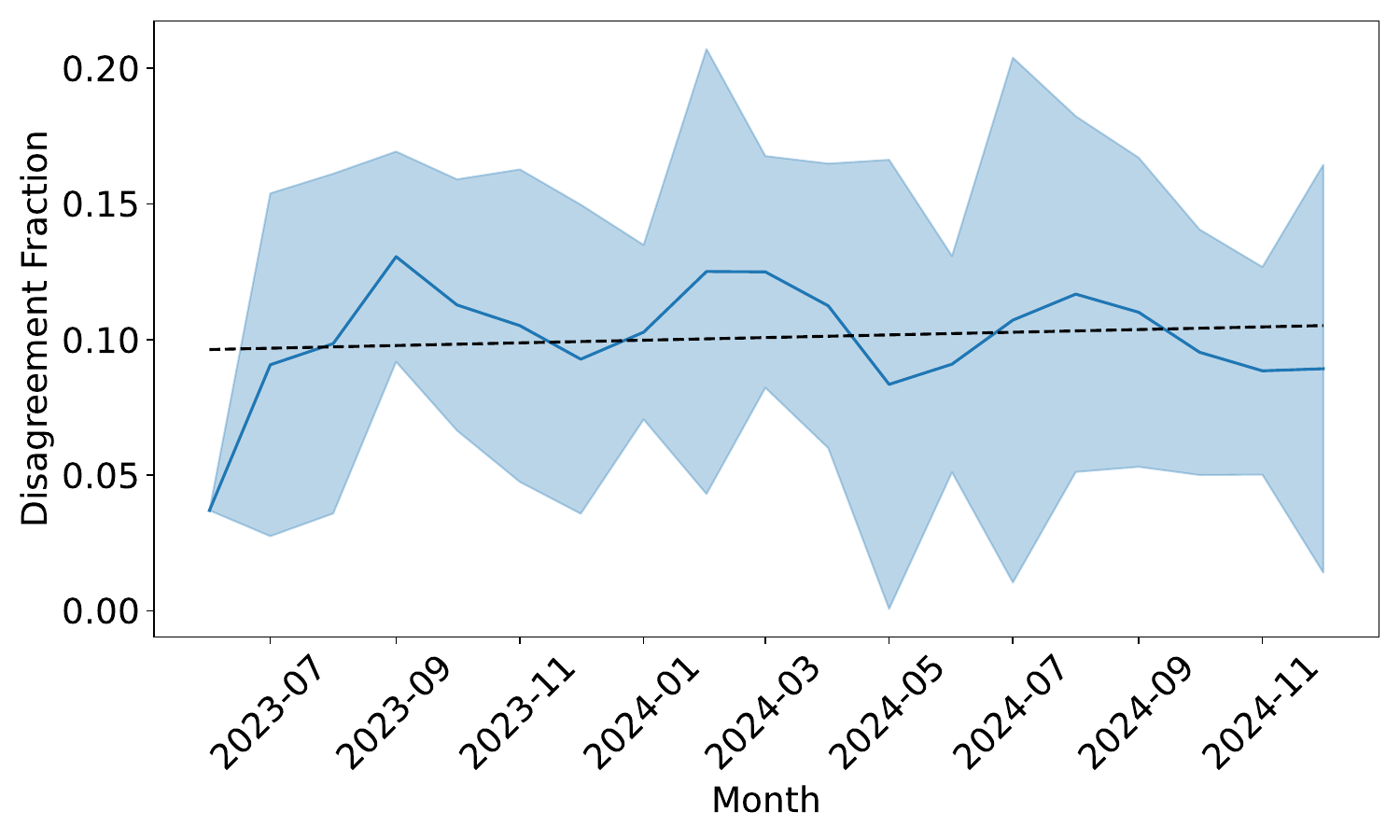}}
\end{minipage}%
\par\medskip

\caption{Disagreement patterns over time for all shows.  $^{***}$ indicates a significant decline ($p < 0.001$) in slope.}
\label{fig:disagreement_over_time_all_shows}
\end{figure*}

\begin{figure*}[ht]
\centering
\begin{minipage}{.24\linewidth}
\centering
\subfloat[Alex Wagner Tonight (MSNBC)]{\label{}\includegraphics[width=\textwidth]{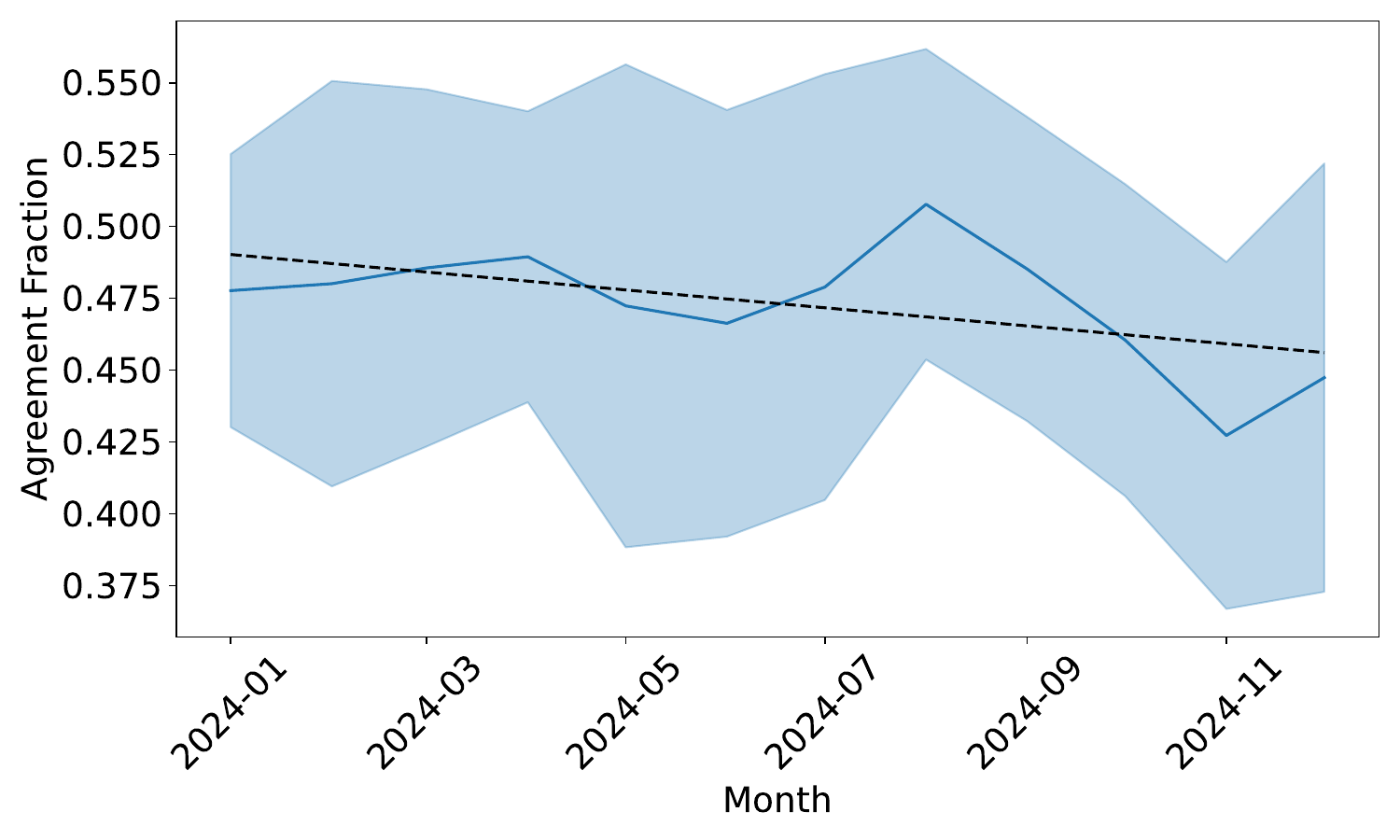}}
\end{minipage}%
\begin{minipage}{.24\linewidth}
\centering
\subfloat[Anderson Cooper (CNN)]{\label{}\includegraphics[width=\textwidth]{img/agreement_timeseries/Anderson_Cooper_agreement_monthly.pdf}}
\end{minipage}%
\begin{minipage}{.24\linewidth}
\centering
\subfloat[Erin Brunett Outfront (CNN)]{\label{}\includegraphics[width=\textwidth]{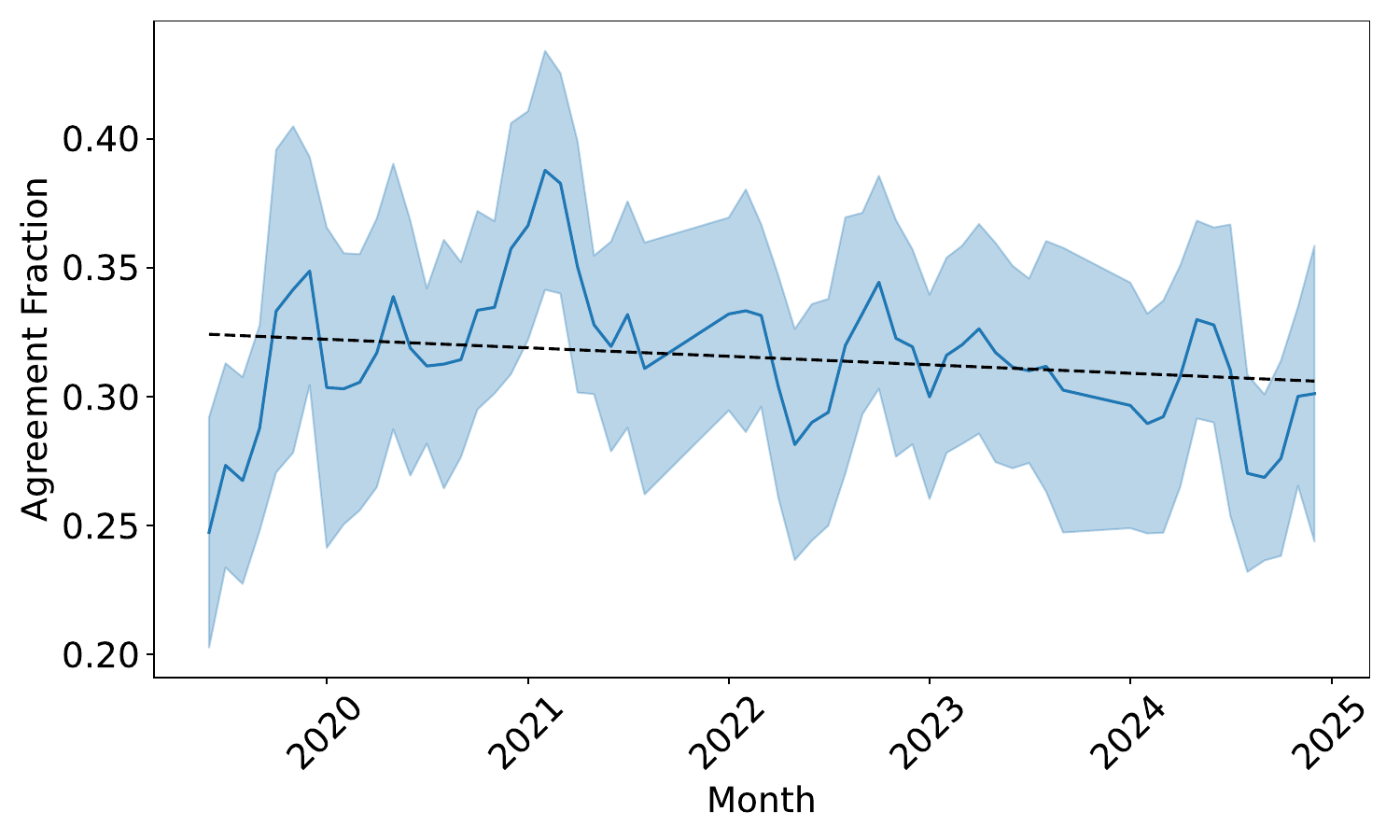}}
\end{minipage}%
\begin{minipage}{.24\linewidth}
\centering
\subfloat[Fareed Zakaria GPS (CNN)]{\label{}\includegraphics[width=\textwidth]{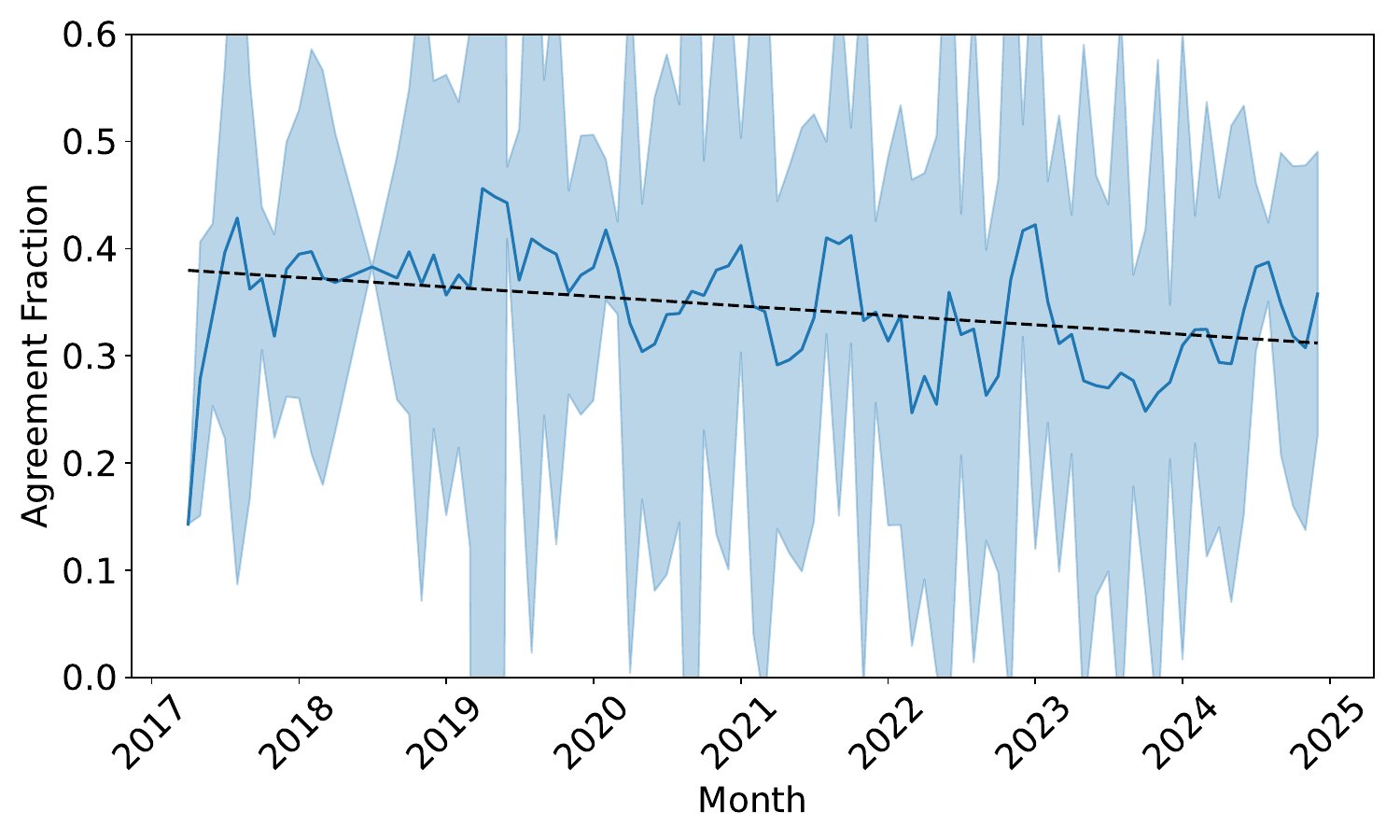}}
\end{minipage}%
\par\medskip

\begin{minipage}{.24\linewidth}
\centering
\subfloat[Gutfeld (Fox)]{\label{}\includegraphics[width=\textwidth]{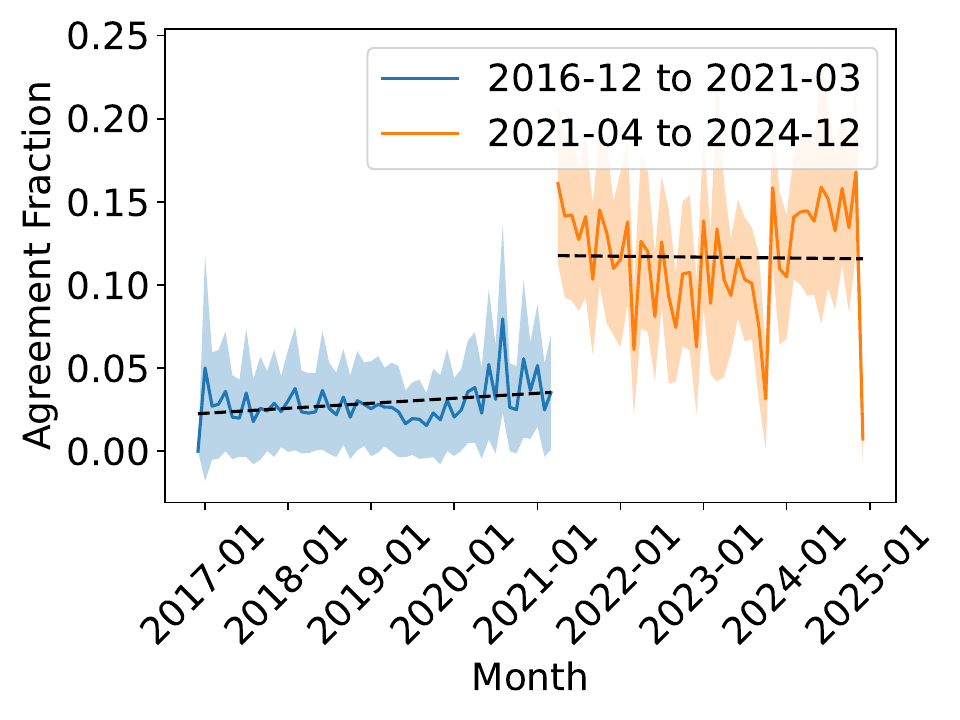}}
\end{minipage}%
\begin{minipage}{.24\linewidth}
\centering
\subfloat[Hannity (Fox) $^{***}$]{\label{}\includegraphics[width=\textwidth]{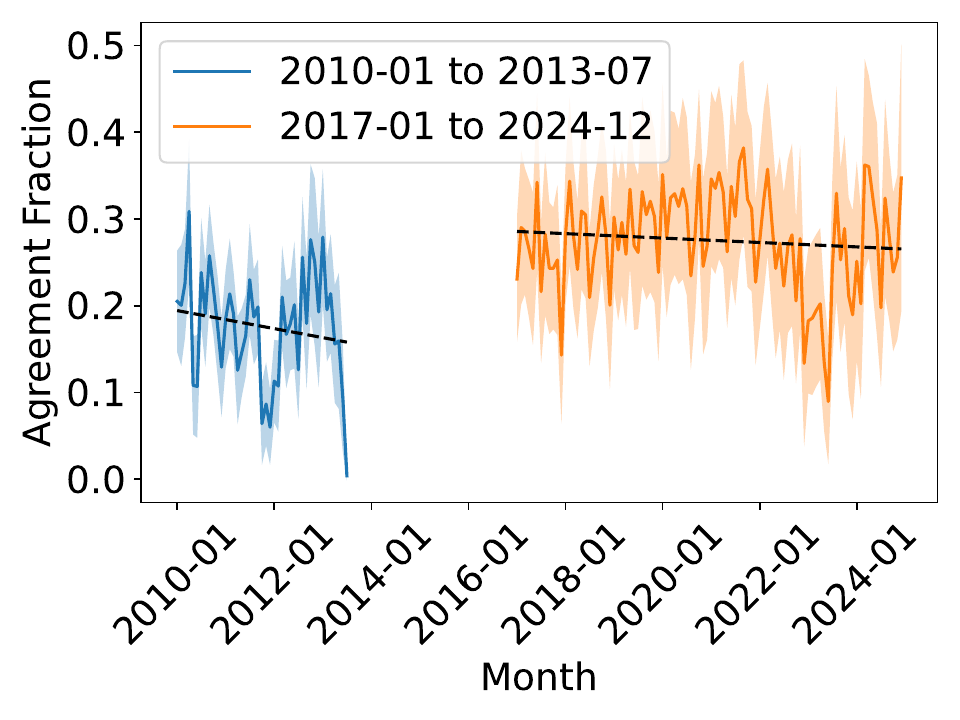}}
\end{minipage}%
\begin{minipage}{.24\linewidth}
\centering
\subfloat[Inside Politics with Dana Bash (CNN)]{\label{Velshi}\includegraphics[width=\textwidth]{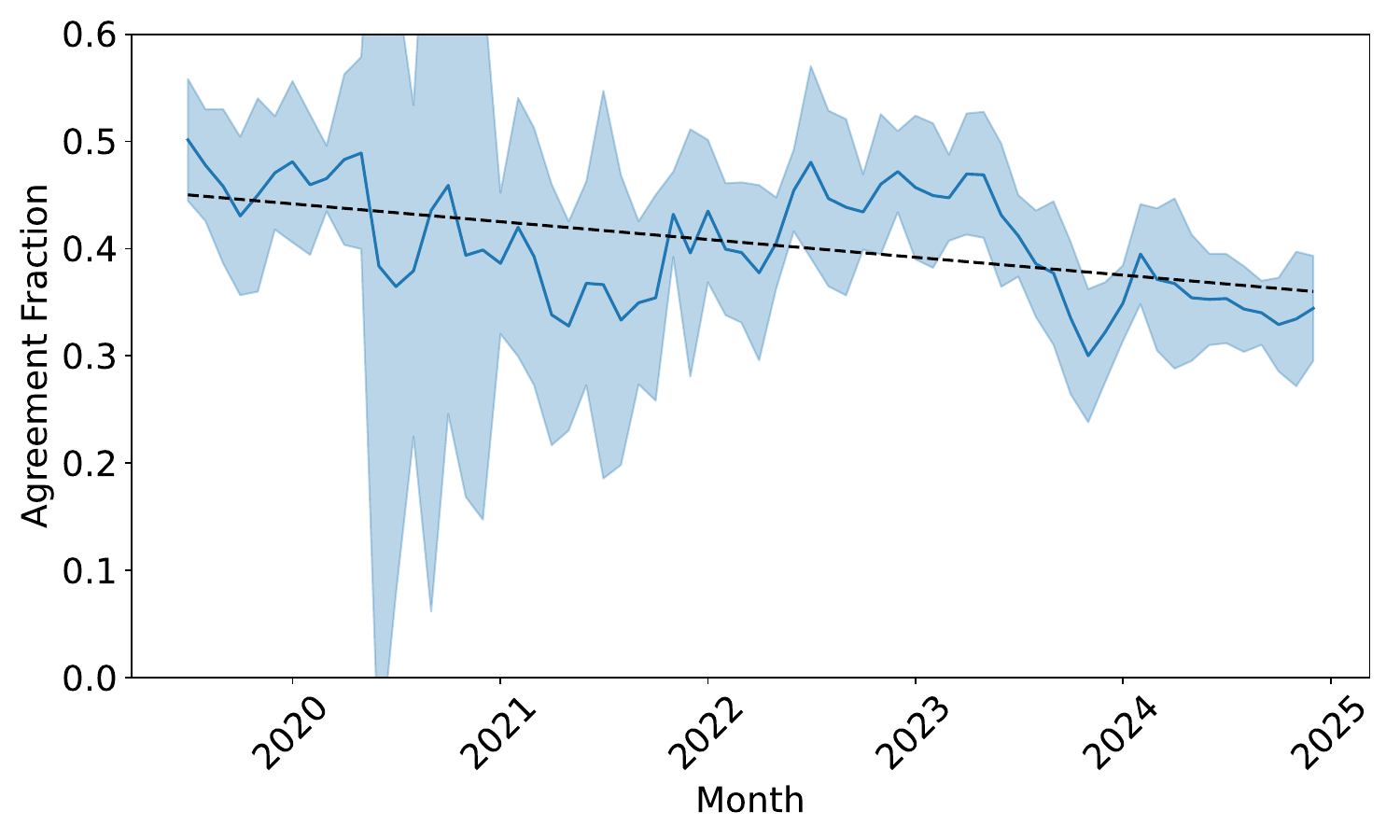}}
\end{minipage}%
\begin{minipage}{.24\linewidth}
\centering
\subfloat[Jesse Waters (Fox)]{\label{Velshi}\includegraphics[width=\textwidth]{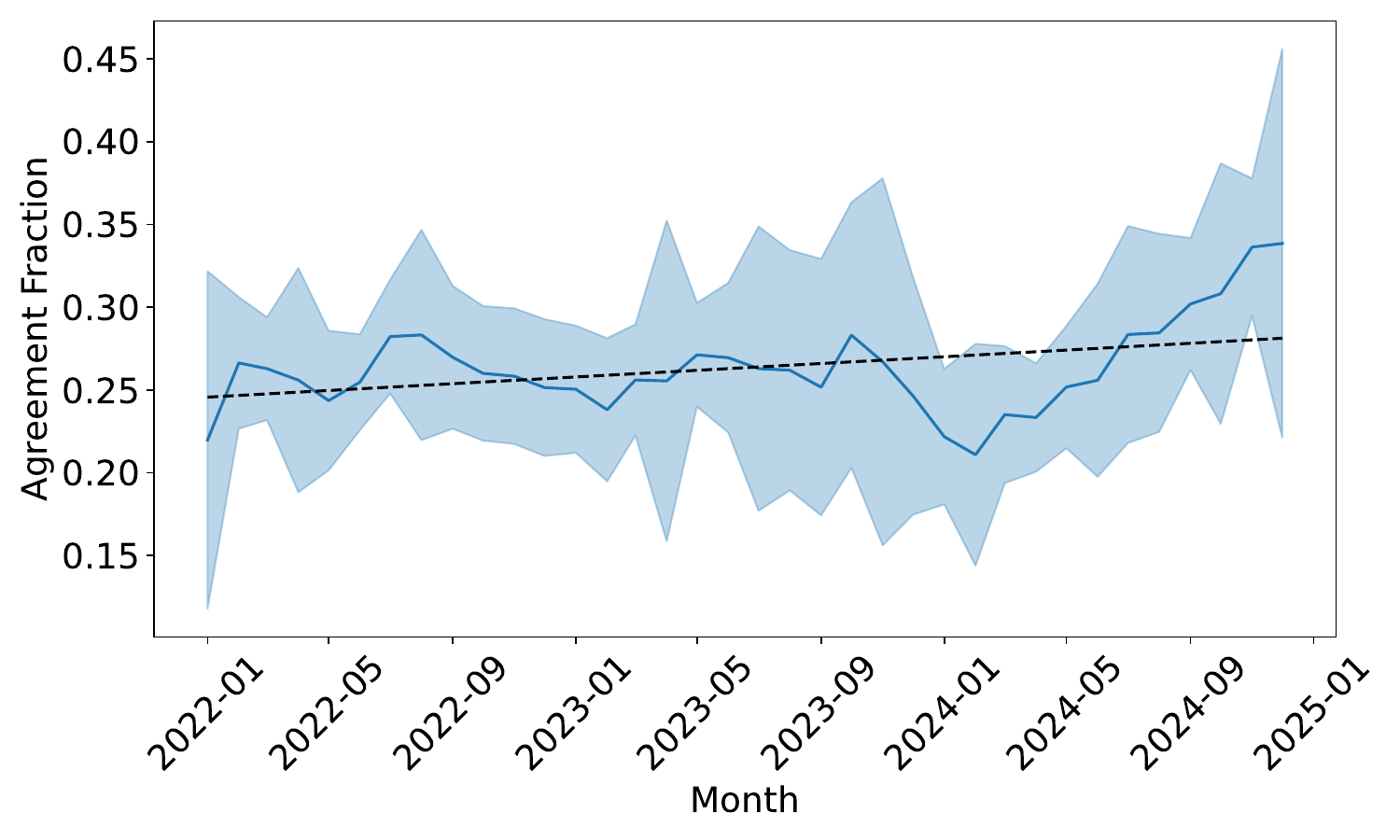}}
\end{minipage}%
\par\medskip

\begin{minipage}{.24\linewidth}
\centering
\subfloat[Laura Coates Live (MSNBC)]{\label{}\includegraphics[width=\textwidth]{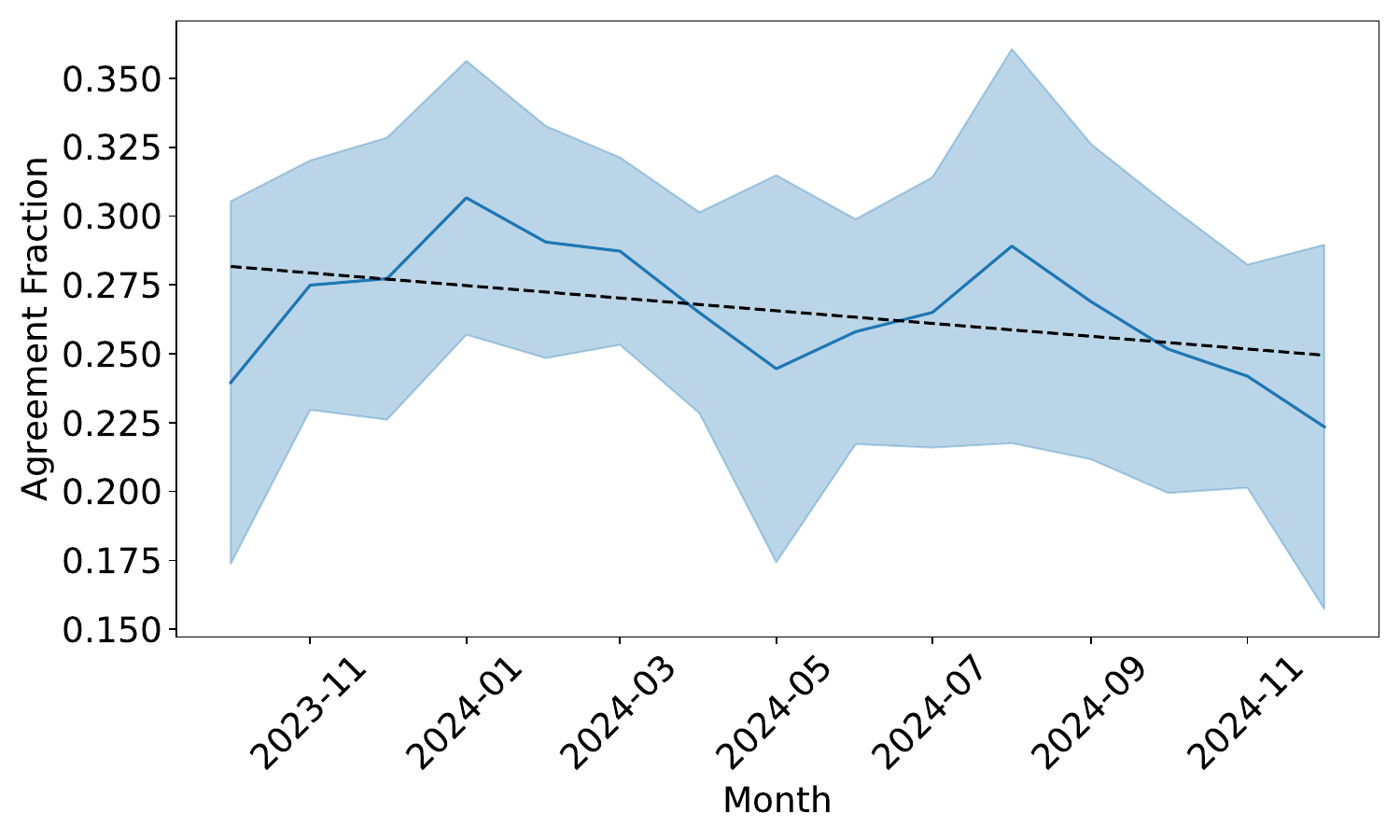}}
\end{minipage}%
\begin{minipage}{.24\linewidth}
\centering
\subfloat[Laura Ingraham (Fox) $^{***}$]{\label{}\includegraphics[width=\textwidth]{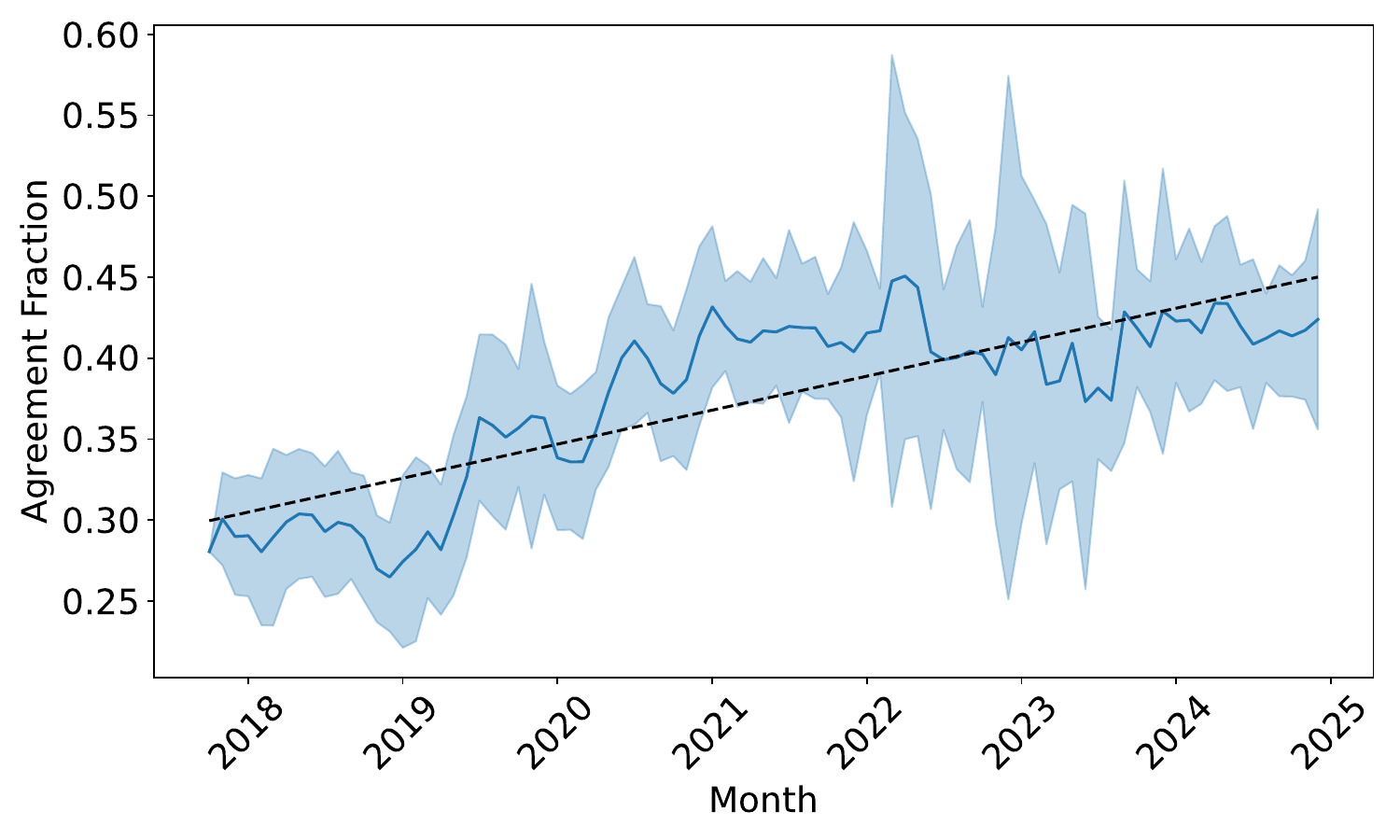}}
\end{minipage}%
\begin{minipage}{.24\linewidth}
\centering
\subfloat[Life Liberty Levin (Fox) $^{***}$]{\label{}\includegraphics[width=\textwidth]{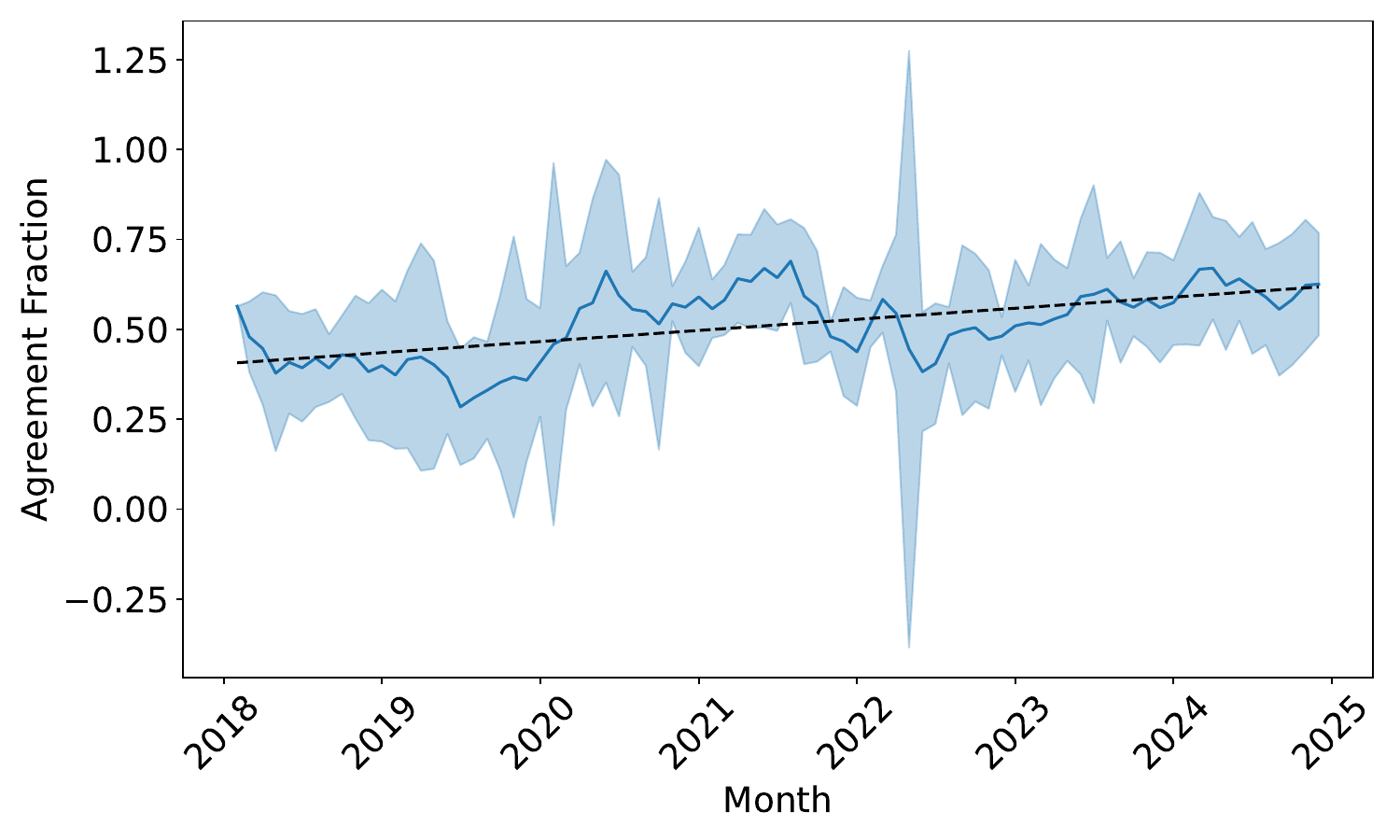}}
\end{minipage}%
\begin{minipage}{.24\linewidth}
\centering
\subfloat[Outnumbered (Fox) $^{***}$]{\label{}\includegraphics[width=\textwidth]{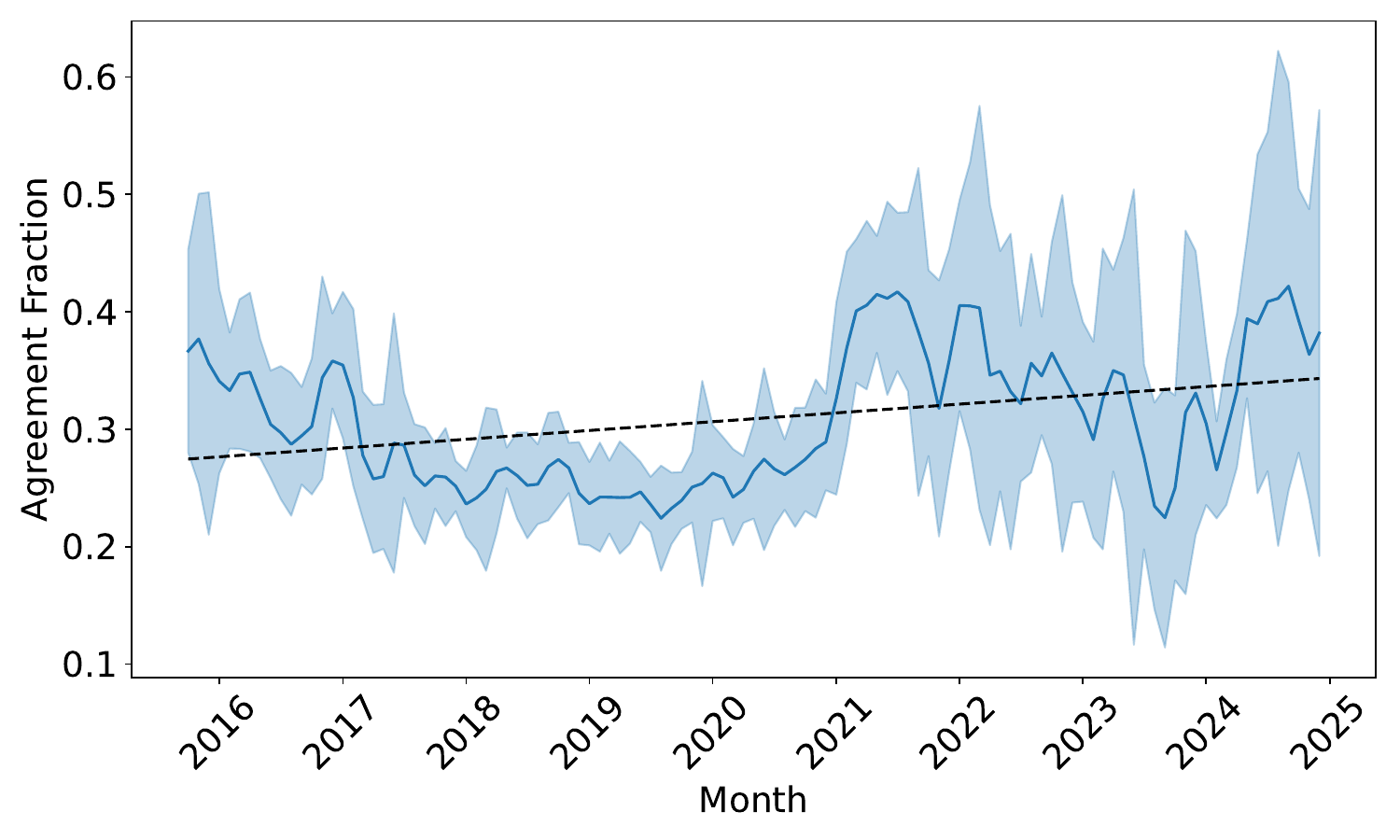}}
\end{minipage}%
\par\medskip

\begin{minipage}{.24\linewidth}
\centering
\subfloat[Rachel Maddow (MSNBC)]{\label{}\includegraphics[width=\textwidth]{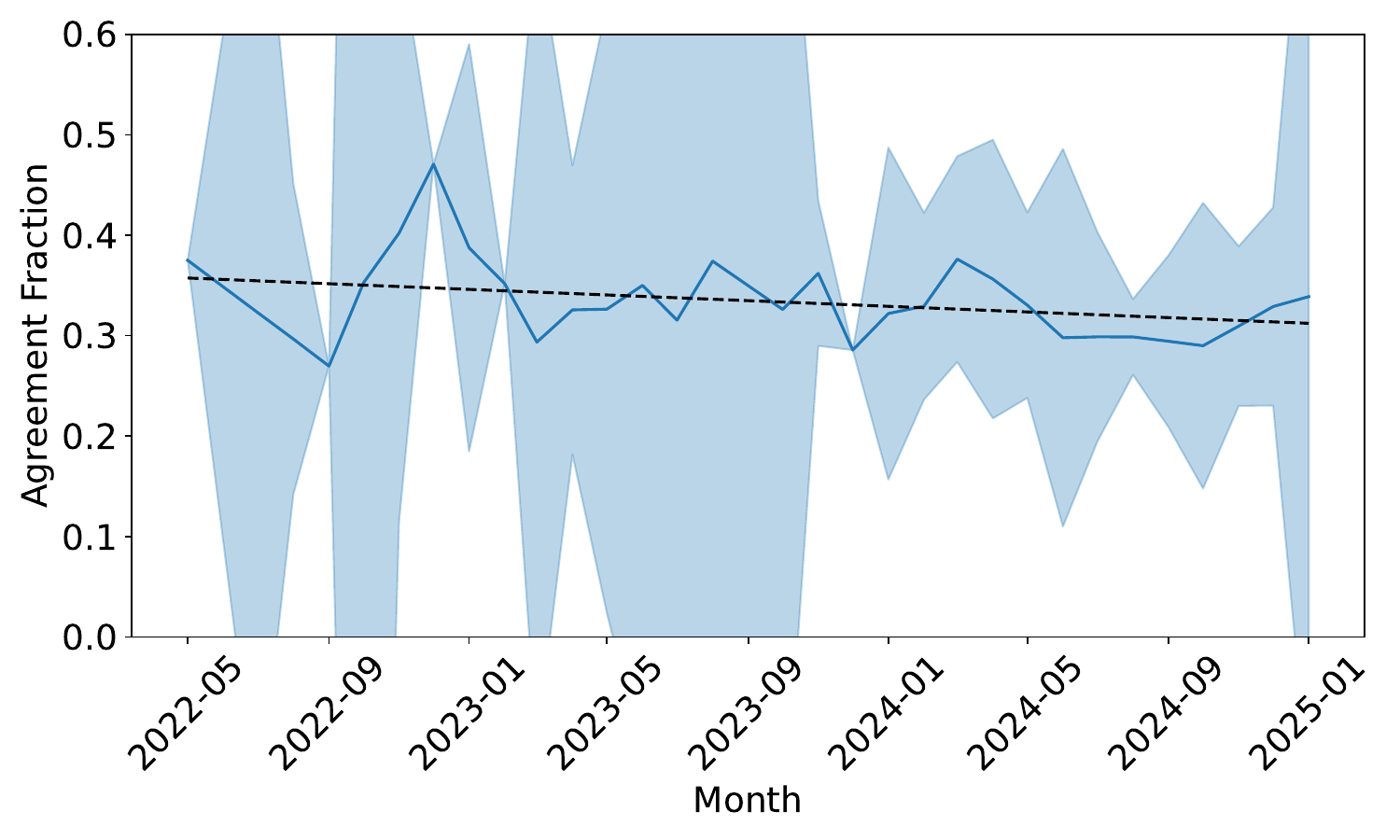}}
\end{minipage}%
\begin{minipage}{.24\linewidth}
\centering
\subfloat[Saturdays Sundays with Jonathan Capehart (MSNBC)]{\label{}\includegraphics[width=\textwidth]{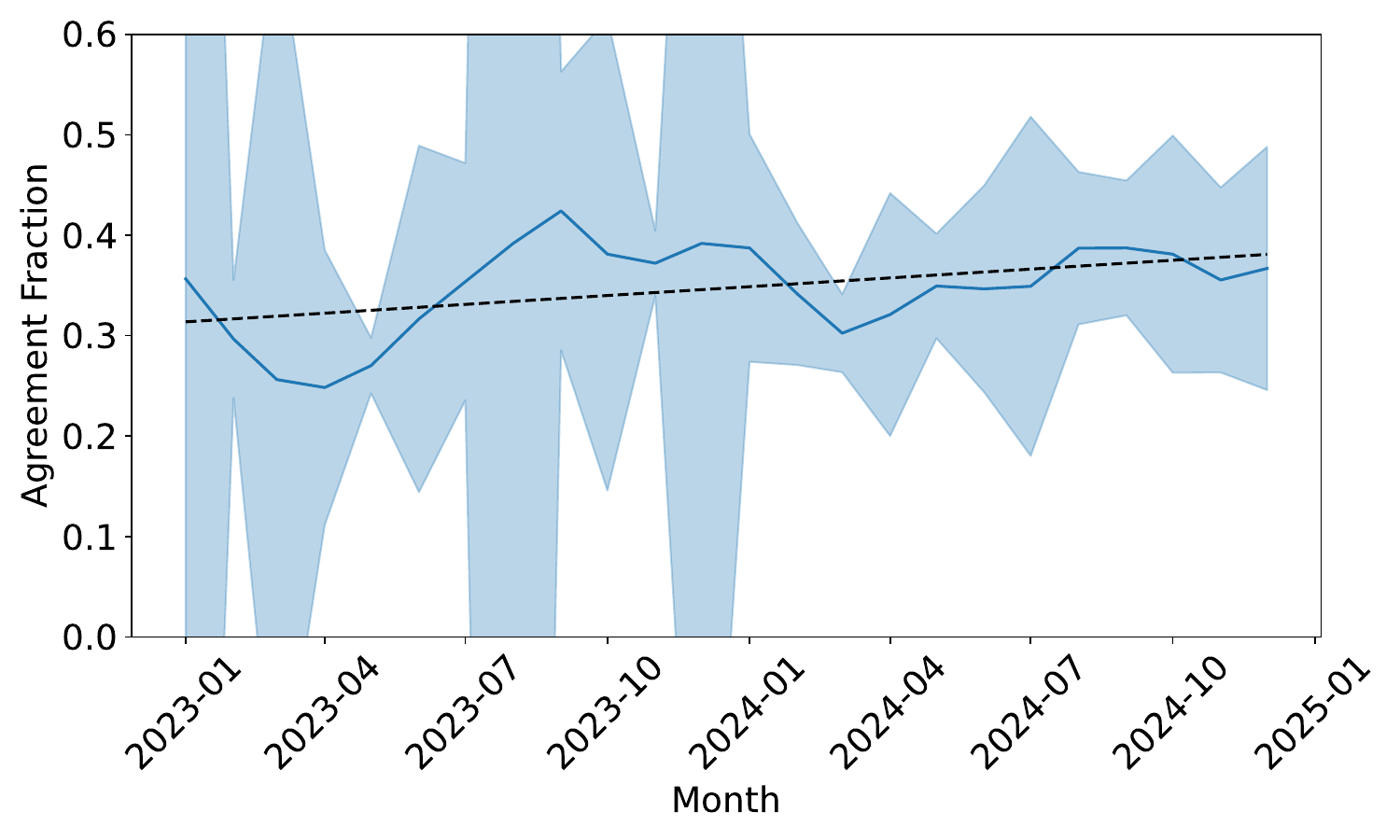}}
\end{minipage}%
\begin{minipage}{.24\linewidth}
\centering
\subfloat[Special Report with Bret Baier (Fox)]{\label{}\includegraphics[width=\textwidth]{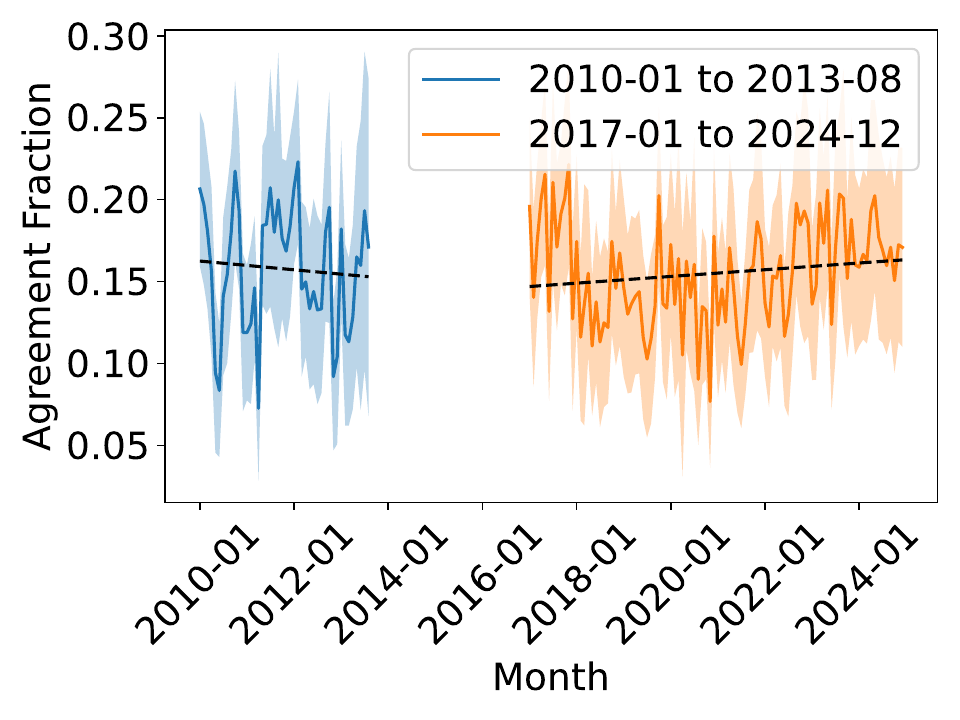}}
\end{minipage}%
\begin{minipage}{.24\linewidth}
\centering
\subfloat[State of the Union (CNN)]{\label{}\includegraphics[width=\textwidth]{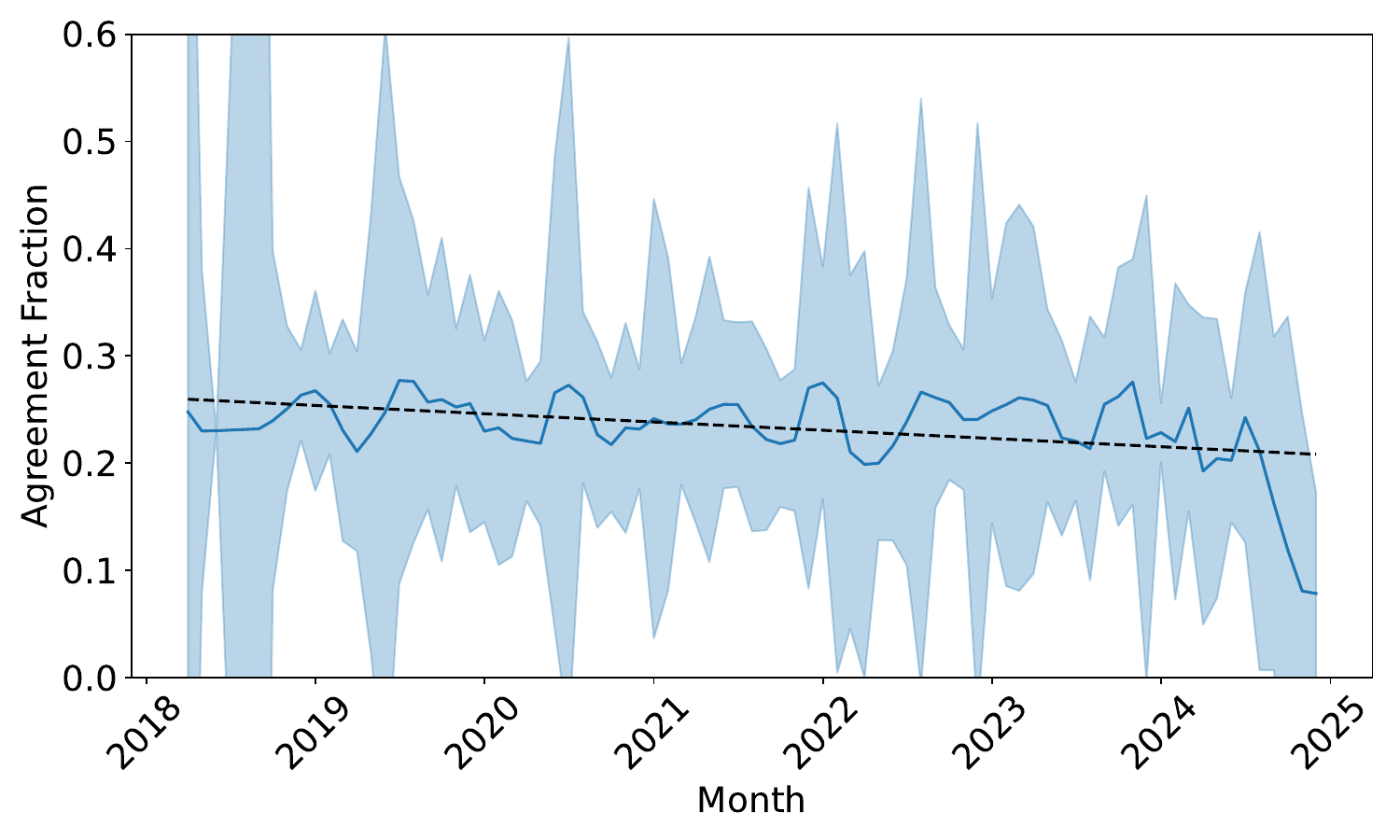}}
\end{minipage}%
\par\medskip

\begin{minipage}{.24\linewidth}
\centering
\subfloat[The Beat with Ari Melber (MSNBC)]{\label{}\includegraphics[width=\textwidth]{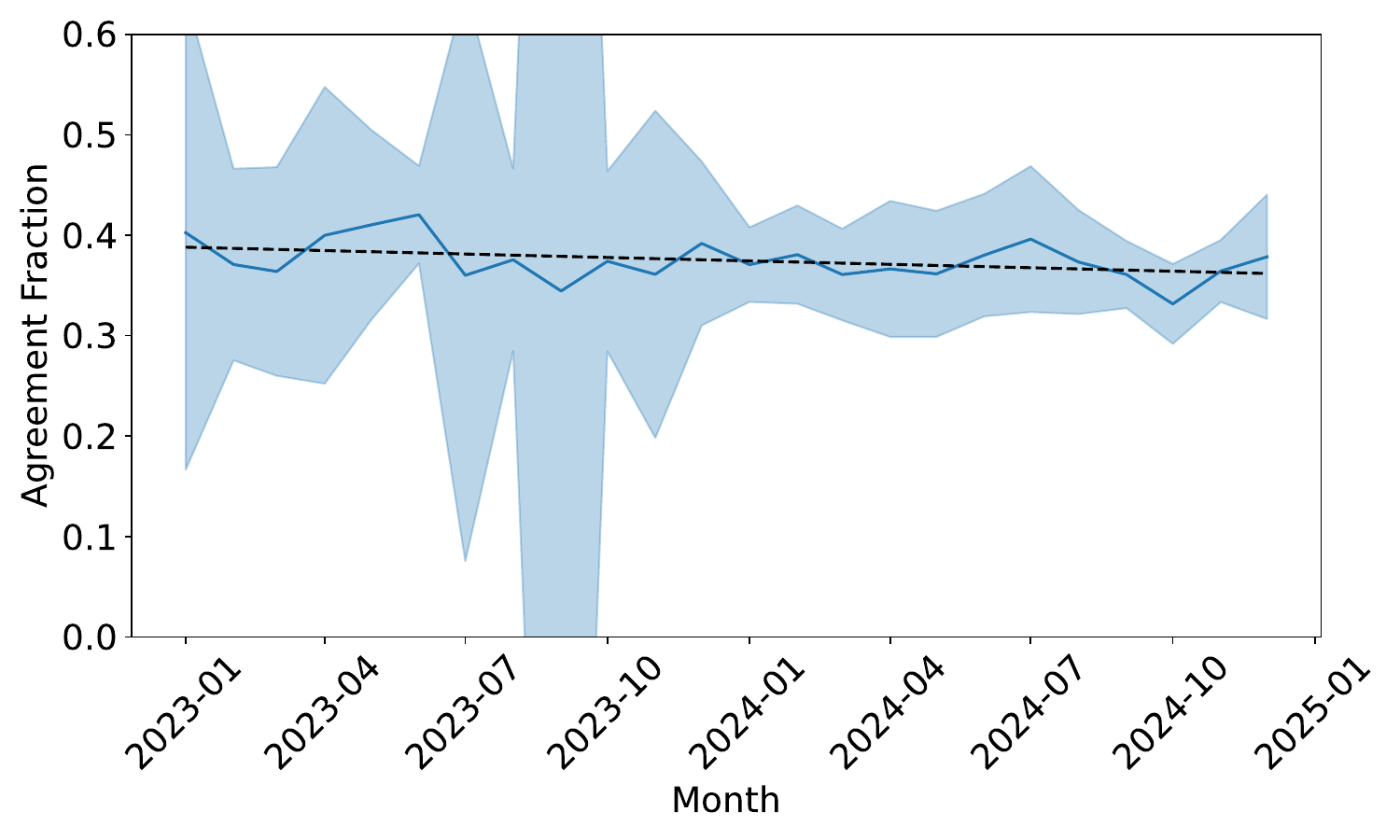}}
\end{minipage}%
\begin{minipage}{.24\linewidth}
\centering
\subfloat[The Lead with Jake Tapper (CNN)]{\label{}\includegraphics[width=\textwidth]{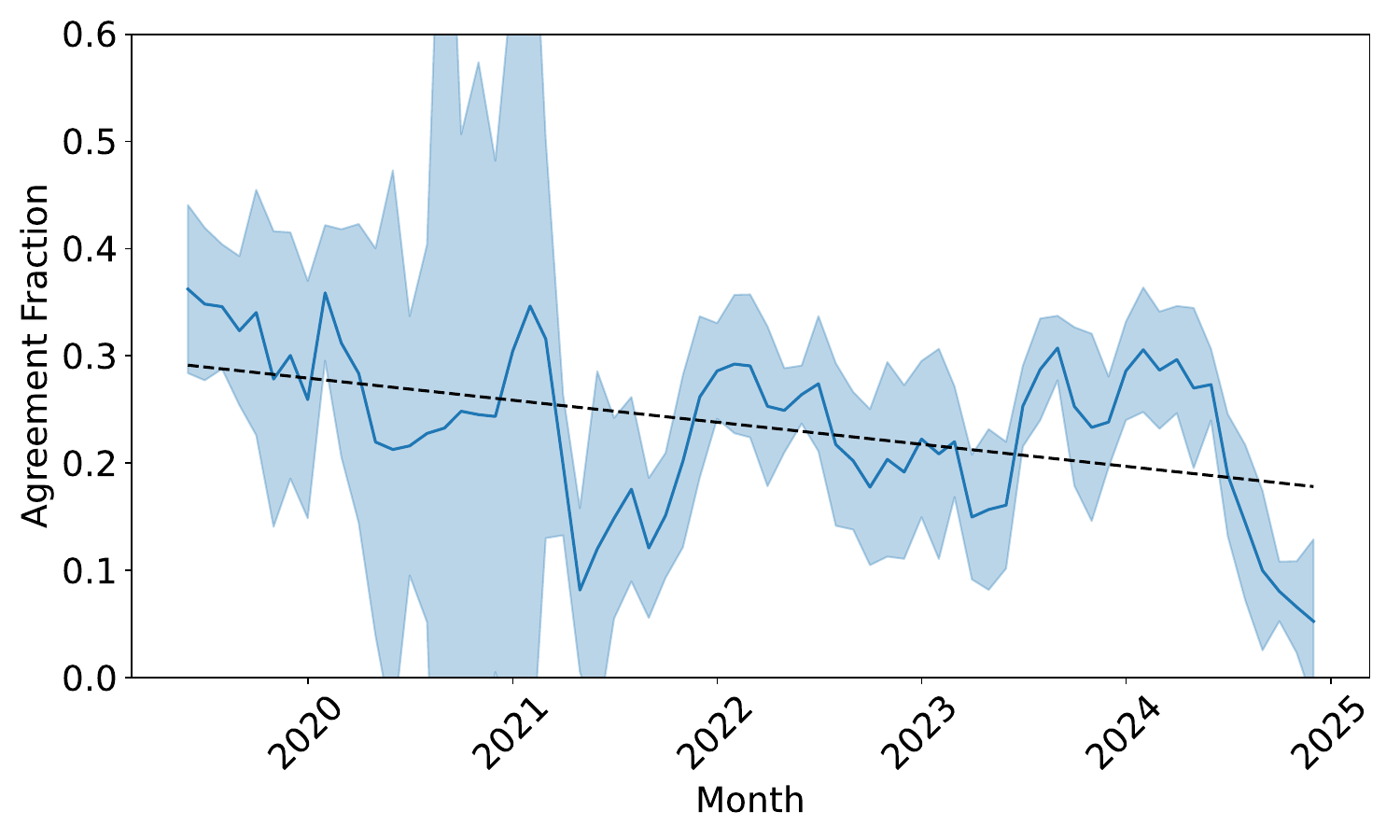}}
\end{minipage}%
\begin{minipage}{.24\linewidth}
\centering
\subfloat[The Savage Nation (MSNBC) $^{***}$]{\label{}\includegraphics[width=\textwidth]{img/agreement_timeseries/The_Savage_Nation_agreement_monthly.pdf}}
\end{minipage}%
\begin{minipage}{.24\linewidth}
\centering
\subfloat[The Source with Kaitlin Collins (CNN)]{\label{}\includegraphics[width=\textwidth]{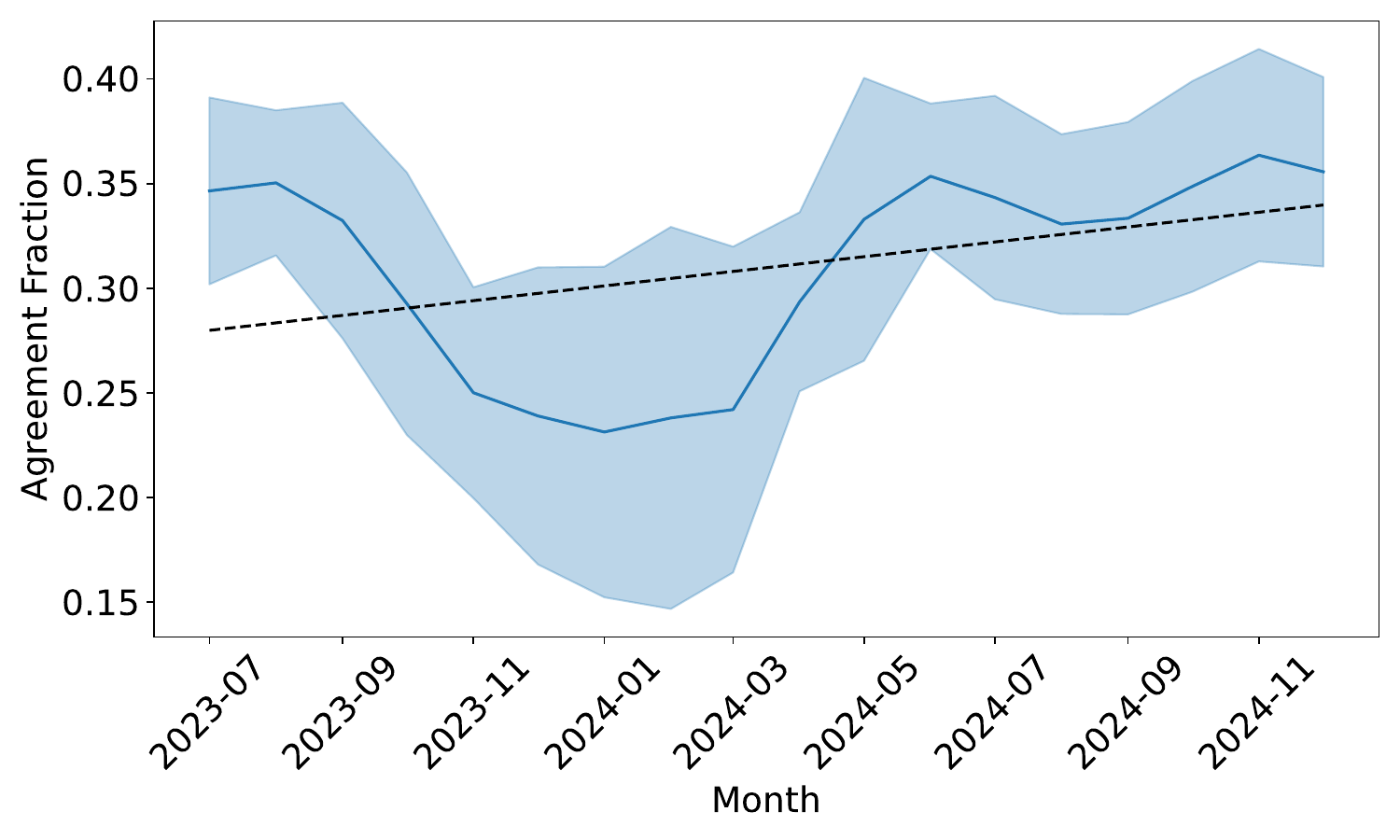}}
\end{minipage}%
\par\medskip

\begin{minipage}{.24\linewidth}
\centering
\subfloat[The Story with Martha MacCallum (Fox)]{\label{}\includegraphics[width=\textwidth]{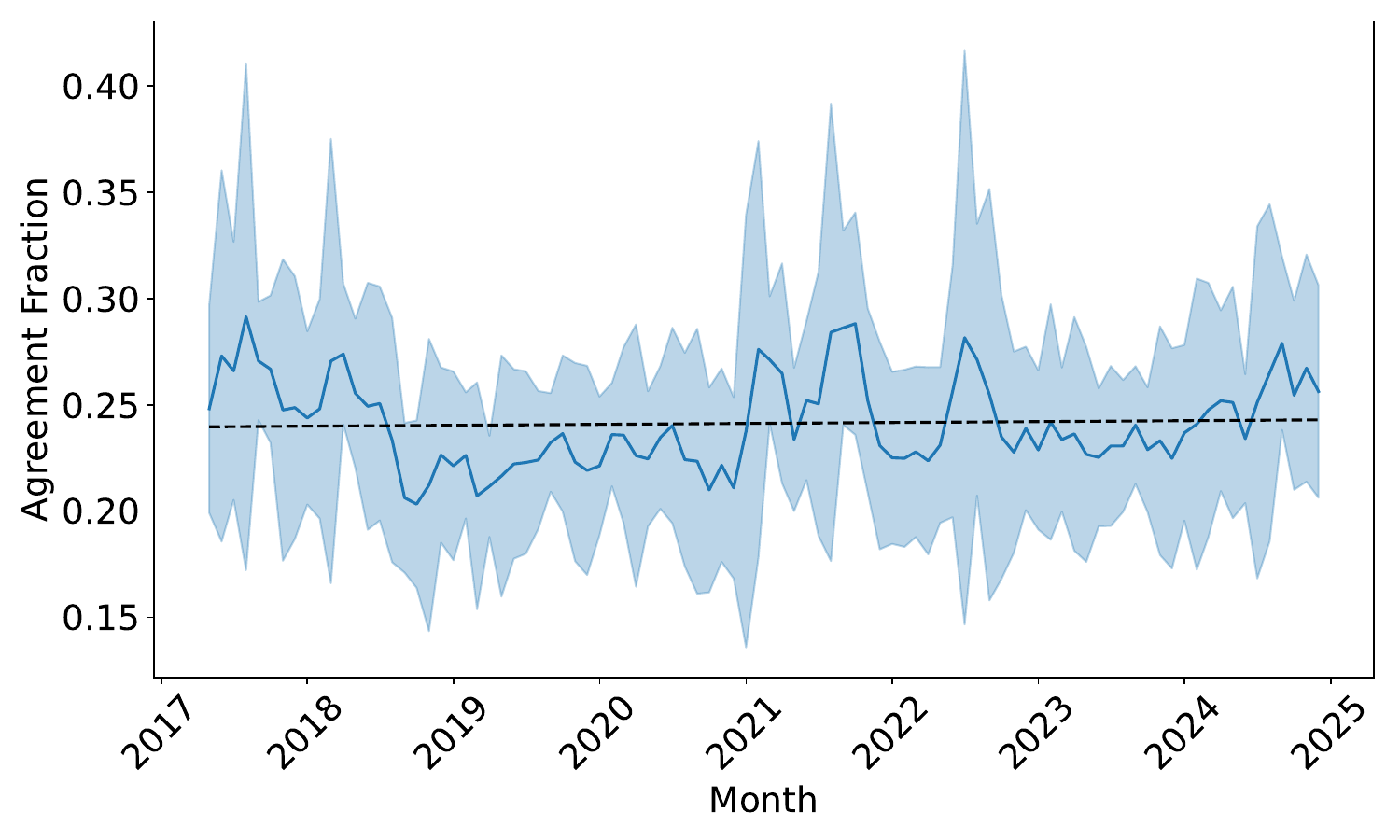}}
\end{minipage}%
\begin{minipage}{.24\linewidth}
\centering
\subfloat[Tucker Carlson (Fox) $^{***}$]{\label{}\includegraphics[width=\textwidth]{img/agreement_timeseries/Tucker_agreement_monthly.pdf}}
\end{minipage}%
\begin{minipage}{.24\linewidth}
\centering
\subfloat[Velshi (MSNBC)]{\label{Velshi}\includegraphics[width=\textwidth]{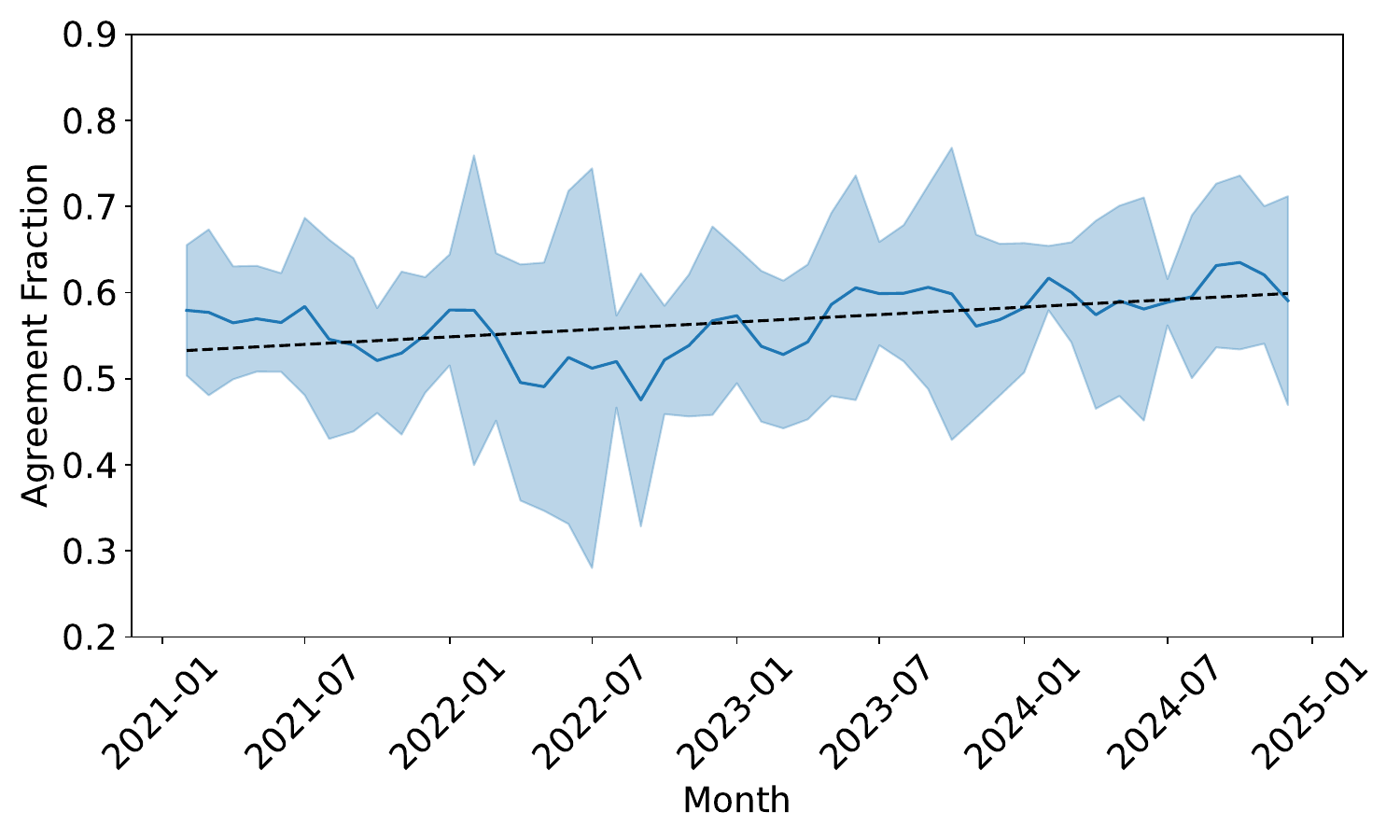}}
\end{minipage}%
\begin{minipage}{.24\linewidth}
\centering
\subfloat[Inside with Jen Psaki (MSNBC) $^{***}$]{\label{}\includegraphics[width=\textwidth]{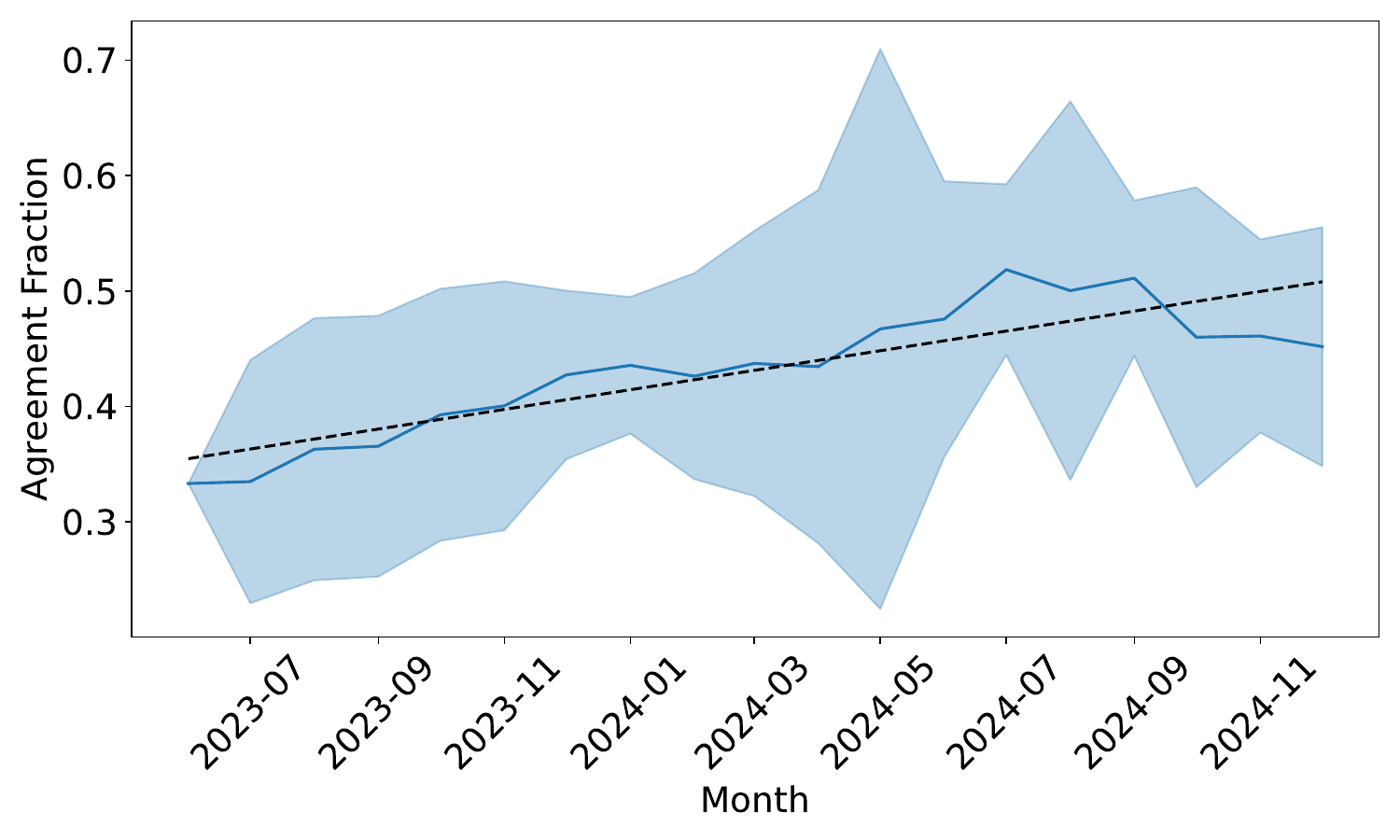}}
\end{minipage}%
\par\medskip

\caption{Agreement patterns over time for all shows.  $^{***}$ indicates a significant increase ($p < 0.001$) in slope.}
\label{fig:agreement_over_time_all_shows}
\end{figure*}

Figures~\ref{fig:disagreement_over_time_all_shows} and~\ref{fig:agreement_over_time_all_shows} show the disagreement and agreement plots for all shows, respectively. 

\begin{table*}[ht]
\centering
\begin{center}
\begin{tabular}{lrrr}
\toprule
Occupation & Disagree Count & Total Pairs & Disagreement Rate \\
\midrule
Entertainer & 393 & 1,929 & 0.204 \\
Manager & 552 & 2,752 & 0.201 \\
Member of congress & 1,598 & 8,036 & 0.199 \\
Rancher & 170 & 855 & 0.199 \\
Insurance & 132 & 689 & 0.192 \\
Director & 1,059 & 5,588 & 0.190 \\
Therapist & 231 & 1,221 & 0.189 \\
Disabled & 301 & 1,597 & 0.188 \\
Senator & 1,599 & 8,590 & 0.186 \\
Airline pilot & 124 & 687 & 0.180 \\
Artist & 547 & 3,041 & 0.180 \\
Principal & 348 & 1,952 & 0.178 \\
Architect & 326 & 1,884 & 0.173 \\
Lawyer/attorney & 6,431 & 37,245 & 0.173 \\
Farmer & 276 & 1,601 & 0.172 \\
Homemaker & 424 & 2,488 & 0.170 \\
Banker & 249 & 1,462 & 0.170 \\
Pastor & 151 & 894 & 0.169 \\
Retired & 5,212 & 30,897 & 0.169 \\
Not employed & 4,992 & 29,599 & 0.169 \\
Consultant & 1,680 & 10,011 & 0.168 \\
Teacher & 605 & 3,611 & 0.168 \\
Legislator & 269 & 1,607 & 0.167 \\
Executive & 1,425 & 8,517 & 0.167 \\
CEO/president & 2,906 & 17,424 & 0.167 \\
Self employed & 683 & 4,116 & 0.166 \\
Author & 140 & 847 & 0.165 \\
Chairman & 464 & 2,809 & 0.165 \\
Engineer & 633 & 3,848 & 0.165 \\
Political scientist & 98 & 598 & 0.164 \\
Contractor & 209 & 1,280 & 0.163 \\
Public relations & 135 & 828 & 0.163 \\
Physician & 909 & 5,585 & 0.163 \\
Accountant & 152 & 955 & 0.159 \\
Firefighter & 224 & 1,410 & 0.159 \\
Partner & 381 & 2,426 & 0.157 \\
Economist & 181 & 1,154 & 0.157 \\
Professor/writer & 1,667 & 10,659 & 0.156 \\
Investor & 147 & 946 & 0.155 \\
Psychologist & 118 & 764 & 0.154 \\
Real estate & 574 & 3,727 & 0.154 \\
Educator & 172 & 1,126 & 0.153 \\
Programmer & 122 & 802 & 0.152 \\
Analyst & 88 & 584 & 0.151 \\
Clerk & 106 & 711 & 0.149 \\
Administrator & 87 & 586 & 0.148 \\
Founder & 317 & 2,155 & 0.147 \\
Producer & 112 & 779 & 0.144 \\
Finance & 134 & 935 & 0.143 \\
Student & 274 & 1,916 & 0.143 \\
Research scientist & 143 & 1,030 & 0.139 \\
Designer & 104 & 833 & 0.125 \\
Unemployed & 125 & 1,042 & 0.120 \\
\bottomrule
\end{tabular}
\caption{Disagreement rates by occupation}
\label{tab:occupation_disagreement}
\end{center}
\end{table*}

\end{document}